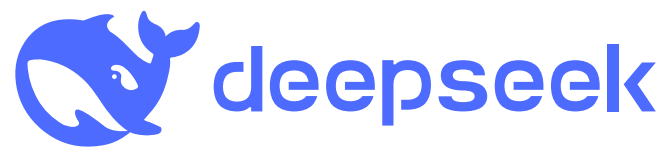

# A Programming Paradigm for Spatiotemporal Composability

Yifan Shi[1,2], Wei Zhang[1], Tianyi Cui[2]

[1]**Peking University** [2]**DeepSeek-AI**

## Abstract

Modern software—from plugin systems to self-evolving agent harnesses—increasingly requires *dynamic composition*, yet its formal foundations remain underdeveloped. We identify two orthogonal dimensions of the problem: *temporal composability*, the ability to completely revert a component's side effects upon removal, and *spatial composability*, the ability to declare and reactively manage inter-component dependencies. We address the two dimensions by lifting classical effect and coeffect concepts to runtime mechanisms. In particular, we formalize *revertible effects*, in which every context transformation carries an inverse that the runtime holds, establishing temporal composability local to one component. We formalize *reactive coeffects*, in which every context change is classified against a component's coeffect specification to drive its activation and deactivation, establishing spatial composability local to one component. We then unify the effect context and the coeffect context into a single context type and mediate every effect and coeffect through it, yielding a discipline we call the *context paradigm*; the mediation induces an observational equivalence up to which the effects of distinct components interleave without disturbing one another. Combining these mechanisms into the notion of a *component*, we give a calculus of dynamic composition whose metatheory carries spatiotemporal composability from a single component to a whole system of interleaved components. We implement these ideas in *Cordis*, a meta-framework of spatiotemporal composability that provides a core library with effect tracking and coeffect resolution, as well as a declarative component loader with configuration reconciliation and hot module replacement.

## Contents

# 1. Introduction

Composition—assembling complex systems from simpler parts—is a foundational principle of software engineering [1]. Traditionally, composition is *static*: function calls, module imports, and class inheritance are resolved at compile time and remain fixed throughout execution. However, modern software increasingly demands *dynamic* composition, where components are loaded, unloaded, and reconfigured at runtime. Plugin architectures [2] and self-evolving agent harnesses both require systems that can safely add and remove functionality on the fly, yet current practice defers to coarse-grained mechanisms [3] that reconfigure only by restarting, discarding runtime state. Despite the growing practical importance of dynamic composition, its theoretical foundations remain underdeveloped, compared to the rich formal frameworks available for static composition.

## 1.1. Dimensions of Composability

To characterize the requirements of dynamic composition, we identify two orthogonal dimensions beyond the well-studied algebraic aspects of composition:

- **Temporal composability** addresses the *time* dimension: upon removal of a component, the modifications the component made to the shared environment must be completely and safely reversed. This requires tracking every resource allocation, event registration, and state mutation the component performs, and guaranteeing their orderly reclamation upon removal.
- **Spatial composability** addresses the *space* dimension: components must be able to declare, discover, and resolve their dependencies on one another in a structured and verifiable manner. This requires managing dependency topology and coordinating component lifecycles in response to dependency changes.

In the static setting, temporal composability reduces to lexical scoping (e.g., RAII [4], bracket patterns [5]), and spatial composability reduces to module import resolution [6]. In the dynamic setting, where components arrive and depart at runtime, both dimensions become significantly harder: temporal composability must handle long-lived, stateful effects whose scope is not lexically bounded; and spatial composability must handle dependencies that appear, disappear, or change identity during execution.

## 1.2. Motivating Examples

### *1.2.1. Plugin Systems*

Plugin systems are a canonical instance of dynamic composition. We use Visual Studio Code (VSCode), one of the most widely-used extensible IDEs, as a representative example.

**Temporal limitation.** VSCode runs all extensions in a shared process called the extension host. Although extensions can be installed dynamically, this host provides no mechanism to unload an individual extension's code at runtime. Once an extension's `activate` function has executed, disabling or uninstalling it requires restarting the entire host, affecting all loaded extensions. Purely declarative extensions such as themes, keybindings, and snippets carry no code and can be removed freely. Among the top 100 extensions by install count, however, 87 contain executable code[1] and will therefore require such a restart upon removal. Although VSCode provides a `deactivate` hook, it serves only as a graceful shutdown callback during the host process' termination, and thus does not enable live removal. Moreover, the hook separates effect disposal from effect creation (in `activate`), violating locality of concern and making complete cleanup difficult to verify.

**Spatial limitation.** VSCode does provide `extensionDependencies` for declaring dependencies between extensions, but it sees little use: among the top 100 extensions by install count, only 7 declare `extensionDependencies` on non-built-in extensions.[1] This scarcity reflects the shape of the extension API, which exposes fixed, surface-level extension points such as commands, views, and language features. Extensions contribute to the host through these points rather than depending on one another, so inter-extension dependencies rarely arise. Moreover, VSCode's mechanism for inter-extension interaction provides no structural contract: it exposes an extension's functionality to others through `vscode.extensions.getExtension(...).exports`, but the returned value is untyped (`any` by default), so the dependent cannot rely on a checked interface. In short, VSCode steers extensions toward a fixed set of host-provided extension points, and offers no safe, structured way for them to depend on one another.

These two limitations are not unique to VSCode; they recur across plugin systems generally [2, 7], differing only in degree.

#### 1.2.2. *Self-Evolving Agent Harnesses*

Modern AI agents rely on runtime agent harnesses [8–10]. These systems may compose diverse tool suites [11] and execution environments, govern permissions and sandboxing, maintain session state and persistence, provide context management and memory systems [12], orchestrate subagents and multi-agent workflows [13], and expose interfaces to users and automation. A future harness may generate and deploy modifications to its own components while continuously serving requests. Model-synthesized reusable tools provide a narrower precursor to component-level self-modification [14]. Each such modification is itself an instance of dynamic composition.

Because these modifications occur continuously and with limited or no human oversight, dynamic composability becomes indispensable. Without temporal composability, each self-modification forces a full restart that discards all process-local accumulated state; at such frequency the cumulative unavailability becomes substantial, and in-flight tasks are disrupted repeatedly; even worse, a faulty self-modification can disable the very process needed to recover. Without spatial composability, each module must itself detect and adapt to changes in the modules it depends on as they appear, disappear, or change identity, and can do so only by ad hoc means; even worse, a naive code-replacement strategy may silently break dependents or introduce circular dependencies that surface only at reload time.

#### 1.2.3. *The Coarse-Grained Workaround*

One reason dynamic composability has received limited formal attention is that operating systems and container orchestrators already provide a coarse-grained substitute. Operating systems yield temporal composability at the granularity of a process; container orchestrators

[1] Data retrieved from the Visual Studio Code Marketplace on June 9, 2026.

[3] yield spatial composability at the granularity of a service. In practice, most software tolerates the lack of fine-grained composability by deferring to these coarse-grained mechanisms: a misbehaving module is handled by restarting the process, and a service dependency is managed by the container orchestrator.

However, this workaround imposes substantial costs. Temporally, each restart discards all process-local accumulated state (e.g., caches, connections, partial computations), and rebuilding it takes seconds to minutes [15]; maintaining availability in the interim requires redundant replicas, incurring resource overhead to compensate for the inability to recover a single component. Spatially, container-level orchestration cannot express dependencies between components sharing an address space, and introduces network overhead for interactions that could be local function calls. Both mechanisms operate at the boundary of processes and containers, yet modern systems increasingly compose at a finer level. This granularity mismatch demands a compositional abstraction that manages effects and dependencies at the same level as the components themselves.

### 1.3. Contributions

The two dimensions of dynamic composability concern, respectively, how computations *modify* and how they *depend on* their environment. These two directions are what *effect systems* [16, 17] and *coeffect systems* [18, 19] formalize: effects provide the formal vocabulary for reasoning about environmental modifications, and coeffects for reasoning about environmental requirements. However, existing formulations restrict reasoning to compile-time analysis over lexically fixed scopes, and do not extend to dynamic scenarios where components arrive and depart at runtime. By lifting effects to a revertible runtime model and coeffects to a reactive dependency resolution mechanism, we obtain a unified formal foundation for dynamic composability, one that is language-agnostic and applicable to any software architecture requiring dynamic composition. We make the following contributions:

1. We formalize **revertible effects** (Section 3.1): every context transformation carries an explicit inverse that the runtime holds, and both tracking and recovery preserve composition, so the context is recovered upon component removal. This establishes local temporal composability.
2. We formalize **reactive coeffects** (Section 3.2): a component declares the coeffects it requires as a specification, and each change of the context is classified against that specification as activating, deactivating, or neutral, driving the component's activation and deactivation. This establishes local spatial composability.
3. We introduce the **context paradigm** (Section 3.3): the effect context and the coeffect context are unified into a single context type, every effect and coeffect is mediated through it, and the mediation induces an observational equivalence up to which the effects of distinct components attain independence.
4. We develop a calculus of dynamic composition (Section 4), which combines the two mechanisms into the notion of a **component** and gives them an operational semantics. The metatheory then carries spatiotemporal composability from a single component to a whole system of interleaved components.
5. We implement these ideas in **Cordis** (Section 5), a meta-framework of spatiotemporal composability that provides a core library realizing the formal model with effect tracking and coeffect resolution, as well as a declarative component loader with configuration reconciliation and hot module replacement.

## 2. Preliminaries

This section provides a concise overview of effect and coeffect systems—the two theoretical pillars underlying our work. We assume familiarity with basic type theory and category theory; the goal here is to fix notation and introduce the key abstractions that Section 3 will operationalize as runtime mechanisms.

### 2.1. Effects

In the simply typed lambda calculus (STLC) [20, 21], a typing judgment $\Gamma \vdash t : T$ states that term $t$ has type $T$ under context $\Gamma$. An *effect system* refines the type to describe what side effects a computation may produce, yielding judgments of the form

$$\Gamma \vdash t : T_{\text{effect}} \tag{1}$$

Here, the result type is annotated with an element of an effect algebra that describes which side effects the computation may produce, enabling compositional reasoning about stateful computations. This approach originates with Lucassen and Gifford [22], who introduced a kinded type system distinguishing types, effects, and regions to discover scheduling constraints in parallel programs.

**Monadic effects.** Moggi [16] first modeled computational effects categorically via monads; Wadler [23] popularized the approach in Haskell. A monad $(T, \eta, \mu)$ on a category $\mathcal{C}$ encapsulates an effectful computation as a value of type $T(A)$, with $\eta : A \to T(A)$ lifting pure values and $\mu : T(T(A)) \to T(A)$ sequencing nested computations. Classic instances include the Maybe monad (for partiality), State monad (for mutable state), and IO monad (for external interaction).

**Algebraic effects.** Plotkin and Power [17, 24] showed that algebraic operations determine monads, establishing a framework in which effect interfaces are decoupled from their implementations. An effect signature $\Sigma$ declares a set of operations (e.g., $\text{get} : () \to S$, $\text{put} : S \to ()$ for state); programs invoke operations freely without committing to a particular interpretation. Plotkin and Pretnar [25] subsequently introduced *effect handlers*, which interpret operations by providing continuation semantics:

$$\text{handle } e \text{ with } \{ \ \text{op}(v, \kappa) \mapsto \ldots \ \} \tag{2}$$

The handler receives the operation argument $v$ and the delimited continuation $\kappa$, which it may invoke zero, one, or multiple times, enabling exceptions, coroutines, and non-determinism within a uniform framework [26]. Languages such as Koka [27, 28], Eff [29], and OCaml 5 [30] have adopted algebraic effects with varying design trade-offs.

### 2.2. Coeffects

Dually to effects, a *coeffect system* [18, 31] enriches the context rather than the type, yielding judgments of the form

$$\Gamma_{\text{coeffect}} \vdash t : T \tag{3}$$

Here, the context is annotated with an element of a coeffect algebra describing what the computation requires from its environment, such as resources to access, permissions to hold, or services to depend on. While effects model a program's impact on the world, coeffects model the world's constraints on the program.

**Comonadic coeffects.** The idea of using comonads to structure context-dependent computation was first developed by Uustalu and Vene [32], who proposed symmetric (semi)monoidal comonads as the dual of Moggi's monadic framework for effects, capturing notions such as dataflow and attribute evaluation. Petricek et al. [18] built on this foundation to propose coeffects as a unified static analysis of context-dependence. A comonad $(D, \varepsilon, \delta)$ captures context-dependent computation: $\varepsilon : D(A) \to A$ extracts the current value from a context, and $\delta : D(A) \to D(D(A))$ duplicates context for nested access. The Environment comonad $D(X) = E \times X$ models dependence on a fixed environment $E$; the Stream comonad $D(X) = \mathbb{N} \to X$ models dependence on temporal data.

**Graded coeffects.** For finer-grained tracking, *graded* coeffect systems use a pre-ordered semiring $\mathcal{S} = (S, \leq, +, \times, 0, 1)$ as the coeffect algebra [33], a discipline later unified with graded effects by Gaboardi et al. [19]. Elements of $S$ annotate each variable binding to quantify its usage: 0 for unused, 1 for linear use, $n$ for bounded use, $\infty$ for unrestricted use. The semiring operations compose coeffects sequentially $(\times)$ and in parallel $(+)$, enabling precise resource tracking, sensitivity analysis [34], and information-flow control [35, 36] within a unified algebraic framework [37].

### 2.3. Relationship to Dynamic Composability

Effect and coeffect systems organize reasoning about computation along two complementary directions: effects describe how a computation *modifies* its environment, whereas coeffects describe how it *depends on* its environment. These two directions correspond to the two dimensions of dynamic composability identified in Section 1:

- **Temporal composability** demands that a component's modifications to the shared environment be revertible upon unloading. The relevant effects are the stateful ones, which durably transform that environment; undoing such a transformation requires it to admit an inverse.
- **Spatial composability** demands that inter-component dependencies be declared and managed reactively. Such dependencies are the very thing coeffects capture, and managing them amounts to resolving each against what the environment supplies.

However, classical effect and coeffect systems are static instruments: effects are tracked within lexically fixed scopes and discharged by compile-time handlers; coeffect annotations are verified against contexts determined before execution. Dynamic composition, by contrast, requires these guarantees to hold for components that arrive and depart at runtime, against contexts that evolve continuously. No fixed lexical scope can delimit a plugin loaded after deployment; no compile-time context can anticipate dependencies that emerge from runtime configuration.

This motivates a shift in perspective: rather than extending static type systems with more annotations, we reify the conceptual structures of effects and coeffects so that a runtime can operate on them directly, establishing dynamically the guarantees these systems provide statically.

## 3. Revertible Effects and Reactive Coeffects

This section lifts the concepts of effects and coeffects introduced in Section 2 to runtime mechanisms, constructing a theory of dynamic composition. The central idea is to turn the *typing contexts* carrying effects and coeffects into *context types*, runtime-operable types that reify the context as a first-class entity. Section 3.1 models an effect as a context transformation paired with an inverse that the runtime holds, establishing temporal composability local to one component; Section 3.2 models a coeffect as a declared dependency against which every context change is classified, establishing its spatial counterpart. Each local guarantee stops where other components enter. Toward the global form of both, Section 3.3 unifies the two contexts into one and introduces the *context paradigm*: every effect and coeffect is mediated through the unified context, and the mediation induces the observational equivalence up to which every later equality is read. Section 3.4 then establishes effect independence and coeffect commutativity, under which the effects of distinct components interleave without disturbing one another.

### 3.1. Revertible Effects

*Temporal composability* is the ability to load and unload components at runtime such that, upon unloading, the shared environment is recovered to its pre-composition state. This requires that every modification a component makes to the environment be both trackable and recoverable. We therefore model an effect as a function of type $\Gamma \to \Gamma \times (\Gamma \to \Gamma)$: applied to the current context, it yields the modified context together with an explicit inverse. Supplying that inverse is what lets the effect be reverted, and returning it to the runtime is what makes the effect trackable. We call such effects *revertible*: by composing these inverses during execution, local temporal composability becomes a structural guarantee.

#### *3.1.1. Effect Context*

Given any impure function $f : X \rightsquigarrow Y$, we transform it into a pure form $f : \Gamma \times X \to \Gamma \times Y$, where $\Gamma$ is the context type. On this pure form, all possible side effects can be represented as transformations on $\Gamma$: for any fixed input $x : X$, the induced map $\gamma \mapsto \mathrm{pr}_1(f(\gamma, x)) : \Gamma \to \Gamma$ captures the side effect of $f$ independently of the return value. Effects on $\Gamma$ therefore live in the monoid of transformations $\Gamma \to \Gamma$ under composition $\circ$, where each monoid axiom has a direct reading as a property of effects:

- *Closure*: the sequential composition of two effects is again an effect;
- *Associativity*: a composite effect is independent of how it is bracketed;
- *Identity*: $\mathrm{id}_\Gamma$, the identity function on $\Gamma$, acts as the unit of composition.

To model effects that can be undone, we pair each transformation $f$ with another transformation $g$ that undoes $f$, and call $g$ a *left inverse* of $f$, abbreviated to *inverse* throughout the paper. Undoing is one-sided: what an inverse is held to is $g \circ f$ and never $f \circ g$. Pairs of transformations carry a multiplication of their own:

**Definition 1.** Define the *twisted composition* of pairs of context transformations by

$$(f_1, g_1) \circ (f_2, g_2) := (f_1 \circ f_2,\ g_2 \circ g_1) \tag{4}$$

As for $\circ$ itself, the left operand acts after the right, and the inverses accumulate in the opposite order. It makes $(\Gamma \to \Gamma) \times (\Gamma \to \Gamma)$ a monoid with unit $(\mathrm{id}_\Gamma, \mathrm{id}_\Gamma)$, the product of the monoid of transformations with its opposite, which we call the twisted composition monoid $\mathfrak{T}_\Gamma$ over $\Gamma$.

To track effects within the context itself, we introduce the following definition:

**Definition 2.** Given a context $\Gamma$, define its *effect context* as:

$$\partial\Gamma := \Gamma \times (\Gamma \to \Gamma) \tag{5}$$

It can be understood as a pair $(\gamma, \varphi)$, where:
- $\gamma : \Gamma$ is the current context state;
- $\varphi : \Gamma \to \Gamma$ is the *accumulator*, the composite of the inverses of the effects performed so far, and the function that recovers the context to its initial state.

In particular, the initial effect context can be represented as $(\gamma_0, \mathrm{id}_\Gamma)$.

We also write $\partial^2\Gamma$ for $\partial(\partial\Gamma) = \partial\Gamma \times (\partial\Gamma \to \partial\Gamma)$; iterating $\partial$ this way yields the tower $\Gamma, \partial\Gamma, \partial^2\Gamma, \cdots$.

Given the presence of the accumulator $\varphi$, all effects performed on $\partial\Gamma$ can be tracked and the context can be recovered. We now give the concrete constructions for tracking and recovery.

**Definition 3.** Define the transformation $\mathrm{track}_\Gamma$ on pairs of context functions:

$$\begin{array}{lcccccc} \mathrm{track}_\Gamma & : & (\Gamma \to \Gamma) \times (\Gamma \to \Gamma) & \to & \partial\Gamma & \to & \partial\Gamma \\ \mathrm{track}_\Gamma & = & (f, g) & \mapsto & (\gamma, \varphi) & \mapsto & (f(\gamma), \varphi \circ g) \end{array} \tag{6}$$

This transformation converts a forward function $f$ together with a candidate inverse $g$ into a transformation of the effect context $\partial\Gamma$. Applying $\mathrm{track}_\Gamma(f, g)$ to a state $(\gamma, \varphi)$ transforms $\gamma$ by $f$ and composes the inverse $g$ onto $\varphi$, thereby tracking the effect of $f$ in the context.

**Theorem 4.** For every $(f, g) \in (\Gamma \to \Gamma) \times (\Gamma \to \Gamma)$, write $f' := \mathrm{track}_\Gamma(f, g)$; then the following diagram commutes, that is,

$$\mathrm{pr}_1 \circ f' = f \circ \mathrm{pr}_1 \tag{7}$$

$$\begin{array}{ccc} \Gamma & \xrightarrow{\;f\;} & \Gamma \\ {\scriptstyle \mathrm{pr}_1}\uparrow & \Big\Downarrow {\scriptstyle \mathrm{track}} & \uparrow{\scriptstyle \mathrm{pr}_1} \\ \partial\Gamma & \xrightarrow[\;f'\;]{} & \partial\Gamma \end{array}$$

*Proof.* For all $(\gamma, \varphi) \in \partial\Gamma$:

$$\begin{aligned} (\mathrm{pr}_1 \circ \mathrm{track}_\Gamma(f, g))(\gamma, \varphi) &= \mathrm{pr}_1(f(\gamma), \varphi \circ g) \\ &= f(\gamma) \\ &= (f \circ \mathrm{pr}_1)(\gamma, \varphi) \end{aligned}$$

□

Theorem 4 ensures that tracking leaves the forward behavior untouched: on the context state, $\mathrm{track}_\Gamma(f, g)$ acts as $f$ does, whatever candidate inverse it carries.

**Theorem 5.** $\mathrm{track}_\Gamma$ is a monoid homomorphism from $\mathfrak{T}_\Gamma$ into $\partial\Gamma \to \partial\Gamma$. That is,
1. $\mathrm{track}_\Gamma(\mathrm{id}_\Gamma, \mathrm{id}_\Gamma) = \mathrm{id}_{\partial\Gamma}$;
2. for all $(f_1, g_1), (f_2, g_2) \in \mathfrak{T}_\Gamma$,

$$\mathrm{track}_\Gamma((f_1, g_1) \circ (f_2, g_2)) = \mathrm{track}_\Gamma(f_1, g_1) \circ \mathrm{track}_\Gamma(f_2, g_2) \tag{8}$$

*Proof.*

1. The unit is carried to the unit, since $\mathrm{track}_\Gamma(\mathrm{id}_\Gamma, \mathrm{id}_\Gamma)(\gamma, \varphi) = (\gamma, \varphi \circ \mathrm{id}_\Gamma) = (\gamma, \varphi)$.
2. For the multiplication, take any $(\gamma, \varphi) \in \partial\Gamma$:

$$\begin{aligned}(\mathrm{track}_\Gamma(f_1, g_1) \circ \mathrm{track}_\Gamma(f_2, g_2))(\gamma, \varphi) &= \mathrm{track}_\Gamma(f_1, g_1)(f_2(\gamma), \varphi \circ g_2) \\ &= (f_1(f_2(\gamma)), \varphi \circ g_2 \circ g_1) \\ &= \mathrm{track}_\Gamma(f_1 \circ f_2, g_2 \circ g_1)(\gamma, \varphi)\end{aligned}$$

□

Theorem 5 ensures that tracking one pair at a time agrees with tracking their twisted composite at once, so a sequence of tracked effects can be reasoned about as a single tracked effect.

**Definition 6.** Define the transformation $\mathrm{recover}_\Gamma$ on $\partial\Gamma$:

$$\begin{aligned}\mathrm{recover}_\Gamma &: \partial\Gamma \to \partial\Gamma \\ \mathrm{recover}_\Gamma &= (\gamma, \varphi) \mapsto (\varphi(\gamma), \mathrm{id}_\Gamma)\end{aligned} \tag{9}$$

This transformation applies the recovery function $\varphi$ to the current state $\gamma$ and resets $\varphi$ to the identity. The following diagram illustrates how recover recovers the context to its initial state after a sequence of effects $f_i' := \mathrm{track}_\Gamma(f_i, g_i)$, $i = 1, \cdots, n$, has been applied to $\partial\Gamma$:

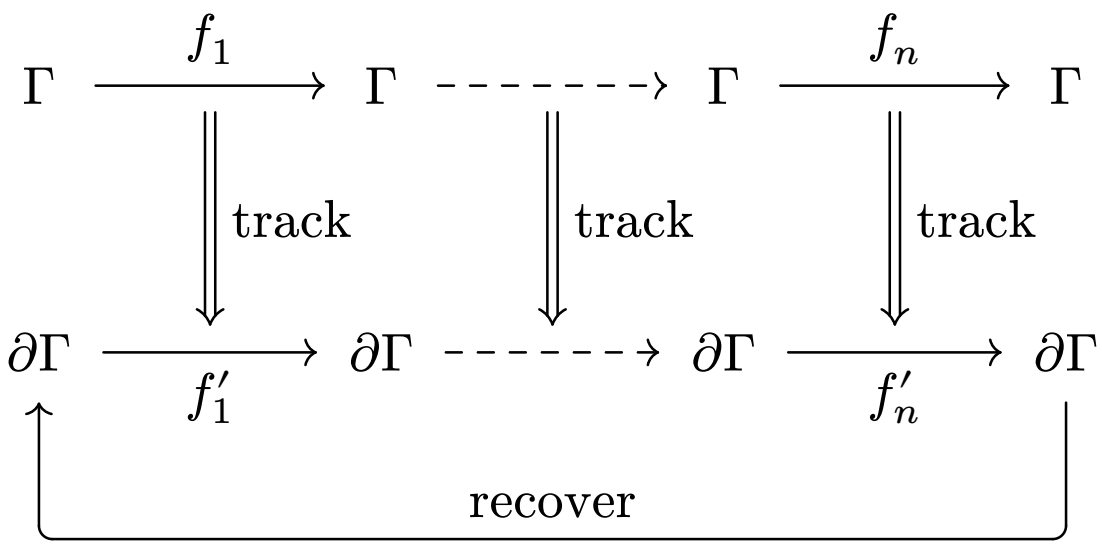

The diagram shows that the tracked effects followed by recover carry the initial effect context back to itself. Each tracking step in fact preserves the result of recovery itself, from whatever state it is taken:

**Theorem 7.** For every $(\gamma, \varphi) \in \partial\Gamma$ and every pair $(f, g)$ with $g(f(\gamma)) = \gamma$,

$$\mathrm{recover}_\Gamma(\mathrm{track}_\Gamma(f, g)(\gamma, \varphi)) = \mathrm{recover}_\Gamma(\gamma, \varphi) \tag{10}$$

*Proof.*

$$\begin{aligned}\mathrm{recover}_\Gamma(\mathrm{track}_\Gamma(f, g)(\gamma, \varphi)) &= \mathrm{recover}_\Gamma(f(\gamma), \varphi \circ g) \\ &= (\varphi(g(f(\gamma))), \mathrm{id}_\Gamma) \\ &= (\varphi(\gamma), \mathrm{id}_\Gamma) = \mathrm{recover}_\Gamma(\gamma, \varphi)\end{aligned}$$

□

Theorem 7 ensures that a tracked effect whose inverse reverts it does not move the result of recovery: recovering after the step returns what recovering before it would have. Recovery reads a state through the quantity $\varphi(\gamma)$ alone, so the guarantee amounts to preserving $\varphi(\gamma)$; we refer to $\varphi(\gamma) = \gamma_0$ as the *soundness invariant* of a state in $\partial\Gamma$. In particular, starting from the initial effect context $(\gamma_0, \mathrm{id}_\Gamma)$, every state reached by tracked effects whose inverses revert them satisfies the invariant, and recovery carries each such state back to $(\gamma_0, \mathrm{id}_\Gamma)$.

The preservation along a sequence follows from Theorem 5 in one application. Let $(f_1, g_1), \cdots, (f_n, g_n)$ be applied in order from $(\gamma, \varphi)$, and write $\delta_0 = \gamma$ and $\delta_i = f_i(\delta_{i-1})$ for the intermediate context states. By Theorem 5, the composite $\mathrm{track}_\Gamma(f_n, g_n) \circ \cdots \circ \mathrm{track}_\Gamma(f_1, g_1)$ is a single tracking step, $\mathrm{track}_\Gamma$ of the twisted composite $(f_n \circ \cdots \circ f_1,\ g_1 \circ \cdots \circ g_n)$. If each inverse reverts its own step, $g_i(\delta_i) = \delta_{i-1}$, then the composite inverse carries $\delta_n$ back to the start: $(g_1 \circ \cdots \circ g_n)(\delta_n) = \delta_0 = \gamma$. The twisted composite therefore meets the hypothesis of Theorem 7 at $\gamma$, and one application of the theorem gives

$$\mathrm{recover}_\Gamma((\mathrm{track}_\Gamma(f_n, g_n) \circ \cdots \circ \mathrm{track}_\Gamma(f_1, g_1))(\gamma, \varphi)) = \mathrm{recover}_\Gamma(\gamma, \varphi) \tag{11}$$

A pair with $g \circ f = \mathrm{id}_\Gamma$ meets the hypothesis at every state.

### *3.1.2. Effect Functions*

The $\mathrm{track}/\mathrm{recover}$ model of the previous section has two limitations:

1. $\mathrm{track}_\Gamma(f, g)$ fixes $g$ before any context state is seen, so one uniform $g$ has to meet the hypothesis of Theorem 7 at every state the effect is applied at. Reverting needs less: an inverse for the one state where $f$ is applied, which may differ from state to state. A per-state inverse cannot be fixed in the argument position before the state is seen; it has to be returned at the point of application.
2. $\mathrm{recover}_\Gamma$ is all-or-nothing: it cannot selectively undo one effect while retaining others.

To address the two issues, we enhance the model at the input and output sides respectively:

1. On the input side, we not only transform $\Gamma$ but also return an inverse function alongside it, so that the inverse is supplied where the effect is applied: $\Gamma \to \Gamma \times (\Gamma \to \Gamma)$, i.e., $\Gamma \to \partial\Gamma$;
2. On the output side, we not only transform $\partial\Gamma$ but also return an inverse function alongside it, so that one effect can be undone while the others are retained: $\partial\Gamma \to \partial\Gamma \times (\partial\Gamma \to \partial\Gamma)$, i.e., $\partial\Gamma \to \partial^2\Gamma$.

The two changes give the input and the output the same shape, i.e., a map from a context to the transformed context paired with an inverse: $\Gamma \to \partial\Gamma$ on the input side and $\partial\Gamma \to \partial^2\Gamma$ on the output side. One type family therefore covers both levels, and we define it at each context as the effect function type $\mathfrak{E}_\Gamma$, refined by a witness to $\mathfrak{E}^*_\Gamma$:

**Definition 8.** Define the effect function $\mathfrak{E}_\Gamma$ and *witnessed* effect function $\mathfrak{E}^*_\Gamma$ as:

$$\begin{aligned}
&\mathfrak{E}_\Gamma := \Gamma \to \Gamma \times (\Gamma \to \Gamma) \\
&\mathfrak{E}^*_\Gamma := (e : \Gamma \to \Gamma \times (\Gamma \to \Gamma)) \\
&\quad \times ((\gamma : \Gamma) \to (\delta : \Gamma) \to (g : \Gamma \to \Gamma) \to ((\delta, g) = e(\gamma) \to g(\delta) = \gamma))
\end{aligned} \tag{12}$$

where $e(\gamma)$ yields a pair $(\delta, g)$ representing:

- $\delta : \Gamma$ is the new context;
- $g : \Gamma \to \Gamma$ is the inverse function of the current effect.

The witness holds each returned inverse to one equation, $g(\delta) = \gamma$: the inverse is required to revert the effect only at the state where it was applied. An element of $\mathfrak{E}^*_\Gamma$ may therefore choose a different inverse at every state. A single $g$ with $g \circ f = \mathrm{id}_\Gamma$ meets the equation at every state at once, so the assignment $(f, g) \mapsto \gamma \mapsto (f(\gamma), g)$ carries such a pair into $\mathfrak{E}^*_\Gamma$, and Theorem 11 shows this assignment to be a homomorphism. The following commutative diagram states the same condition: the inverse $e$ returns reverts the transformation at the state where $e$ was applied:

$$\begin{array}{ccc} \Gamma & \underset{g}{\overset{f}{\rightleftarrows}} & \Gamma \\ & {\scriptstyle e}\searrow \quad \Uparrow{\scriptstyle \mathrm{pr}_2} \quad \nearrow{\scriptstyle \mathrm{pr}_1} & \\ & \partial\Gamma & \end{array}$$

Since effect functions $\mathfrak{E}_\Gamma$ are no longer endomorphisms on the context, they cannot be directly composed. We therefore define a new operation for effect composition:

**Definition 9.** Given functions $f, g \in \mathfrak{E}_\Gamma$, define their *effect composition* $f \diamond g$ as:

$$\begin{aligned} f \diamond g \quad &: \quad \Gamma \;\to\; \partial\Gamma \\ f \diamond g \quad &= \quad \gamma \;\mapsto\; \begin{array}{l} \textbf{let } (\delta, s) = g(\gamma) \textbf{ in} \\ \textbf{let } (\varepsilon, t) = f(\delta) \textbf{ in} \\ (\varepsilon, s \circ t) \end{array} \end{aligned} \tag{13}$$

**Theorem 10.** Effect composition carries the monoid structure of $\mathfrak{T}_\Gamma$ over to $\mathfrak{E}_\Gamma$. That is,

1. $(\mathfrak{E}_\Gamma, \diamond)$ is a monoid with unit $\eta_\Gamma := \gamma \mapsto (\gamma, \mathrm{id}_\Gamma)$;
2. the assignment $(f, g) \mapsto \gamma \mapsto (f(\gamma), g)$ is a monoid homomorphism from $\mathfrak{T}_\Gamma$ into $\mathfrak{E}_\Gamma$.

*Proof.*

1. Associativity and the unit laws follow componentwise from those of $\circ$.
2. Write $e_i = \gamma \mapsto (f_i(\gamma), g_i)$; then $(e_1 \diamond e_2)(\gamma) = (f_1(f_2(\gamma)), g_2 \circ g_1)$, which is the image of $(f_1, g_1) \circ (f_2, g_2)$, and $(\mathrm{id}_\Gamma, \mathrm{id}_\Gamma)$ maps to $\eta_\Gamma$. □

**Theorem 11.** Witnessing survives effect composition, and a uniform inverse witnesses at every state. That is,

1. $\mathfrak{E}^*_\Gamma$ is a submonoid of $\mathfrak{E}_\Gamma$;
2. the homomorphism of Theorem 10 carries every pair with $g \circ f = \mathrm{id}_\Gamma$ into $\mathfrak{E}^*_\Gamma$.

*Proof.*

1. The unit lies in $\mathfrak{E}^*_\Gamma$ since $\mathrm{id}_\Gamma(\gamma) = \gamma$. For closure, take $f, g \in \mathfrak{E}^*_\Gamma$ and any $\gamma \in \Gamma$, and let $(\delta, s) = g(\gamma), (\varepsilon, t) = f(\delta)$, so that $(f \diamond g)(\gamma) = (\varepsilon, s \circ t)$. Then $s(\delta) = \gamma$ and $t(\varepsilon) = \delta$, therefore $(s \circ t)(\varepsilon) = s(\delta) = \gamma$.
2. $g \circ f = \mathrm{id}_\Gamma$ gives $g(f(\gamma)) = \gamma$ at every $\gamma$, so the image of such a pair is witnessed at every state. □

Just as track lifts a pair of transformations on $\Gamma$ to $\partial\Gamma$, we define effect to lift $\mathfrak{E}_\Gamma$ to $\mathfrak{E}_{\partial\Gamma}$:

**Definition 12.** Define the effect function transformation $\mathrm{effect}_\Gamma$ as:

$$\begin{aligned} \mathrm{effect}_\Gamma \quad &: \quad \mathfrak{E}_\Gamma \;\to\; \partial\Gamma \;\to\; \partial^2\Gamma \\ \mathrm{effect}_\Gamma \quad &= \quad e \;\mapsto\; (\gamma, \varphi) \;\mapsto\; \begin{array}{l} \textbf{let } (\delta, g) = e(\gamma) \textbf{ in} \\ ((\delta, \varphi \circ g), \mathrm{track}_\Gamma(g, \mathrm{pr}_1 \circ e)) \end{array} \end{aligned} \tag{14}$$

Since $\mathrm{effect}_\Gamma(e)$ is itself $\mathfrak{E}_{\partial\Gamma}$, what it returns is an inverse in the sense of Definition 8 read one level up. That inverse is itself a track of the pair obtained by swapping the two directions of the effect. The ordinary tracking rule applies once more: undoing the effect is an effect in its own right, transforming the state by $g$, and the way to undo *that* is to perform the effect again, which is what $\mathrm{pr}_1 \circ e$ does. The inverse therefore composes onto the accumulator it is handed, exactly as track prescribes.

We can now prove properties for effect analogous to those of track.

**Theorem 13.** effect preserves the $\diamond$ operation. That is, $\forall f, g \in \mathfrak{E}_\Gamma$:

$$\mathrm{effect}_\Gamma(f) \diamond \mathrm{effect}_\Gamma(g) = \mathrm{effect}_\Gamma(f \diamond g) \tag{15}$$

*Proof.* Take any $(\gamma, \varphi) \in \partial\Gamma$, and let $(\delta, s) = g(\gamma)$ and $(\varepsilon, t) = f(\delta)$, so that $(f \diamond g)(\gamma) = (\varepsilon, s \circ t)$ and $\mathrm{pr}_1 \circ (f \diamond g) = (\mathrm{pr}_1 \circ f) \circ (\mathrm{pr}_1 \circ g)$. Then

$$\begin{aligned}(\mathrm{effect}_\Gamma(f) \diamond \mathrm{effect}_\Gamma(g))(\gamma, \varphi) &= ((\varepsilon, \varphi \circ s \circ t), \mathrm{track}_\Gamma(s, \mathrm{pr}_1 \circ g) \circ \mathrm{track}_\Gamma(t, \mathrm{pr}_1 \circ f)) \\ &= ((\varepsilon, \varphi \circ s \circ t), \mathrm{track}_\Gamma(s \circ t, (\mathrm{pr}_1 \circ f) \circ (\mathrm{pr}_1 \circ g))) \\ &= \mathrm{effect}_\Gamma(f \diamond g)(\gamma, \varphi)\end{aligned}$$

where the first step unfolds Definition 12 at $(\gamma, \varphi)$ and at $(\delta, \varphi \circ s)$, the second is Theorem 5, and the third folds Definition 12. □

How the two levels relate is what the following diagram shows. Its upper triangle is the witness condition of $e$, according to Definition 8, and its lower triangle is the question of whether $e'$ is witnessed the way $e$ is.

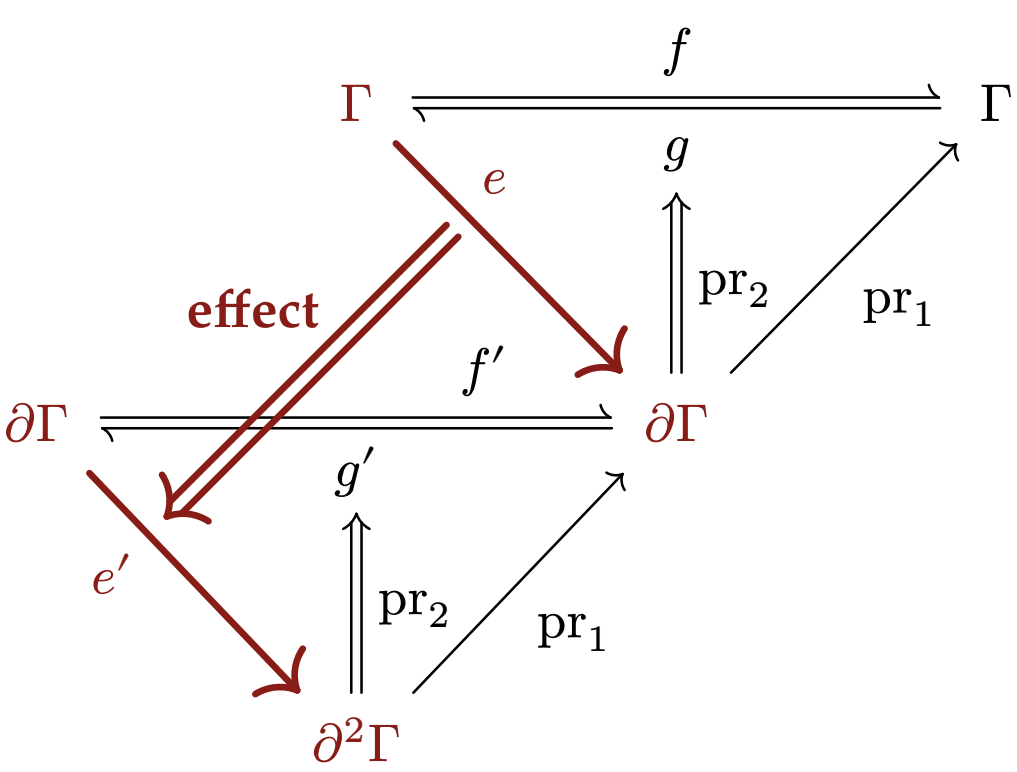

Between the levels, the projection $\mathrm{pr}_1$ relates each lifted map to the map it lifts, as it does for $\mathrm{track}_\Gamma$ in Theorem 4.

**Theorem 14.** Let $e \in \mathfrak{E}_\Gamma$, write $f := \mathrm{pr}_1 \circ e$, and let $e' := \mathrm{effect}_\Gamma(e)$ with forward map $f' := \mathrm{pr}_1 \circ e'$. Then

1. $\mathrm{pr}_1 \circ f' = f \circ \mathrm{pr}_1$;
2. for each $(\gamma, \varphi) \in \partial\Gamma$, the lifted inverse $g' := \mathrm{pr}_2(e'(\gamma, \varphi))$ and the inverse $g := \mathrm{pr}_2(e(\gamma))$ witnessed there satisfy $\mathrm{pr}_1 \circ g' = g \circ \mathrm{pr}_1$.

*Proof.*

1. By Definition 12, $f'(\gamma, \varphi) = (f(\gamma), \varphi \circ g)$, whose state is $f(\gamma) = (f \circ \mathrm{pr}_1)(\gamma, \varphi)$.
2. This is Theorem 4 applied to $g' = \mathrm{track}_\Gamma(g, f)$. □

Whether the lower triangle closes is settled by computing what the lifted inverse returns:

**Theorem 15.** Let $e \in \mathfrak{E}^*_\Gamma$ and write $f := \mathrm{pr}_1 \circ e$. Fix $(\gamma, \varphi) \in \partial\Gamma$, let $(\delta, g) = e(\gamma)$, and write $(\Delta, g')$ for the value of $\mathrm{effect}_\Gamma(e)$ at $(\gamma, \varphi)$. Then

$$g'(\Delta) = (\gamma, \varphi \circ g \circ f) \tag{16}$$

The state is recovered exactly. The accumulator is restored as well, equivalently $\mathrm{effect}_\Gamma(e) \in \mathfrak{E}^*_{\partial\Gamma}$, if and only if $g \circ f = \mathrm{id}_\Gamma$; and in every case $(\varphi \circ g \circ f)(\gamma) = \varphi(\gamma)$, so the soundness invariant is preserved.

*Proof.* By Definition 12, $\Delta = (\delta, \varphi \circ g)$ and $g' = \mathrm{track}_\Gamma(g, f)$, so

$$g'(\Delta) = (g(\delta), \varphi \circ g \circ f) = (\gamma, \varphi \circ g \circ f)$$

using $g(\delta) = \gamma$. Membership in $\mathfrak{E}^*_{\partial\Gamma}$ requires this to equal $(\gamma, \varphi)$ at every input; taking $\varphi = \mathrm{id}_\Gamma$ turns the equality of accumulators into $g \circ f = \mathrm{id}_\Gamma$, and that condition conversely gives the equality of accumulators for every $\varphi$. Finally $(\varphi \circ g \circ f)(\gamma) = \varphi(g(\delta)) = \varphi(\gamma)$. □

The lower triangle therefore closes only when the inverse witnessed at $\gamma$ reverts $f$ at every state, so $\mathrm{effect}_\Gamma$ does not carry $\mathfrak{E}^*_\Gamma$ into $\mathfrak{E}^*_{\partial\Gamma}$. What holds in every case is agreement at $\gamma$: $\mathrm{recover}_\Gamma(g'(\Delta)) = \mathrm{recover}_\Gamma(\gamma, \varphi)$, which is the whole of what Theorem 7 assumes of an accumulator, so reverting leaves the recovery target untouched.

#### *3.1.3. Effect Iterators*

What a component loads by is not one effect but a sequence of them, and what its unloading reverts is the whole sequence. Reverting effects in the reverse order of application requires nothing further, because each inverse then meets the state its own application produced:

**Theorem 16.** Let $e_1, \cdots, e_n \in \mathfrak{E}^*_\Gamma$ be applied in order from $(\gamma_0, \mathrm{id}_\Gamma)$ and reverted in the reverse order. Then

1. each revert recovers the context state its application ran against;
2. every intermediate state satisfies the soundness invariant.

*Proof.* Each step is an application or a revert. An application carries $(\gamma, \varphi)$ to $(\delta, \varphi \circ g)$ with $g(\delta) = \gamma$, so it preserves $\varphi(\gamma)$ by Theorem 7, whose hypothesis is exactly the witness of $\mathfrak{E}^*_\Gamma$. Reverting in the reverse order hands each inverse the state its own application produced, so by Theorem 15 that revert recovers the preceding state exactly and preserves $\varphi(\gamma)$ as well; neither conclusion depends on the accumulator the inverse receives. □

The sequence itself deserves a reification. An *effect iterator* performs it one effect at a time, each of whose iterations yields the modified context, an inverse, and a continuation:

**Definition 17.** Define the effect iterator $\mathfrak{I}_\Gamma$ and *witnessed* effect iterator $\mathfrak{I}^*_\Gamma$ as the following recursive types:

$$\begin{aligned}
\mathfrak{I}_\Gamma &\coloneqq \mu\mathfrak{I}.\ \Gamma \to \Gamma \times (\Gamma \to \Gamma) \times \mathsf{Maybe}(\mathfrak{I}) \\
\mathfrak{I}^*_\Gamma &\coloneqq \mu\mathfrak{I}.\ (e : \Gamma \to \Gamma \times (\Gamma \to \Gamma) \times \mathsf{Maybe}(\mathfrak{I})) \\
&\qquad \times ((\gamma : \Gamma) \to (\delta : \Gamma) \to (g : \Gamma \to \Gamma) \to (o : \mathsf{Maybe}(\mathfrak{I})) \to ((\delta, g, o) = e(\gamma) \to g(\delta) = \gamma))
\end{aligned} \tag{17}$$

where $e(\gamma)$ yields a triple $(\delta, g, o)$ representing:

- $\delta$ is the new context;
- $g$ is the inverse function of the current effect;
- $o$ indicates the continuation:
  - $\mathsf{Nothing}$ signals iteration termination;
  - $\mathsf{Just}(i)$ provides the next iteration.

The witness holds each iteration to the constraint Definition 8 places on a single effect, and the continuation a witnessed iterator yields is again witnessed.

The effect iterator transformation $\text{effect}^{\text{iter}}_\Gamma$ extends $\text{effect}_\Gamma$ to the iterator structure through recursive invocation:

**Definition 18.** Define the effect iterator transformation $\text{effect}^{\text{iter}}_\Gamma$ as:

$$
\begin{array}{llllll}
\text{effect}^{\text{iter}}_\Gamma & : & \mathfrak{I}_\Gamma & \to & \partial\Gamma & \to \quad \partial^2\Gamma \\
\text{effect}^{\text{iter}}_\Gamma & = & i & \mapsto & (\gamma, \varphi) & \mapsto \begin{array}{l}
\textbf{let } (\delta, g, o) = i(\gamma) \textbf{ in} \\
\textbf{let } t = \text{track}_\Gamma(g, \text{pr}_1 \circ i) \textbf{ in} \\
\textbf{match } o \\
\quad \mid \mathsf{Nothing} \Rightarrow ((\delta, \varphi \circ g), t) \\
\quad \mid \mathsf{Just}(i') \Rightarrow \textbf{let } (s, r) = \text{effect}^{\text{iter}}_\Gamma(i')(\delta, \varphi \circ g) \textbf{ in} \\
\qquad\qquad\qquad (s, t \circ r)
\end{array}
\end{array}
\tag{18}
$$

At each iteration, the inverse $g$ is composed onto $\varphi$ in application order, so the accumulator $\varphi \circ g_1 \circ \cdots \circ g_k$ reverts the effects in LIFO order when applied (Theorem 16). Because $\text{effect}^{\text{iter}}_\Gamma$ lands in the same $\partial\Gamma \to \partial^2\Gamma$ as $\text{effect}_\Gamma$ does, an iterator is an effect in its own right and can be used wherever an effect can, and Section 4 reads a component's whole loading as one iterator. The $\mathsf{Maybe}(\mathfrak{I})$ continuation makes a boundary available between any two consecutive iterations, at which the context is whatever the iterations so far have made it and the accumulator recovers those and nothing more. In this sense the effect iterator is a reified delimited continuation, the structure that mainstream languages expose through the `yield` operator [38], so the model maps directly onto the generators they already provide.

A plain effect function is the degenerate case: an $e \in \mathfrak{E}_\Gamma$ embeds as the iterator whose first iteration already yields Nothing,

$$\gamma \mapsto \textbf{let } (\delta, g) = e(\gamma) \textbf{ in } (\delta, g, \mathsf{Nothing}) \tag{19}$$

and the embedding carries $\mathfrak{E}^*_\Gamma$ into $\mathfrak{I}^*_\Gamma$, the two witnesses asking the same equation. Every notion defined at iterators below is read at an effect function through this embedding.

Together, these constructions constitute *revertible effects*: each effect function in $\mathfrak{E}^*_\Gamma$ explicitly provides its own inverse, effect tracks the effect on $\partial\Gamma$, and the $\diamond$ operation composes effect functions while preserving revertibility. What they deliver is *local temporal composability*, local in that the guarantee is read of one component's effects taken by themselves. We take that to be the following criterion: for every sequence of effect functions a component applies, the accumulator recovers the context it began at (Theorem 7), and reverting the sequence hands each inverse the state its own application ran against (Theorem 16). Loading a component is running one iterator and accumulating its inverses in $\varphi$; unloading it is applying $\varphi$. Two things the criterion leaves out, and both arrive once several components are in play: reverting out of the order the accumulator imposes, and a sequence that interleaves the effects of others. Both are supplied by independence, a condition on the effects rather than a property of the construction (Section 3.4).

### 3.2. Reactive Coeffects

*Spatial composability* is the ability for components to declare dependencies on one another and for the system to resolve, provide, and withdraw those dependencies at runtime. This requires that dependency satisfaction be re-evaluated whenever the shared context changes, so that a component activates when its dependencies become available and deactivates when they are withdrawn. We therefore model dependencies of a component as a specification and classify each change to the context, against that specification, as activating, deactivating, or neutral. Classifying against the specification is what detects a change in satisfaction; responding to that classification is what drives activation and deactivation. We call such coeffects *reactive*: by classifying context changes and driving activation and deactivation from them, local spatial composability becomes a structural guarantee.

### 3.2.1. Coeffect Context

Traditional inversion-of-control (IoC) containers [39] typically model dependencies as simple key-value mappings. This section formalizes IoC as a coeffect context that synergizes with revertible effects to provide a mathematical foundation for dynamic composition.

**Definition 19.** Given a type family $\mathcal{V} : K \to \mathrm{Type}$, define the *coeffect context* as the dependent partial function type:

$$\Sigma := (k : K) \rightharpoonup \mathcal{V}_k \tag{20}$$

where $\sigma : \Sigma$ is a finite partial function assigning to each $k \in \mathrm{dom}(\sigma) \subseteq K$ a value of type $\mathcal{V}_k$. We write:

- $\sigma(k)$ for application (defined when $k \in \mathrm{dom}(\sigma)$);
- $\sigma[k \mapsto v]$ for the table binding $v$ at $k$ and agreeing with $\sigma$ elsewhere;
- $\sigma \setminus k$ for restriction (defined when $k \in \mathrm{dom}(\sigma)$);
- $k \in \mathrm{dom}(\sigma)$ for membership.

The use of a type family $\mathcal{V}$ ensures that each dependency key $k$ is associated with a specific value type $\mathcal{V}_k$, providing static type safety for dependency access. Extension and restriction carry preconditions, imposed by the operations below: a dependency cannot be provided twice ($k \notin \mathrm{dom}(\sigma)$ for extension) nor revoked if absent ($k \in \mathrm{dom}(\sigma)$ for restriction). A violated precondition is signaled as an error and produces no transition, so the effect algebra, which describes the transitions that do occur, applies to these operations unchanged. A reader preferring to internalize the failure may read every $\Sigma \rightharpoonup \Sigma$ below as $\Sigma \to \mathsf{Maybe}(\Sigma)$ and compose in the $\mathsf{Maybe}$ monad (Section 2.1), at the cost of replacing each identity by the partial identity on the operation's domain. Based on this context structure, we define two core operations:

**Definition 20.** The get and set operations on $\Sigma$ are defined as:

$$\begin{array}{lccccc} \mathrm{get} & : & (k : K) & \to & \Sigma & \rightharpoonup & \mathcal{V}_k \\ \mathrm{get} & = & k & \mapsto & \sigma & \mapsto & \sigma(k) \\ \mathrm{set} & : & (k : K) \times \mathcal{V}_k & \to & \Sigma & \rightharpoonup & \Sigma \times (\Sigma \rightharpoonup \Sigma) \\ \mathrm{set} & = & (k, v) & \mapsto & \sigma & \mapsto & (\sigma[k \mapsto v],\, \lambda\sigma'.\sigma' \setminus k) \end{array} \tag{21}$$

where $\mathrm{get}(k)$ requires $k \in \mathrm{dom}(\sigma)$ and $\mathrm{set}(k, v)$ requires $k \notin \mathrm{dom}(\sigma)$ as preconditions.

Notably, $\mathrm{set}(k, v)$ has type $\mathfrak{E}^*_\Sigma$, i.e., an effect function on the coeffect context. We can therefore directly apply the effect machinery from Section 3.1: $\mathrm{effect}_\Sigma$ provides automatic tracking and recovery of dependency registrations. This is the synergy between reactive coeffects and revertible effects: coeffect operations are effects, and effects are revertible.

#### 3.2.2. *Specification and Notification*

The preceding definitions describe how individual dependencies are registered and accessed. Accessing an absent dependency, however, is a runtime failure. A component should therefore activate only once all the dependencies it declares are present, rather than accessing them optimistically and failing when one is missing. This raises two questions: whether a component's declared dependencies are jointly satisfied, and how the system should respond when that status changes. The coeffect context Σ carries a natural observational structure that makes both questions tractable: for any coeffect specification $d \subseteq K$, define the *satisfaction predicate*:

$$\sigma \vDash d := \forall k \in d.\ k \in \mathrm{dom}(\sigma) \tag{22}$$

This predicate is decidable (since $\mathrm{dom}(\sigma)$ is finite). Since all mutations to $\sigma$ pass through effect functions (whose inverses recover the previous domain), changes to satisfaction are detectable at each effect boundary. This is the algebraic basis of reactivity: the effect system guarantees that every coeffect change is observed.

**Definition 21.** A *coeffect specification* is:

$$\mathfrak{D}_\Sigma := \mathsf{Set}(K) \tag{23}$$

representing the set of dependencies a component declares from the environment.

What makes this specification reactive is how it classifies state transitions. Any effect that transforms $\sigma$ to $\sigma'$ can be classified by a specification $d \in \mathfrak{D}_\Sigma$ according to whether $d$'s satisfaction status is altered:

**Definition 22.** Given a coeffect specification $d \subseteq K$ and states $\sigma, \sigma' \in \Sigma$, define:

$$\mathrm{notify}_d(\sigma, \sigma') := \begin{cases} \text{activating} & \text{if } \sigma \nvDash d \wedge \sigma' \vDash d \\ \text{deactivating} & \text{if } \sigma \vDash d \wedge \sigma' \nvDash d \\ \text{neutral} & \text{otherwise} \end{cases} \tag{24}$$

An activating transition triggers the execution of the component's effects, tracked as Section 3.1 prescribes, and a deactivating transition triggers recovery by applying the accumulator. The activation and deactivation so triggered receive their operational semantics in Section 4.

What set and notify deliver together is *local spatial composability*, local in the same sense as before, the guarantee being read of one component's coeffects taken by themselves. We take that to be the following criterion: a component activates only at a state satisfying its specification, so it never reads a binding that is absent, and every change to the context is classified against that specification, so a loss of satisfaction is detected where it happens and drives a deactivation. Both halves are immediate from the definitions above, satisfaction being a precondition checked where the component would activate and $\mathrm{notify}_d$ being defined at every transition; one direction of the coeffect ordering comes with the first half, a component activating only after the components that provide its declared keys. Two things the criterion leaves out, and both arrive once several components are in play: withdrawing a binding only after the deactivations it causes have finished, and keeping the bindings an activation reads unmoved while the activation runs. Both are conditions on other components rather than on the one acting, so they belong to the global form of the guarantee, which Section 4.3.3 establishes.

#### *3.2.3. Isolation and Interception*

The basic coeffect context $\Sigma$ models a flat dependency table. In practice, however, the system may need to bind distinct values to the same logical dependency for different components. This section extends the coeffect context with two mechanisms: *coeffect isolation* (the same key resolves differently in different contexts) and *coeffect interception* (cross-cutting behavior on dependency access).

**Realization.** The two mechanisms differ from get and set in what they act on. A provision writes the shared table every component reads, so it is an effect on that table and carries an inverse to withdraw it. Isolation and interception instead adjust how a key is *resolved* for the components under one context, leaving the table itself as it stands. Typing an operation as an effect fixes its *denotation*, a successor state paired with an inverse, but not its *realization*, which determines how that inverse is carried out.

**Definition 23.** An effect function on a context admits two *realizations*:
- *In-place* realization mutates the context and returns a nontrivial inverse; the successor aliases the input, and recovery runs the inverse to undo the mutation.
- *Derived* realization leaves the input intact and returns a fresh context deriving from it, with the identity as its inverse; recovery discards the derived context. A context derived from another is what the recursive structure of Definition 28 carries.

In a purely functional setting the two coincide, and an imperative host may choose either per operation; Section 5.1.2 implements both. Isolation and interception are given derived realization outright: each produces a fresh context whose own table differs from the inherited one, so each is typed below as a map from context to context rather than as an effect function. Nothing in the shared table changes, so there is no effect to track and nothing for Definition 12 to lift, and recovery discards the derived context along with the adjustment it carried. Assignment on a derived table overrides whatever the inherited table held at the key, which is why neither operation carries a precondition.

**Coeffect Isolation.** By introducing *isolation realms*, coeffect isolation allows the same dependency to bind to different values in different contexts. This has broad applications in multi-tenant systems, testing environments, and component sandboxes.

**Definition 24.** Define the coeffect context with isolation as:

$$\Sigma^{\mathrm{iso}} := (K \rightharpoonup R) \times ((r : R) \rightharpoonup \mathcal{V}_r) \tag{25}$$

It can be represented as a pair $(\rho, \sigma)$, where:
- $\rho : K \rightharpoonup R$ is the isolation realm table, assigning a realm identifier to each isolated key; a key outside $\mathrm{dom}(\rho)$ resolves to its own realm, so we write $\rho(k) = k$ there ($R \supseteq K$);
- $\sigma : (r : R) \rightharpoonup \mathcal{V}_r$ is the dependency table, a partial dependent function from realm identifiers to typed values.

The two-layer mapping structure decouples the logical layer from the storage layer, making dependency access context-aware. When accessing a key $k$, the system first resolves $\rho(k)$ to obtain a realm identifier $r$, then accesses $\sigma(r)$ for the actual value.

**Definition 25.** The get, set, and isolate operations on $\Sigma^{\mathrm{iso}}$ are:

$$
\begin{array}{rcccccc}
\text{get} & : & (k:K) & \to & \Sigma^{\text{iso}} & \rightharpoonup & \mathcal{V}_{\rho(k)} \\
\text{get} & = & k & \mapsto & (\rho,\sigma) & \mapsto & \sigma(\rho(k)) \\
\text{set} & : & (k:K)\times\mathcal{V}_{\rho(k)} & \to & \Sigma^{\text{iso}} & \rightharpoonup & \Sigma^{\text{iso}}\times(\Sigma^{\text{iso}}\rightharpoonup\Sigma^{\text{iso}}) \\
\text{set} & = & (k,v) & \mapsto & (\rho,\sigma) & \mapsto & ((\rho,\sigma[\rho(k)\mapsto v]),\ \lambda(\rho',\sigma').(\rho',\sigma'\setminus\rho'(k))) \\
\text{isolate} & : & K\times R & \to & \Sigma^{\text{iso}} & \to & \Sigma^{\text{iso}} \\
\text{isolate} & = & (k,r) & \mapsto & (\rho,\sigma) & \mapsto & (\rho[k\mapsto r],\sigma)
\end{array}
\tag{26}
$$

where $\text{get}$ and $\text{set}$ carry the preconditions of Definition 20 transported along $\rho$, namely $\rho(k)\in\text{dom}(\sigma)$ and $\rho(k)\notin\text{dom}(\sigma)$. The context that $\text{isolate}(k,r)$ derives assigns the realm $r$ to $k$ and inherits the dependency table unchanged, so a key already isolated is reassigned rather than refused.

The coeffect isolation mechanism essentially implements a runtime ad-hoc polymorphism system. Through isolation realm identifiers, the same dependency key can resolve to entirely different values in different contexts, and this polymorphism can be dynamically adjusted at runtime. Compared to traditional dependency injection, coeffect isolation provides finer-grained control, enabling customized isolation for specific components; $\text{set}$ remains an effect function ($\mathfrak{E}^*_{\Sigma^{\text{iso}}}$) and thus inherits revertibility, whereas $\text{isolate}$ needs none, deriving a context instead of writing the shared table.

**Coeffect Interception.** The second mechanism, *coeffect interception*, attaches cross-cutting metadata to dependency access, adding behavior without modifying the dependency value. This metadata can be either context-carried or component-declared, so we extend both the coeffect context and the coeffect specification:

**Definition 26.** Define the coeffect context and specification with interception as:

$$
\begin{aligned}
\Sigma^{\text{inter}} &:= ((k:K)\to\mathcal{M}_k)\times((k:K)\rightharpoonup(\mathcal{M}_k\to\mathcal{V}_k)) \\
\mathfrak{D}^{\text{inter}} &:= (k:K)\rightharpoonup\mathcal{M}_k
\end{aligned}
\tag{27}
$$

The context $\Sigma^{\text{inter}}$ is a pair $(\iota,\sigma)$: $\iota$ is the *context-carried* metadata installed on the context itself, empty ($\epsilon_k$) by default; and $\sigma$ maps each key $k$ to a *provider function* from metadata $\mathcal{M}_k$ to value $\mathcal{V}_k$. A specification $d\in\mathfrak{D}^{\text{inter}}$ carries the *component-declared* metadata, assigning each key its metadata $d(k)$, with $\text{dom}(d)$ serving as the dependency set. Each key equips its metadata with a monoid $(\mathcal{M}_k,\oplus_k,\epsilon_k)$: the merge $\oplus_k$ is associative with identity $\epsilon_k$ (the empty metadata).

**Definition 27.** The $\text{get}$, $\text{set}$, and $\text{intercept}$ operations on $\Sigma^{\text{inter}}$ are:

$$
\begin{array}{rcccccc}
\text{get} & : & (k:K)\times\mathcal{M}_k & \to & \Sigma^{\text{inter}} & \rightharpoonup & \mathcal{V}_k \\
\text{get} & = & (k,\mu) & \mapsto & (\iota,\sigma) & \mapsto & \sigma(k)(\mu\oplus_k\iota(k)) \\
\text{set} & : & (k:K)\times(\mathcal{M}_k\to\mathcal{V}_k) & \to & \Sigma^{\text{inter}} & \rightharpoonup & \Sigma^{\text{inter}}\times(\Sigma^{\text{inter}}\rightharpoonup\Sigma^{\text{inter}}) \\
\text{set} & = & (k,\psi) & \mapsto & (\iota,\sigma) & \mapsto & ((\iota,\sigma[k\mapsto\psi]),\ \lambda(\iota',\sigma').(\iota',\sigma'\setminus k)) \\
\text{intercept} & : & (k:K)\times\mathcal{M}_k & \to & \Sigma^{\text{inter}} & \to & \Sigma^{\text{inter}} \\
\text{intercept} & = & (k,\nu) & \mapsto & (\iota,\sigma) & \mapsto & (\iota[k\mapsto\iota(k)\oplus_k\nu],\sigma)
\end{array}
\tag{28}
$$

where $\text{get}$ and $\text{set}$ carry the preconditions of Definition 20 on the provider table, namely $k\in\text{dom}(\sigma)$ and $k\notin\text{dom}(\sigma)$. The context that $\text{intercept}(k,\nu)$ derives merges $\nu$ onto the metadata inherited at $k$ and inherits the provider table unchanged.

When a component with specification $d$ accesses key $k$, the system evaluates $\sigma(k)(d(k) \oplus_k \iota(k))$: the component-declared metadata is merged with the context-carried metadata $\iota$, and the provider function is applied to the result. This merge follows each key's own semantics (e.g. scalar fields are overwritten, set-valued fields unioned) and is right-biased, so $\iota(k)$ takes priority and can override the component's declaration, letting an enclosing context constrain how a component uses a coeffect without modifying that component (e.g. Section 6.3).

### 3.3. The Context Paradigm

Section 3.1 and Section 3.2 each act on a context, the first as the carrier of effects and the second as the carrier of coeffects. Section 3.3.1 constructs a unified context carrying both, gives each of its keys a set of operations, and establishes the context paradigm by constraining the stages of an effect iterator (Definition 30). Section 3.3.2 then makes the operations the standard of comparison: two context states are observationally equivalent when no sequence of operations distinguishes them, and every equality of Section 3.1 is re-read up to that equivalence.

#### *3.3.1. Unified Context*

For a context $\Gamma$, the effect context $\partial\Gamma$ (Section 3.1) provides a higher-level abstraction, carrying the previous-level context and that level's accumulator (Definition 2). Making this structure recursive and combining it with the coeffect context $\Sigma$ yields the following type:

**Definition 28.** The context type $\Gamma_\infty$ is defined as:

$$\Gamma_\infty := \mu\Gamma.\ \Gamma \times (\Gamma \to \Gamma) \times \Sigma \tag{29}$$

where the three projections are:
- $\Gamma$: the current context state (recursive);
- $\Gamma \to \Gamma$: the accumulator, which reverts this level's effects;
- $\Sigma$: the coeffect context carrying dependency information.

Under this definition, $\mathsf{effect}$ maps $\mathfrak{E}_{\Gamma_\infty}$ to itself, unifying the $\partial$-tower into a single self-similar type. The coeffect context $\Sigma$ is structurally integrated: dependency operations ($\mathsf{set}$, $\mathsf{get}$) act on $\Sigma$, and the accumulator holds their inverses. Since the type family $\mathcal{V}$ underlying $\Sigma$ is unconstrained, any state the system needs to share across components can be encoded as a dependency with an appropriate value type—$\Sigma$ subsumes all shared mutable states, not just inter-component dependencies. Every interaction between a component and its environment passes through this single entity.

Passing through one entity is a discipline only where there is nothing else to pass through, so what a component may do with a bound value has to be fixed as well. A key therefore carries more than a value type:

**Definition 29.** A *coeffect* at a key $k$ is a pair $(\mathcal{V}_k, \mathcal{A}_k)$, where $\mathcal{V}_k$ is the value type of Definition 19 and $\mathcal{A}_k$ is a set of *coeffect operations*, the operations the value bound at $k$ provides to a component holding it. An operation $a \in \mathcal{A}_k$ carries an argument type $X_a$ and an outcome type $B_a$, and acts on the value alone:

$$a : X_a \to \mathcal{V}_k \rightharpoonup \mathcal{V}_k \times (\mathcal{V}_k \rightharpoonup \mathcal{V}_k) \times B_a \tag{30}$$

its first two constituents forming an effect function on $\mathcal{V}_k$ witnessed as Definition 8 requires, and its third an *outcome*. The operations induce the equivalence $\simeq_k$ on $\mathcal{V}_k$ up to which values at *k* are compared (Section 3.3.2). An operation acts on the coeffect context through its *lift*

$$a^\Sigma(x)(\sigma) := \mathbf{let}\ (v, g, b) = a(x)(\sigma(k))\ \mathbf{in}\ (\sigma[k \mapsto v],\ \lambda\sigma'.\sigma'[k \mapsto g(\sigma'(k))],\ b) \tag{31}$$

defined when $k \in \mathrm{dom}(\sigma)$, whose first two constituents are an effect function on $\Sigma$.

Typing an operation of *k* on $\mathcal{V}_k$ is what confines it to the binding at *k*: the lift reads and writes that binding and leaves every other key as it stands, so no side condition is needed to say so. Where isolation is in force the binding it reaches is the one the realm resolves to (Definition 24), two keys sharing a realm sharing one binding. An operation whose behavior turns on another key reads that key's value into its argument $X_a$, and the reactive discipline of Section 3.2.2 is what holds the binding in place for as long as the component that read it runs (Theorem 70), the value moving only by operations of that key. The pair is not yet the whole of a coeffect: once the independence of two operations has been defined, Section 3.4.2 completes the pair with a third constituent, a witness certifying that the operations of $\mathcal{A}_k$ are pairwise independent.

What a component performs is a sequence of stages in which each may depend on what the ones before it yielded: an operation on a value some key binds, or the provision of a binding of its own. Iterators of that shape, one stage per iteration, are the form the discipline takes at an effect function.

**Definition 30.** For key sets $P \subseteq S \subseteq K$, the *context-mediated* iterators $\mathfrak{I}^{\mathcal{A}}_{\Sigma}(S, P)$ form the least set of iterators on $\Sigma$ that contains the unit $\sigma \mapsto (\sigma, \mathrm{id}_\Sigma, \mathsf{Nothing})$ and, whenever each named continuation is Nothing or Just of a member, contains

$$\begin{aligned} &\sigma \mapsto \mathbf{let}\ (\delta, s, b) = a^\Sigma(x)(\sigma)\ \mathbf{in}\ (\delta,\ s,\ c_b) \quad \text{for } k \in S,\, a \in \mathcal{A}_k,\, x : X_a,\, (c_b)_{b \in B_a} \\ &\sigma \mapsto \mathbf{let}\ (\delta, s) = \mathrm{set}(k, v)(\sigma)\ \mathbf{in}\ (\delta,\ s,\ c) \quad\ \ \text{for } k \in P,\, v : \mathcal{V}_k,\, c \end{aligned} \tag{32}$$

An *operation stage* performs one coeffect operation and chooses what follows it by the outcome, so an argument may depend on the outcomes already obtained. A *provision stage* installs one binding, at a key no operation can create, an operation presupposing the binding it acts on, and yields the restriction set pairs with the extension (Definition 20). The stages *occurring in* a member are its own and those of every iterator its continuations reach.

Membership in this class is the formal content of mediating every interaction through the context. What falls outside it is a map reading anything else, whether a key it performs no stage at or a location no key binds; an allocator drawing handles from a counter the context does not carry is the second case, and becomes context-mediated once the counter is bound at a key of its own.

**Hierarchical composition.** The recursive structure of $\Gamma_\infty$ supports hierarchical control: a parent context aggregates multiple child-level effects, forming a tree-shaped control structure that maintains modularity while enabling unified cross-level management. The effect transformation realizes a literal "plug-in" metaphor:

- Loading a component corresponds to executing its effects (plugging in);
- Unloading a component corresponds to reverting its effects (unplugging, without affecting other running components);

- Components at different levels of the hierarchy are independently loadable and unloadable; a parent context aggregates and manages the effects of all its children, enabling arbitrarily nested composition.

### *3.3.2. Observational Equivalence*

The recovery guarantee of Section 3.1 asserts an equality of states (Theorem 7), which is an idealization, because the physical state cannot be recovered as it stood. For example, `free` releases a block to the allocator without restoring the layout the heap had before `malloc`; and a generative name is not restored by the inverse that discards it, since the next creation draws a fresh one [40]. The equalities of Section 3 are therefore to be read up to an equivalence $\simeq$, and we take $\simeq$ to be an *observational equivalence*: two states are related when no observer can distinguish them. Comparing behavior rather than representation is the established route to program equivalence [41], and the relation such a comparison yields depends on what the observer is given to work with [42]. What an observer of a context is given is the coeffects it carries, and what an observer of a value is given is the operations of its key (Definition 29), so the relation at each key is generated from those operations, and the relation on a context is assembled from the relations at its keys. Both constructions are the business of this subsection, and quotienting by the result is what makes the independence of Section 3.4.1 attainable.

An observer of a value runs the operations of its key and reads their outcomes.

**Definition 31.** Let $V$ carry a set $\mathcal{A}$ of operations in the sense of Definition 29. A *test* over $\mathcal{A}$ is a finite word whose letters are forward maps and yielded inverses of the effect functions $a(x)$, over every $a \in \mathcal{A}$ and every argument $x : X_a$, each letter applied to the value the letters before it left; its *outcomes* are those the letters that are forward maps yield along the way, and it is undefined where a precondition fails. Values $v, v' : V$ are *indistinguishable*, written $v \approx_{\mathcal{A}} v'$, when every test over $\mathcal{A}$ is defined at both or at neither and yields the same outcomes at both. The equivalence at a key is indistinguishability under its own operations:

$$\simeq_k \quad := \quad \approx_{\mathcal{A}_k} \tag{33}$$

An operation *respects* an equivalence when, at related values, it is defined at both or at neither and, where defined, yields related successors, inverses carrying related values to related values, and equal outcomes.

**Lemma 32.** Each $\simeq_k$ is an equivalence that every operation of $\mathcal{A}_k$ respects, and it is the coarsest such relation. That is,

1. $\approx_{\mathcal{A}}$ is an equivalence, and every operation of $\mathcal{A}$ respects it;
2. every equivalence that every operation of $\mathcal{A}$ respects is contained in $\approx_{\mathcal{A}}$.

*Proof.*

1. Agreement of tests, in definedness and in outcomes, is reflexive, symmetric, and transitive. Let $v \approx_{\mathcal{A}} v'$ and let $a \in \mathcal{A}$ be applied to an argument. Prefixing a test by one letter is again a test, so the values the forward map reaches are indistinguishable, as are the values any one yielded inverse reaches from indistinguishable arguments; the one-letter test gives definedness at both or neither and equality of the outcome.
2. Let $R$ be such an equivalence and $vRv'$. Each letter of a test is a forward map or a yielded inverse of an operation, and respect carries $R$ along either, keeping the values reached related and the outcomes equal at every letter. Hence every test agrees at $v$ and $v'$. □

Clause (2) doubles as the proof principle for $\simeq_k$: to relate two values, exhibit an equivalence the operations respect that contains the pair.

**Definition 33.** Two coeffect contexts are related at a set $S \subseteq K$ of keys when they bind the same keys of $S$ to related values, and two states of a context when their coeffect projections are:

$$\begin{aligned} \sigma \simeq_S \sigma' \quad &:= \quad \mathrm{dom}(\sigma) \cap S = \mathrm{dom}(\sigma') \cap S \wedge \forall k \in \mathrm{dom}(\sigma) \cap S.\ \sigma(k) \simeq_k \sigma'(k) \\ \gamma \simeq_S \gamma' \quad &:= \quad \sigma_\gamma \simeq_S \sigma_{\gamma'} \end{aligned} \tag{34}$$

writing $\sigma_\gamma$ for the coeffect projection of $\gamma$ (Definition 28). Each $\simeq_k$ is an equivalence on $\mathcal{V}_k$ by Lemma 32, so $\simeq_S$ is an equivalence on coeffect contexts and on states, each of the three properties holding key by key. The subscript is dropped where $S = K$, so that $\simeq$ is the finest of these relations and $\simeq_S$ forgets the keys outside $S$ as well.

The part of a state that no key binds is thereby forgotten, and forgetting it is what lets Theorem 7 be read up to $\simeq$ at all: the heap layout and the generative name of the examples above lie outside the relation unless some key binds them. A restriction forgets more, comparing only what the keys of $S$ bind, and Section 4 reads each claim about one component at the restriction that component's own declarations name (Definition 48). What Section 3.2.2 needs of $\simeq$ follows rather than being assumed. Related states have the same domain, so they agree on the satisfaction predicate $\sigma \vDash d$ and on the classification $\mathrm{notify}_d$ of Definition 22, and reactivity is a property of $\Sigma/\simeq$.

Substituting $\simeq$ for $=$ throughout is not by itself enough, because an effect function returns an inverse as well as a state, and two states that $\simeq$ identifies have to yield inverses $\simeq$ identifies as well.

**Definition 34.** The relation of Definition 33 on states and each $\simeq_k$ on the values of its key are the base cases, and on any other base type $\simeq_S$ is equality. The relation extends along the type formers:

$$\begin{aligned} &\text{for } f, g : X \to Y, && f \simeq_S g && := && (\gamma : X) \to (\gamma' : X) \to (\gamma \simeq_S \gamma' \to f(\gamma) \simeq_S g(\gamma')) \\ &\text{for } a, b : X_1 \times \cdots \times X_n, && a \simeq_S b && := && (a_1 \simeq_S b_1) \wedge \cdots \wedge (a_n \simeq_S b_n) \\ &\text{for } x, y : \mathsf{Maybe}(X), && x \simeq_S y && := && \begin{cases} z \simeq_S z' & \text{if } x = \mathsf{Just}(z) \text{ and } y = \mathsf{Just}(z') \\ \top & \text{if } x = y = \mathsf{Nothing} \\ \bot & \text{otherwise} \end{cases} \end{aligned} \tag{35}$$

On a recursive type the clauses are read coinductively, $\simeq_S$ being the greatest relation satisfying its unfolding, as on $\mathfrak{I}_\Gamma$. A map or an iterator *respects* $\simeq_S$ when it is related to itself, written $f \simeq_S f$.

**Lemma 35.** On maps and on iterators $\simeq_S$ is a partial equivalence: it is symmetric and transitive, so two related members each respect it.

*Proof.* On the base types $\simeq_S$ is an equivalence, by Definition 33 and Lemma 32. For functions, symmetry is symmetry of the base relations applied at input and output, and transitivity reads the middle map at $\gamma' \simeq_S \gamma'$: from $f \simeq_S g$, $g \simeq_S h$, and $\gamma \simeq_S \gamma'$ follow $f(\gamma) \simeq_S g(\gamma')$ and $g(\gamma') \simeq_S h(\gamma')$, whence $f(\gamma) \simeq_S h(\gamma')$; products inherit both componentwise and $\mathsf{Maybe}$ by cases. For iterators, the inverse $R^{-1}$ and the composite $R \circ R$ of the greatest relation $R$ satisfy the unfolding again, by the two arguments above read coinductively, so both are contained in $R$. Then $f \simeq_S g$ gives $f \simeq_S f$ by symmetry and transitivity, which is respect. $\square$

A map respecting $\simeq_S$ is one that descends to $\Gamma/\simeq_S$, and two maps related by $\simeq_S$ are two that descend to the same map there, each respecting it by Lemma 35. Reflexivity is the one property the function former does not preserve: $f \simeq_S f$ demands related outputs at every pair of related inputs and not at the equal ones alone, so it holds of a map exactly where the map descends, which is why respect is a condition rather than a given. Respect at two key sets is two conditions of which neither implies the other, since $\simeq \subseteq \simeq_S$ weakens the hypothesis and the conclusion together. A map that branches on a key outside $S$ is the case that separates them, respecting $\simeq$ and failing to respect $\simeq_S$.

**Definition 36.** Define the effect function *witnessed up to* $\simeq_S$ as:

$$\begin{aligned} \mathfrak{E}^S_\Gamma &:= (e : \Gamma \to \Gamma \times (\Gamma \to \Gamma)) \times (e \simeq_S e) \\ &\quad \times ((\gamma : \Gamma) \to (\delta : \Gamma) \to (g : \Gamma \to \Gamma) \to ((\delta, g) = e(\gamma) \to g(\delta) \simeq_S \gamma)) \end{aligned} \tag{36}$$

The clause $e \simeq_S e$ carries every constituent's respect, its instance at $\gamma \simeq_S \gamma$ relating each yielded inverse to itself. We write $\mathfrak{E}^*_\Gamma$ for $\mathfrak{E}^K_\Gamma$, and taking $\simeq$ to be equality on $\Gamma$ recovers Definition 8, every map being equal to itself. The key set is where a component's declarations enter: what Section 4 holds an effect function to is $\mathfrak{E}^S_\Gamma$ at the keys that component names (Definition 48).

The iterator of Section 3.1.3 is witnessed the same way, its continuation compared by Definition 34 and witnessed again by the recursion:

**Definition 37.** Define the effect iterator *witnessed up to* $\simeq_S$ as:

$$\begin{aligned} \mathfrak{I}^S_\Gamma &:= \mu\mathfrak{I}.\ (e : \Gamma \to \Gamma \times (\Gamma \to \Gamma) \times \mathsf{Maybe}(\mathfrak{I})) \times (e \simeq_S e) \\ &\quad \times ((\gamma : \Gamma) \to (\delta : \Gamma) \to (g : \Gamma \to \Gamma) \to (o : \mathsf{Maybe}(\mathfrak{I})) \to ((\delta, g, o) = e(\gamma) \to g(\delta) \simeq_S \gamma)) \end{aligned} \tag{37}$$

The embedding of Section 3.1.3 carries $\mathfrak{E}^S_\Gamma$ into $\mathfrak{I}^S_\Gamma$, and taking $\simeq$ to be equality on $\Gamma$ and $S = K$ recovers the witnessed $\mathfrak{I}^*_\Gamma$ of Definition 17.

**Lemma 38.** With $\mathfrak{E}^*_\Gamma$ read as in Definition 36, every equality of states asserted in Section 3.1 holds with $=$ replaced by $\simeq$, and the accumulator of every state reachable from $(\gamma_0, \mathrm{id}_\Gamma)$ respects $\simeq$. The same holds of any equivalence substituted for $\simeq$, the proof using no property of the relation beyond transitivity and respect.

*Proof.* An accumulator is a composition of inverses, each respecting $\simeq$, the clause $e \simeq e$ of Definition 36 read at $\gamma \simeq \gamma$, and a composition of maps respecting $\simeq$ respects $\simeq$, the base case being $\mathrm{id}_\Gamma$. The proofs of Section 3.1 then go through unchanged, respect being what carries a relation through an inverse: from $g_2(\delta_2) \simeq \delta_1$ and $g_1(\delta_1) \simeq \gamma$ respect gives $(g_1 \circ g_2)(\delta_2) \simeq \gamma$, which is the step each composition of inverses takes, and the soundness invariant of Theorem 7 reads $\varphi(\gamma) \simeq \gamma_0$ by that step. ☐

The stage shape of Definition 30 settles which readings of $\simeq$ a member admits, and with them its membership in the witnessed iterators, which is what lets a claim about one component be read at the keys that component names.

**Lemma 39.** Let $i \in \mathfrak{I}^{\mathcal{A}}_\Sigma(S', P)$ and let $S \subseteq K$ contain every key at which a stage of $i$ occurs. Then $i$ lies in $\mathfrak{I}^S_\Sigma$ (Definition 37); in particular $i$ and every inverse it yields respect $\simeq_S$, and the witnesses hold at equality.

*Proof.* By induction on the construction of Definition 30. The unit yields its argument, $\mathrm{id}_\Sigma$, and Nothing everywhere. At an operation stage performing $a \in \mathcal{A}_k$, let $\sigma \simeq_S \sigma'$; then $k \in S$, so $\sigma(k) \simeq_k \sigma'(k)$. $a$ respects $\simeq_k$ (Lemma 32), so it is defined at both or at neither and yields $\simeq_k$-related values, equal outcomes, and inverses carrying $\simeq_k$-related values to $\simeq_k$-related values, and its lift reads and writes $k$ alone; the states reached are therefore $\simeq_S$-related, the inverses are $\simeq_S$-related and respect $\simeq_S$, and the equal outcomes select one continuation, to which the induction hypothesis applies. At a provision stage at $k \in P \subseteq S$, the precondition $k \notin \mathrm{dom}(\sigma)$ holds at both or at neither, the states reached bind $k$ to the one value the stage carries, and the restriction it yields respects $\simeq_S$, two related states binding $k$ alike. The witness of each stage holds at equality, the lift of an operation's inverse restoring the value Definition 29 witnesses it to and the restriction reverting the extension (Definition 20), and equality gives $\simeq_S$. □

A stage reads the key it performs at and nothing else, so the hypothesis is met by taking $S$ to be the keys the member names.

### 3.4. Attaining Independence

On the strength of Section 3.3, this section supplies the condition that extends the local guarantees to a system of interleaved components. Section 3.4.1 defines the condition, namely *independence* of two effect functions: every transformation of one commutes with every transformation of the other, forward maps and yielded inverses alike. Section 3.4.2 then reduces independence of context-mediated effect functions to commutativity at single coeffects and refines the coeffect with a witness of that commutativity, parallel to the witness of an effect function.

#### *3.4.1. Effect Independence*

Reverting an effect at the state its own application produced is what Theorem 16 covers; reverting one at any other state is what this subsection covers. Two situations call for the latter. An inverse may be run while later effects are still in place, which is what removing one component from a running system amounts to; and one sequence may interleave the effects of several components, each holding the inverses of its own, so that the inverses of one component are separated by the applications of another. In both an inverse meets a state that foreign effects have moved, and whether it still reverts what it was built to revert is a question of commutation: what has to commute is every transformation one effect can perform with every transformation the other can perform, forward map and yielded inverse alike. A single accumulator settles neither situation, $\varphi$ being a composite that runs every inverse it holds in one order and all at once.

An iterator gives rise to several maps: the forward map of each iteration it can reach, and the inverse each yields at each context state. Commutation of two iterators relates the two collections rather than two single maps, and those collections are what the definitions below quantify over.

**Definition 40.** For an iterator $i \in \mathfrak{I}_\Gamma$, let $\mathrm{reach}(i)$ be the least set of iterators containing $i$ and closed under continuation. The *transformation monoid* $\mathfrak{M}(i)$ is the submonoid of $\Gamma \to \Gamma$ generated by the forward maps and the yielded inverses of every iterator in $\mathrm{reach}(i)$, and the *generators* of $\mathfrak{M}(i)$ are the elements of that generating set:

$$\begin{aligned}\mathrm{reach}(i) &:= \bigcap\{S \mid i \in S \wedge \forall i' \in S, \gamma \in \Gamma.\ i'(\gamma) = (-, -, \mathsf{Just}(i'')) \Rightarrow i'' \in S\} \\ \mathfrak{M}(i) &:= \langle \{\mathrm{pr}_1 \circ i' \mid i' \in \mathrm{reach}(i)\} \cup \{\mathrm{pr}_2(i'(\gamma)) \mid i' \in \mathrm{reach}(i), \gamma \in \Gamma\}\rangle\end{aligned} \tag{38}$$

Write $\mathrm{len}(i)$ for the supremum of $|C|$ over the chains $C \subseteq \mathrm{reach}(i)$ that continuation orders. Through the embedding of Section 3.1.3, $\mathfrak{M}(e)$ at an effect function $e$ is generated by the forward map of $e$ together with every inverse $e$ yields; an effect induced by a pair $(f, g) \in \mathfrak{T}_\Gamma$ has $f$ and $g$ for its generators, the inverse it yields being $g$ at every state.

**Lemma 41.** Commutation is settled on the generators, and $\diamond$ enlarges no transformation monoid. That is,

1. if every generator of $\mathfrak{M}(e_1)$ commutes with every generator of $\mathfrak{M}(e_2)$, then every element of $\mathfrak{M}(e_1)$ commutes with every element of $\mathfrak{M}(e_2)$;
2. $\mathfrak{M}(e_1 \diamond e_2) \subseteq \langle \mathfrak{M}(e_1) \cup \mathfrak{M}(e_2)\rangle$.

*Proof.*

1. The maps commuting with every generator of $\mathfrak{M}(e_2)$ form a submonoid of $\Gamma \to \Gamma$, since $\mathrm{id}_\Gamma$ lies in it and $f \circ f'$ does where $f$ and $f'$ do. That submonoid contains the generators of $\mathfrak{M}(e_1)$ by hypothesis and hence contains $\mathfrak{M}(e_1)$. Fixing $f \in \mathfrak{M}(e_1)$, the maps commuting with $f$ likewise form a submonoid containing the generators of $\mathfrak{M}(e_2)$ and hence $\mathfrak{M}(e_2)$.
2. By Definition 9 the forward map of $e_1 \diamond e_2$ is $(\mathrm{pr}_1 \circ e_1) \circ (\mathrm{pr}_1 \circ e_2)$ and the inverse it yields at any state is $s \circ t$ for an $s$ yielded by $e_2$ and a $t$ yielded by $e_1$. Every generator of $\mathfrak{M}(e_1 \diamond e_2)$ is therefore a composite of generators of the two. ☐

**Definition 42.** Iterators $i, j \in \mathfrak{I}_\Gamma$ are *independent* when

1. every transformation of one commutes with every transformation of the other,
$$\forall f \in \mathfrak{M}(i), g \in \mathfrak{M}(j).\quad f \circ g = g \circ f \tag{39}$$
2. neither one's transformations disturb what the other yields, inverse and continuation alike,
$$\forall i' \in \mathrm{reach}(i), g \in \mathfrak{M}(j), \gamma \in \Gamma.\quad \mathrm{pr}_{2,3}(i'(g(\gamma))) = \mathrm{pr}_{2,3}(i'(\gamma)) \tag{40}$$
and the same with $i$ and $j$ exchanged.

A family $(i_l)_{l \in L}$ is *pairwise independent* when $i_l$ and $i_{l'}$ are independent for every $l \neq l'$. A family may repeat an iterator, and holding one independent of itself is holding $\mathfrak{M}(i)$ commutative.

Read at effect functions through the embedding of Section 3.1.3, clause (2) compares the inverse alone, the continuation being Nothing at every state, and the result below is given at effect functions; Section 4 applies the definition at the iterators themselves. For effects induced by pairs $(f_1, g_1)$ and $(f_2, g_2)$, clause (1) is by Lemma 41(1) the commutation of the four pairs $f_1, f_2$; $g_1, g_2$; $f_1, g_2$; and $g_1, f_2$, and clause (2) holds outright, an induced effect yielding one inverse at every state. Commutation under $\diamond$ is a different property. What $e_1 \diamond e_2 = e_2 \diamond e_1$ equates is the composite forward map of the two orders with each other and the composite inverse of the two orders with each other, each inverse entering the composite at the state its own application produced; independence instead relates each transformation of one effect to each transformation of the other, a forward map paired with a foreign inverse included.

Under independence an inverse may be run at a state later effects have moved, and it withdraws there its own contribution and nothing else, whatever order the inverses are applied in:

**Theorem 43.** Let $e_1, \cdots, e_n \in \mathfrak{E}^*_\Gamma$ be pairwise independent and applied in order from $\gamma_0$, and let each $g_i$ be the inverse $e_i$ yields where it is applied. Applying the $n$ inverses at the state the sequence reaches, in the order of any permutation of $\{1, \cdots, n\}$, reaches $\gamma_0$.

*Proof.* Write $f_i := \mathrm{pr}_1 \circ e_i$, let $\delta_i := f_i(\delta_{i-1})$ with $\delta_0 := \gamma_0$, so that $g_i = \mathrm{pr}_2(e_i(\delta_{i-1}))$. Fix $j$ and write $\delta'_i := \big(f_i \circ \cdots \circ f_{j+1}\big)\big(\delta_{j-1}\big)$ for the states of the sequence with $e_j$ omitted, so that $\delta'_j = \delta_{j-1}$. Two claims hold for every $u$ with $j \le u \le n$.

(1) $\delta_u = f_j(\delta'_u)$ and $g_j(\delta_u) = \delta'_u$. The first equation is an induction on $u$: at $u = j$ it reads $\delta_j = f_j\big(\delta_{j-1}\big)$, which is the definition of $\delta_j$, and for the inductive step, $\delta_{u+1} = f_{u+1}(\delta_u) = f_{u+1}\big(f_j(\delta'_u)\big) = f_j\big(f_{u+1}(\delta'_u)\big) = f_j\big(\delta'_{u+1}\big)$, the middle equality being clause (1) of Definition 42 for $e_{u+1}$ and $e_j$, which are distinct effects of the family since $u+1 > j$. For the second equation, clause (1) carries $g_j$ out through the forward maps applied after $e_j$, leaving the witness of $e_j$ to be used at the one state it holds at:

$$g_j(\delta_u) = \big(g_j \circ f_u \circ \cdots \circ f_{j+1}\big)\big(\delta_j\big) = \big(f_u \circ \cdots \circ f_{j+1}\big)\big(g_j\big(f_j\big(\delta_{j-1}\big)\big)\big) = \delta'_u$$

the last equality resting on $g_j\big(f_j\big(\delta_{j-1}\big)\big) = \delta_{j-1}$, which is the witness Definition 8 requires of $e_j$ at $\delta_{j-1}$.

(2) Each $e_i$ with $i > j$ yields at $\delta'_{i-1}$ the same inverse $g_i$ it yields at $\delta_{i-1}$: by (1) the state $\delta_{i-1}$ is $f_j(\delta'_{i-1})$, and $f_j \in \mathfrak{M}\big(e_j\big)$, so clause (2) of Definition 42 for $e_i$ and $e_j$ gives $\mathrm{pr}_2\big(e_i\big(f_j(\delta'_{i-1})\big)\big) = \mathrm{pr}_2(e_i(\delta'_{i-1}))$.

The theorem follows by downward induction on $n$. Let the permutation begin with $j$. By (1) applying $g_j$ at $\delta_n$ reaches $\delta'_n$, the state the sequence with $e_j$ omitted reaches, and by (2) the inverses the remaining effects yielded there are the $g_i$ in hand. That sequence is pairwise independent, being a subfamily, so the induction hypothesis applies to it and to the rest of the permutation; the empty sequence reaches $\gamma_0$. □

Section 4.3.2 carries this conclusion to a trace of a whole system, where the steps of other fibers intervene between an effect and its revert.

### *3.4.2. Coeffect Commutativity*

The commutation Definition 42 requires is read up to $\simeq$ by Lemma 38, a yielded continuation compared as Definition 34 compares two iterators, and reading it that way is what makes it attainable at all: two operations may leave values that $\simeq_k$ identifies and still count as commuting. Of two operations it requires one thing more than of the effect functions their lifts induce, an operation yielding an outcome as well.

**Definition 44.** Operations $a$ and $a'$ are *independent* when their lifts are independent as effect functions (Definition 42) at every pair of arguments, and neither one's transformations disturb the outcome the other yields:

$$\forall x : X_a, g \in \mathfrak{M}(a'^{\Sigma}), \sigma \in \Sigma. \quad \mathrm{pr}_3(a^{\Sigma}(x)(g(\sigma))) = \mathrm{pr}_3(a^{\Sigma}(x)(\sigma)) \tag{41}$$

and the same with $a$ and $a'$ exchanged, writing $\mathfrak{M}(a^{\Sigma})$ for the submonoid generated by the forward maps and yielded inverses of the lifts $a^{\Sigma}(x)$ over every argument, as Definition 40 generates $\mathfrak{M}$. A key $k$ is *commutative* when any two operations of $\mathcal{A}_k$ are independent, an operation being held independent of itself as well.

Across distinct keys the condition holds outright.

**Theorem 45.** Operations at distinct keys are independent.

*Proof.* Let $a$ lie in $\mathcal{A}_k$ and $a'$ in $\mathcal{A}_{k'}$ with $k \neq k'$. By Definition 29 every generator of $\mathfrak{M}(a^\Sigma)$ is of the form $\sigma \mapsto \sigma[k \mapsto u(\sigma(k))]$ for a map $u$ on $\mathcal{V}_k$, being either the lift of a forward map or the lift of a yielded inverse, and likewise for $a'$ at $k'$. Two such maps commute, each reading and writing one key alone and the two keys differing, and Lemma 41(1) extends the commutation from the generators to the two monoids. For the second condition, what $a^\Sigma$ yields at $\sigma$, inverse and outcome alike, is determined by $\sigma(k)$, which every generator of $\mathfrak{M}(a'^\Sigma)$ leaves as it stands. □

The condition therefore turns on the pairs at one key, and the proof of their independence is made a constituent of the coeffect itself, as the proof that an inverse reverts is a constituent of the effect function (Definition 8):

**Definition 46.** A coeffect at $k$ (Definition 29) is *witnessed* when it carries, as a third constituent beside $\mathcal{V}_k$ and $\mathcal{A}_k$, a proof that $k$ is commutative (Definition 44).

The two witnesses are parallel: each certifies the condition its consumers would otherwise have to assume, the returned inverse reverting there and the operations commuting here, and each is supplied where the definition is written rather than checked where it is used. By Theorem 45 the proof concerns the operations of $k$ alone, so the obligation falls on the component providing the key and on no component consuming it; the examples below are how it is discharged. From here on every coeffect is witnessed, and Section 4 reads every key of $K$ so.

A key whose value is a table of entries is commutative when each registration takes an entry of its own, registration of a route or of an event listener being the representative case. The operation draws an identifier for the entry it adds and the inverse it yields removes that entry, so two registrations name two entries whatever they register: either order leaves a table that answers every test alike, and either registration can be withdrawn while the other stands. Replicated data types are designed to this condition and attach a unique tag to each addition for this very reason, a set whose additions and removals name a bare element having no such property [43]. A key whose value is an ordered chain is not commutative, since a middleware inserted before another sees a different request, and neither order can be withdrawn without disturbing the other.

The allocator of the opening example divides by what its interface publishes. Where the handles it hands out are compared by no operation of the key, no test observes them, so $\simeq_k$ relates two heaps up to a renaming of handles and allocation is commutative; a renaming is an equivalence the operations respect, contained in $\simeq_k$ by Lemma 32(2), and it is how CompCert relates the memory states of a program and of its translation [44]. Where the addresses are outcomes compared by equality, the outcome of a further allocation separates the two orders of allocation, and the key is not commutative. POSIX draws the same line across its own allocators: `mmap` may return any unused address and `creat` may assign any unused inode, whereas `open` is required to return the lowest available descriptor, and that requirement alone is what stops two descriptor allocations from commuting [45].

Definition 31 turns each of these divisions into a design choice. $\simeq_k$ is indistinguishability under the tests the operations of $k$ generate, so an interface publishing fewer outcomes admits fewer tests and coarsens the relation, and withholding an outcome its callers do not need can carry a key from one side of a division to the other. The scalable commutativity rule applies the same move across the POSIX interface, and reads commutativity as indistinguishability through an interface rather than equality of internal states [45].

Independence of two context-mediated iterators (Definition 30) turns on their keys alone:

**Theorem 47.** Let $i_1 \in \mathfrak{I}^{\mathcal{A}}_{\Sigma}(S_1, P_1)$ and $i_2 \in \mathfrak{I}^{\mathcal{A}}_{\Sigma}(S_2, P_2)$ with $P_1 \cap S_2 = P_2 \cap S_1 = \varnothing$, and let every key at which operations of both occur be commutative (Definition 44). Then $i_1$ and $i_2$ are independent (Definition 42).

*Proof.* Every iterator $\mathrm{reach}(i_l)$ contains is the unit or a stage, whose forward map and yielded inverses are the constituents of the operation or the set its head performs, so the generators of $\mathfrak{M}(i_l)$ are among those of the stages occurring in $i_l$, together with $\mathrm{id}_{\Sigma}$.

For clause (1) of Definition 42 it is enough, by Lemma 41(1), that those generators commute pairwise. Each is *key-local*: defined by the binding at one key alone, presence included, and writing that binding alone, being the lift of an operation's forward map or of an inverse it yields (Definition 29), the extension a provision stage takes, or the restriction it yields (Definition 20). Two key-local maps at distinct keys commute, each leaving what the other reads and writes as it stands. This settles every pair involving a provision-stage generator, whose key lies in one $P_l$ and hence outside the other member's every key by hypothesis, and every pair of operation generators at distinct keys, which is Theorem 45; a pair of operation generators at one key is covered by that key's commutativity.

For clause (2), take $i' \in \mathrm{reach}(i_1)$, $g \in \mathfrak{M}(i_2)$, and $\sigma \in \Sigma$. The unit yields $(\mathrm{id}_{\Sigma}, \mathsf{Nothing})$ everywhere. A provision stage at $k \in P_1$ yields the restriction at $k$ and its one continuation whatever the state, and is defined at $\sigma$ and $g(\sigma)$ alike, its precondition reading the presence of $k$, which every generator of $\mathfrak{M}(i_2)$ leaves as it stands. An operation stage at $k \in S_1$ yields what $a^{\Sigma}(x)$ yields at $\sigma(k)$, inverse and outcome; where no operation of $i_2$ occurs at $k$ the generators of $\mathfrak{M}(i_2)$ leave $\sigma(k)$ as it stands, and where one does the key is commutative by hypothesis, and independence of its operations, applied to one generator of $g$ at a time, yields the same inverse and the same outcome at $g(\sigma)$. Equal outcomes select one continuation, so the yields agree. □

With every coeffect witnessed, the commutativity hypothesis of Theorem 47 is supplied at every key (Definition 46), and only the disjointness $P_1 \cap S_2 = P_2 \cap S_1 = \varnothing$ remains to be checked of a pair; Section 4 reads that disjointness off the two components' declarations.

A component's effect function is the lift of a context-mediated iterator along the coeffect projection (Definition 56), and independence transfers to that lift, whose transformations move the projection alone. The assumption Section 3.4.1 leaves open is met that way, the witness of each coeffect supplying the commutativity Theorem 47 consumes, and with it the temporal composability of a whole system of components.

What the decomposition divides is a computation's commuting part from its order-sensitive part. The commuting part is carried by the effects: a component performs them in whatever order its task calls for, and Theorem 43 reverts them in whatever order the system finds convenient, no two components constraining each other. The order-sensitive part is carried by the coeffects, since a key whose operations do not commute is one whose order has to be imposed from outside the effects, and two places are available for imposing it. Within one component the accumulator imposes it, reverting in LIFO order whatever the effects (Theorem 16). Across components a declared coeffect imposes it, one component providing what another declares and the provision preceding the declaration's satisfaction (Section 3.2.2). Composability is thereby had at the grain of components rather than of single effects, which is the scale Section 4 works at.

One limit of the theorem is worth naming, and it is the hypothesis this section opened on: binding every shared location at a key is the paradigm's discipline and not a property of the construction, so a location the system cannot reify as a coeffect lies outside the boundary of Section 6.1 and outside the theorem with it.

# 4. A Calculus of Dynamic Composition

This section gives the theory of Section 3 an operational semantics. It decomposes a running system into *components*, each a triple of a coeffect specification, a provision, and a witnessed effect function. The instantiations of components are *fibers*, and the calculus supplies the rules that move them: orchestration rules, by which the orchestrator inserts and retires fibers, and lifecycle rules, by which the system activates and deactivates them unprompted. The metatheory then establishes temporal and spatial composability in their global form, the guarantees Section 3 reads of one component holding of every fiber of an arbitrary interleaving.

## 4.1. Components and Fibers

This section fixes the objects the rules act on: the *component*; the *fiber*, an instantiation of a component carrying a lifecycle state of its own; and the *registry*, which holds the fibers a state carries and from which the coeffect context is read off.

**Components.** A component is given as a triple, its coeffect side split into what it reads from the environment and what it provides to it.

**Definition 48.** A *component* over a context $\Gamma$ carrying both effects and coeffects (Definition 28) is defined as:

$$\mathfrak{C}_\Gamma := (d : \mathfrak{D}_\Gamma) \times (p : \mathfrak{P}_\Gamma) \times \mathfrak{I}_\Gamma^{d \cup p} \tag{42}$$

representing a triple $(d, p, e)$, where:

- $d : \mathfrak{D}_\Gamma$ is the coeffect specification of Definition 21, declaring the dependencies required from the environment;
- $p : \mathfrak{P}_\Gamma := \mathsf{Set}(K)$ is the *coeffect provision*, declaring the coeffect keys the component may provide, and no key outside $p$ is one its effect function installs a binding at;
- $e : \mathfrak{I}_\Gamma^{d \cup p}$ is the witnessed effect function, an effect iterator (Definition 17) witnessed up to $\simeq_{d \cup p}$ (Definition 37), defining the effects contributed when the component is active together with the inverses that withdraw them; a plain effect function enters through the embedding of Section 3.1.3.

Subscripts are taken on $\Gamma$ throughout, the coeffect context being one of its projections (Definition 28), so the $\mathfrak{D}_\Sigma$ of Definition 21 is written $\mathfrak{D}_\Gamma$ here.

**Fibers.** One component may be instantiated many times over, and each instantiation is activated and deactivated over time, carrying a lifecycle state of its own. We name such an instantiation a *fiber*. A fiber records the component that produced it, the fiber it was instantiated under, the coeffects it provides, and where in its lifecycle it stands.

**Definition 49.** Fix a set $\mathfrak{N}$ of fiber names. A *fiber* instantiating the component $(d, p, e) \in \mathfrak{C}_\Gamma$ is a tuple $\langle d, p, e, \pi, \sigma, \tau, \theta \rangle$, where

- $d : \mathfrak{D}_\Gamma$, $p : \mathfrak{P}_\Gamma$, and $e : \mathfrak{I}_\Gamma^{d \cup p}$ are the coeffect specification, provision, and effect function of Definition 48;
- $\pi : \mathfrak{N} \cup \{\mathsf{root}\}$ is the *parent*, the fiber this one was instantiated under, or the root marker $\mathsf{root}$;
- $\sigma : \Sigma$ is the fiber's own coeffect table (Definition 19), empty until it activates and written by its effects as they run;
- $\tau : \{\bot, \top\}$ is the *retirement* flag, $\bot$ in a fresh fiber and $\top$ once the orchestrator has retired the fiber;
- $\theta : \Theta_\Gamma$ is the *lifecycle state*:

$$\Theta_\Gamma := \mathsf{Inactive} \mid \mathsf{Reloading}(i, g, \omega) \mid \mathsf{Active}(g, \omega) \mid \mathsf{Unloading}(g, \omega) \tag{43}$$

  where $i : \mathfrak{I}_\Gamma^{d \cup p}$ is the remaining effect iterator, $g : \Gamma \to \Gamma$ the accumulator built so far, and $\omega : d \to \mathfrak{N}$ the *committed view*.

A fiber is *installed* when its lifecycle state carries an accumulator and a committed view:

$$\mathrm{installed}_n(\gamma) := \theta_n \neq \mathsf{Inactive} \tag{44}$$

and an installed fiber *resolves $k$ to $m$* when $\omega_n(k) = m$.

A *transition* is what moves a fiber from one lifecycle state to another, and between transitions the fiber rests at one of the two *settled* states, Inactive or Active, contributing nothing at the first and its effects at the second. A transition runs in one of two directions: an *activation* executes $e$, accumulating side effects on the context, and a *deactivation* applies the accumulator to recover the context. A transition in a real runtime is spread over an interval rather than taken in one step, so each direction has a state of its own that the fiber occupies while the transition runs, Reloading for an activation and Unloading for a deactivation. The committed view $\omega$ sends each key the fiber declares to the name of the fiber that provided it when the transition committed. Section 4.2 draws the four states as a state machine (Figure 1) and supplies the rules on its edges.

**Registry.** A state holds its fibers under their names, and both the identity of a fiber and the coeffect context of Section 3.2 are read off that arrangement.

**Definition 50.** Write $\mathfrak{F}_\Gamma$ for the set of fibers over $\Gamma$. A state $\gamma \in \Gamma$ carries a *registry*

$$F_\gamma : \mathfrak{N} \rightharpoonup \mathfrak{F}_\Gamma \tag{45}$$

a finite partial function whose parent pointers form a tree rooted at $\mathsf{root}$, together with whatever else in $\Gamma$ no fiber's $\sigma$ names. We write $\gamma(n)$ for $F_\gamma(n)$, and abbreviate a field of $\gamma(n)$ by subscripting it with $n$ where the state is clear, so that $d_n, p_n, e_n, \pi_n, \sigma_n, \tau_n, \theta_n$ are the fields of Definition 49 and $g_n, \omega_n$ the accumulator and committed view that $\theta_n$ carries; $\gamma[\theta_n \mapsto \theta']$, $\gamma[n \mapsto \langle \cdots \rangle]$, and $\gamma \setminus n$ are the states differing from $\gamma$ in one field, one fiber, and the presence of one fiber respectively.

A fiber's name is what gives it an identity that survives its own mutation: every rule below rewrites the lifecycle state of one fiber and leaves the others alone, so the rule has to say which one, and two fields refer to fibers rather than describe them, the parent $\pi$ and the committed view $\omega$. Names are atoms: no rule computes one, inspects its structure, or relates two of them by anything but equality, and introducing a fiber simply draws one not already in use. This is the discipline of dynamically created local names [40], used here for fiber identity.

Each fiber owning a table means the coeffect context is derived rather than stored: it is what the active fibers jointly provide.

$$\sigma_\gamma := \bigcup\{\sigma_m \mid m \in \operatorname{dom}(F_\gamma),\, \theta_m = \mathsf{Active}(-,-)\} \tag{46}$$

The union is well defined because a fiber's own table holds only the keys of its provision, $\operatorname{dom}(\sigma_n) \subseteq p_n$ (Definition 48), and the provisions of distinct fibers are disjoint, O-Insert admitting no fiber whose provision meets an existing one (Section 4.2.1), so each $k \in \operatorname{dom}(\sigma_\gamma)$ lies in the table of exactly one Active fiber, whose name we write $\operatorname{provider}_k(\gamma) \in \mathfrak{N}$ and call the *provider* of $k$. Each key therefore has one possible provider, fixed by the provisions and not by the state. No rule writes a table directly: the bindings a fiber provides are the provision stages its own effect function performs, which land in $\sigma_n$ and so are already part of the state $e_n$ returns, and they leave again with the accumulator, and the operations it performs at a declared key act on the value its provider's table holds (Definition 56). An effect function is built from such stages and from nothing else, so a location no key binds is one no fiber touches.

The disjointness the union rests on is where this chapter parts company with Section 3.2.3, and it simplifies the formalization rather than the systems it models. The isolation of Definition 24 lets one key resolve through a realm table, so that two fibers may provide the same key in different realms; a calculus carrying realms would relax disjointness to disjointness within a realm, resolving a declared key against the realm of the fiber declaring it (Section 4.4 supplies that reading). We read every key at one shared realm instead, and a system that wants several providers of one key keeps two routes: realms, and the broker of Section 6.2, one fiber providing the key and dispatching among implementations registered with it. Within the calculus, the disjointness restricts how often a component may be instantiated: one with a non-empty provision has one fiber at a time, so the many instantiations below are of components providing nothing, which is the common case of a component that only consumes, or that instantiates others.

With $\sigma_\gamma$ in hand, the satisfaction relation of Section 3.2.2 applies unchanged, $\gamma \vDash d$ abbreviating $\sigma_\gamma \vDash d$. A key lies in $\operatorname{dom}(\sigma_\gamma)$ exactly when some Active fiber has installed it, its provision being the keys it may install rather than the ones it has, so $\gamma \vDash d$ already requires that every declared key have an Active provider. Taking the union over Active fibers alone is what lets a fiber cease to provide before it has withdrawn anything, which Section 4.2.2 turns into the ordering discipline, and it fixes how a transition in progress reads: a Reloading or Unloading fiber reads its coeffects through the $\omega$ it holds and provides none of its own, so a key its transition has already written is not yet one a dependent may activate against.

The two declarations of Definition 48 are the two directions of one *interface*, $d$ what a component reads from the environment and $p$ what it writes to it, and the superscript on the effect function's type holds the witness to that interface. A fiber holds its own bindings whether or not they are published, so the projection the witness is read at is a second reading of the same tables.

**Definition 51.** Write

$$\sigma_\gamma^S := \bigcup\{\sigma_m|_S \mid m \in \operatorname{dom}(F_\gamma)\} \tag{47}$$

for the bindings the registry records at a set $S \subseteq K$ of keys, over every fiber and not the Active ones alone; disjointness of provisions makes it a function, and $\sigma_\gamma$ is the restriction of $\sigma_\gamma^K$ to the Active fibers. This is the coeffect projection Definition 33 is read at throughout this section, so that $\gamma \simeq_S \delta$ is $\sigma_\gamma^S \simeq \sigma_\delta^S$ and $\mathfrak{I}_\Gamma^S$ (Definition 37) is the effect functions witnessed at those keys.

The keys of both declarations are read off the tables rather than off $\sigma_\gamma$, and the witness condition is why: a binding a transition has written stays in the fiber's table before the fiber is Active, and that binding is what the inverse is held to remove, so the projection the witness is read at has to hold it wherever the fiber's lifecycle stands. The same reading keeps the relation where the control fields cannot move it: a write to a lifecycle state can move a table into or out of $\sigma_\gamma$ with every binding left as it stands, whereas $\sigma^S_\gamma$ moves only where some binding does. What the witness is thereby held to restore is the two declarations and nothing else.

### 4.2. The Calculus

This section gives the calculus: nine rules generating two relations. An *orchestration* rule, prefixed O- and written $\gamma \Rightarrow \delta$, is an action the orchestrator may perform; its premises say when the action is legal, not when it occurs. A *lifecycle* rule, prefixed L- and written $\gamma \longrightarrow \delta$, is a step the system takes unprompted whenever its premises hold. A sequence of steps interleaves the two. Eight of the nine lie on the edges of Figure 1; O-Retire writes the retirement flag alone and applies at every lifecycle state, so it would be a self-loop at each of the four nodes, and the figure omits it.

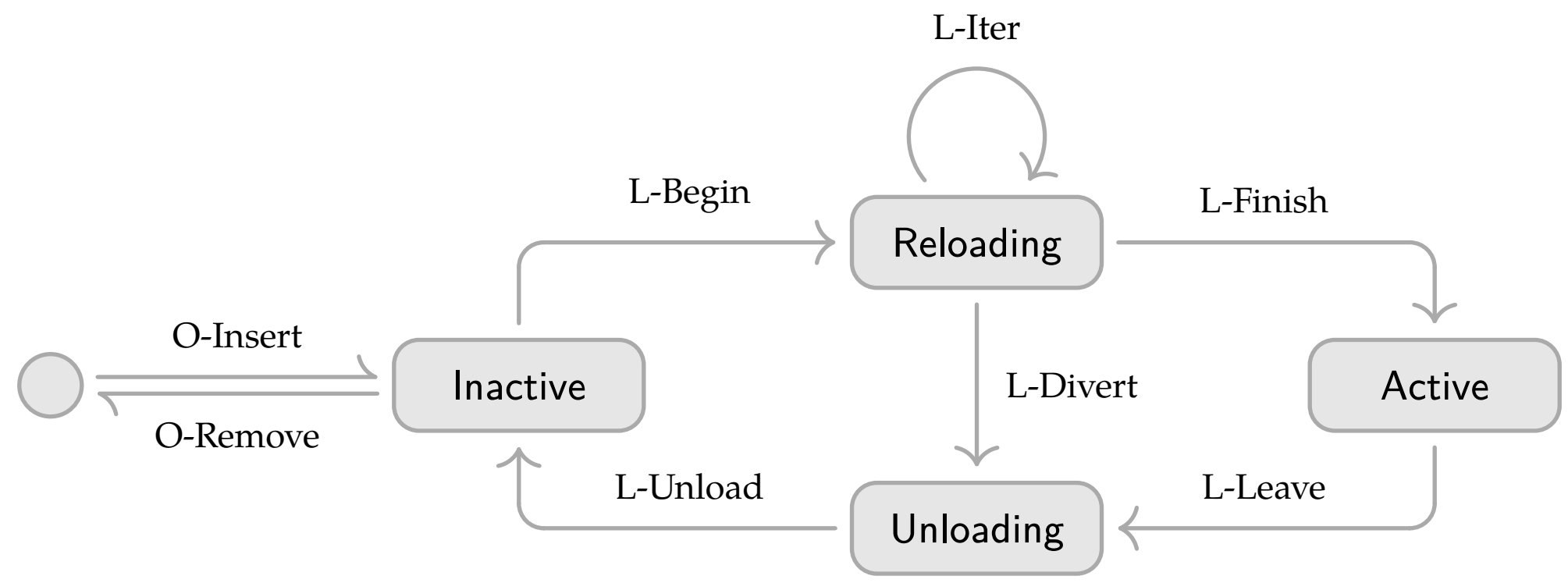

Figure 1 | The component lifecycle, with the empty node marking a fiber absent from the registry

#### 4.2.1. *Orchestration*

Insertion and retirement are the only external inputs: the orchestrator requests that a fiber exist or stop existing, and never sets its lifecycle state directly.

$$\frac{n \notin \mathrm{dom}(F_\gamma) \quad \pi \in \mathrm{dom}(F_\gamma) \cup \{\mathsf{root}\} \quad (d,p,e) \in \mathfrak{C}_\Gamma \quad \forall m \in \mathrm{dom}(F_\gamma).\ p \cap p_m = \varnothing}{\gamma \Rightarrow \gamma[n \mapsto \langle d,p,e,\pi,\varnothing,\bot,\mathsf{Inactive}\rangle]}\ \text{O-Insert}$$

$$\frac{n \in \mathrm{dom}(F_\gamma)}{\gamma \Rightarrow \gamma[\tau_n \mapsto \top]}\ \text{O-Retire}$$

$$\frac{\tau_n = \top \quad \theta_n = \mathsf{Inactive} \quad \sigma_n = \varnothing \quad \forall m.\ \pi_m \neq n}{\gamma \Rightarrow \gamma \setminus n}\ \text{O-Remove}$$

O-Retire is unconditional on the fiber's state because retiring is a request, and the lifecycle rules are what carry it out. Retirement is separated from removal for the same reason: a retired fiber that is still Active must first be deactivated, and removing it earlier would discard the accumulator and leak. The premise $\forall m.\ \pi_m \neq n$ keeps the tree well-formed by removing children before their parent, and $\sigma_n = \varnothing$ admits only an entry holding no bindings, so that a removal discards none: a deactivation leaves the entry so (Corollary 69), and it is what lets Theorem 68 count a removal as no change to the tables. The last premise of O-Insert is where the single-source discipline is imposed: a key has one possible provider because the orchestrator may not admit a second component declaring it.

**Instantiation.** A component may instantiate another while installing its effects, which is what a plugin host does when a plugin loads plugins of its own. The rules so far leave the registry to the orchestration rules alone, so such an instantiation has nowhere to happen. One primitive gives it somewhere.

**Definition 52.** An iteration of $e_n$ may *instantiate* a component $(d, p, e) \in \mathfrak{C}_\Gamma$. In place of a state map it takes the O-Insert of that component with $\pi = n$, and it yields as its inverse the O-Retire of the fiber so instantiated. The rule draws the name, subject to the freshness premise of O-Insert, and hands it to the effect function.

The inverse retires rather than removes, and the reason is that an inverse has to apply wherever it is reached. O-Remove carries premises, so an inverse built from it can fail to: a parent whose child is still $\mathsf{Active}$ could not run its accumulator, and no rule would move the child, since Definition 53 does not read the fiber tree. O-Retire has $n \in \mathrm{dom}(F_\gamma)$ as its only premise. The entry it leaves behind at the state the instantiation was taken is retired, $\mathsf{Inactive}$, and holds an empty table, which is the vestigial entry of Lemma 62: it differs from the absence of the fiber in control fields alone, and no rule tells the two apart.

Retiring a child sets $\tau$ and so takes its target view to $\bot$, after which the ordinary rules carry it back to $\mathsf{Inactive}$. The parent is not made to wait, O-Retire being unconditional, so L-Unload applies to the parent whether or not the child has left. A grandchild is reached one level at a time, the child's own accumulator retiring what the child instantiated. Theorem 73 covers this cascade and the one the guard of Section 4.2.2 imposes along coeffects together.

### 4.2.2. *Lifecycle*

The six lifecycle rules divide by the direction they move a fiber in: an activation carries a fiber toward a target view it does not yet hold, and a deactivation carries one away from a committed view that is no longer its target.

**Target views.** The rules compare each fiber against a *target*, namely whether it ought to be running and against which resolution of its dependencies. The target is not a property of the fiber alone, since the keys a fiber declares are resolved against the whole state, so it is a predicate on that state.

**Definition 53.** The *target view* of $n$ at $\gamma$ maps each declared key to its provider, so it is a total map $d_n \to \mathfrak{N}$, and is $\bot$ when $n$ ought not to be running at all:

$$\mathrm{target}_n(\gamma) := \begin{cases} \bot & \text{if } \tau_n \vee \neg(\gamma \vDash d_n) \\ (k \in d_n) \mapsto \mathrm{provider}_k(\gamma) & \text{otherwise} \end{cases} \tag{48}$$

A state is *quiescent* when every fiber has settled at its target view, no transition left in progress:

$$\mathrm{quiet}(\gamma) := \forall n \in \mathrm{dom}(F_\gamma).\ \begin{cases} \mathrm{target}_n(\gamma) = \bot & \text{if } \theta_n = \mathsf{Inactive} \\ \mathrm{target}_n(\gamma) = \omega_n & \text{if } \theta_n = \mathsf{Active}(-, \omega_n) \\ \bot & \text{otherwise} \end{cases} \tag{49}$$

The target answers to two things and to nothing else: retirement, through $\tau_n$, and coeffect resolution, through $\gamma \vDash d_n$ and $\text{provider}_k$, each declared key being read off $\sigma_\gamma$ at the one shared realm of Section 4.1.

The committed view of Definition 49 has the same type as the target view, and the lifecycle is driven by comparing them: $\omega_n$ is the resolution $n$ activated against, $\text{target}_n(\gamma)$ the one it should be running against, and every rule below fires on their agreeing or differing. This is the reactive discipline of Section 3.2: a transition is initiated whenever the target view changes, regardless of which of the two moved it. Recording a provider rather than a value is what makes the comparison usable, since a different fiber providing an equal value would otherwise compare equal. The value a component reads is reached through the view, since the provider's table holds that value, and the implementation holds the map in `fiber.committed` and a hash of it in `fiber.target` (Section 5.1.3).

**Activation.** An activation may execute multiple effects in sequence, and the deactivation must revert them. The effect iterator $e_n$ models such an activation (Section 3.1.3), each of its iterations yielding the modified context, an inverse, and a continuation, and a component's whole activation is one run of $e_n$: L-Begin enters Reloading, L-Iter takes one iteration, and L-Finish lands the last.

$$\frac{\theta_n = \mathsf{Inactive} \quad \omega = \text{target}_n(\gamma) \neq \bot}{\gamma \longrightarrow \gamma[\theta_n \mapsto \mathsf{Reloading}(e_n, \text{id}_\Gamma, \omega)]} \text{ L-Begin}$$

$$\frac{\theta_n = \mathsf{Reloading}(i, g, \omega) \quad \text{target}_n(\gamma) = \omega \quad i(\gamma) = (\delta, h, \mathsf{Just}(i'))}{\gamma \longrightarrow \delta[\theta_n \mapsto \mathsf{Reloading}(i', g \circ h, \omega)]} \text{ L-Iter}$$

$$\frac{\theta_n = \mathsf{Reloading}(i, g, \omega) \quad \text{target}_n(\gamma) = \omega \quad i(\gamma) = (\delta, h, \mathsf{Nothing})}{\gamma \longrightarrow \delta[\theta_n \mapsto \mathsf{Active}(g \circ h, \omega)]} \text{ L-Finish}$$

Each iteration composes the newly yielded inverse onto the accumulator as $g \circ h$, following Definition 18, so that the accumulator applies the inverses in last-in-first-out order.

**Deactivation.** A deactivation enters from either stage of the lifecycle: L-Divert takes out a fiber whose transition is still in progress, and L-Leave one that is Active. It cannot be taken in one step. A component being torn down because its provider is going away is running its own teardown code, which may need the very coeffect that is being withdrawn; closing a connection pool typically means handing the connections back to whatever provided them. The consumer must therefore still be able to read the key throughout its own deactivation, and the provider's withdrawal must take effect only afterwards, which gives content to the ordering Section 3.2 requires of dependencies and dependents. A deactivation taken in one step would remove the provisions and run the inverse together, with no interval between them for a consumer's teardown to occupy. The rules below therefore separate the decision from the act, and the act is guarded by the following condition.

**Definition 54.** The fiber $n$ is *relied upon* at $\gamma$ when some other installed fiber resolves a key to it:

$$\text{relied}_n(\gamma) := \exists m \in \text{dom}(F_\gamma),\, k \in d_m.\ m \neq n \wedge \text{installed}_m(\gamma) \wedge \omega_m(k) = n \tag{50}$$

$$\frac{\theta_n = \mathsf{Reloading}(i, g, \omega) \quad \mathrm{target}_n(\gamma) \neq \omega \quad (\delta, h) = (\gamma, \mathrm{id}_\Gamma) \vee i(\gamma) = (\delta, h, -)}{\gamma \longrightarrow \delta[\theta_n \mapsto \mathsf{Unloading}(g \circ h, \omega)]} \text{ L-Divert}$$

$$\frac{\theta_n = \mathsf{Active}(g, \omega) \quad \mathrm{target}_n(\gamma) \neq \omega}{\gamma \longrightarrow \gamma[\theta_n \mapsto \mathsf{Unloading}(g, \omega)]} \text{ L-Leave}$$

$$\frac{\theta_n = \mathsf{Unloading}(g, \omega) \quad \neg\, \mathrm{relied}_n(\gamma) \quad g(\gamma) = \delta}{\gamma \longrightarrow \delta[\theta_n \mapsto \mathsf{Inactive}]} \text{ L-Unload}$$

L-Divert may fall between any two consecutive iterations of a transition, routing the fiber into Unloading with the inverses accumulated so far rather than applying them on the spot. Routing through Active instead would let the fiber provide its coeffects for the length of one step and oblige its dependents to activate against a component that is already leaving. The first of L-Divert's two alternatives *aborts* the iteration the fiber is holding, which only an iteration boundary makes possible, so the granularity at which a divert may fall is that of the iterator; the second lets that iteration land, serving the host Section 4.4 admits, in which an iteration in flight cannot be declined.

L-Leave records the decision to deactivate without acting on it, which stops the fiber providing its coeffects while leaving its own committed view and everyone else's intact. L-Unload applies the accumulator, discards the committed view, and leaves the fiber Inactive; it is the only rule in the calculus that applies an accumulator, an L-Divert routing here rather than applying one of its own.

The two requirements are then carried by different parts of the form: the consumer's reading by the committed view, which L-Unload discards as its last act, and the deferral of the withdrawal by the premise $\neg\, \mathrm{relied}_n(\gamma)$, which we call the *guard* and which holds a provider's withdrawal back until every consumer that resolves a key to $n$ has gone. For a fiber L-Divert takes out of its first transition the guard is vacuous, a fiber that has never been Active providing nothing and appearing in no committed view. Theorem 70 establishes both requirements.

The guard is imposed per binding rather than per fiber: $\mathrm{relied}_n(\gamma)$ tests whether some committed view names $n$, so a fiber that declares none of $n$'s keys is no obstacle, and neither is one that resolved a key of $n$'s in another realm (Section 3.2.3). Under the single-source discipline of O-Insert the per-binding reading coincides with the coarser test $\exists m \neq n,\, k \in d_m.\ \mathrm{installed}_m(\gamma) \wedge k \in p_n$, a key having one possible provider there.

A guard of this kind ordinarily deadlocks. What keeps it from doing so is Unloading together with $\sigma_\gamma$ being the union over Active fibers alone: once L-Leave or L-Divert has marked $n$, its table leaves $\sigma_\gamma$, so no target view can name $n$ any longer, and every consumer that committed to $n$ is itself on its way out. Theorem 73 turns that into the claim that the guard always releases.

The guard orders deactivations along coeffects and not along the fiber tree: a parent may run its inverse while a child of it is still Unloading, since relied speaks only of committed views. Parent and child are accordingly ordered more weakly than Theorem 70 orders a provider and its consumer, and a parent and a child whose effects meet at a shared key are governed by the pairwise independence of Lemma 66 instead.

The rules are nondeterministic: several fibers may hold a committed view differing from their target view, and the relation commits to no order among them. They are also *reactive only*, in that no rule mentions a scheduler; the steps are any sequence of rule applications, so a theorem proved over all such sequences holds for every scheduling policy a runtime might adopt.

### 4.2.3. *Confinement*

With instantiation the one exception in hand, the discipline an effect function is held to can be given. It bounds what an application of an iteration writes, so that the rule applying it accounts for every other change, and what it reads, so that a fiber sees the coeffects it declared and no more of the registry. Bounding the writes is what lets Section 4.3 read Table 1 as a complete inventory of them.

**Definition 55.** A map $f : \Gamma \to \Gamma$ is *confined to $n$* when for every $\gamma \in \Gamma$ with $n \in \mathrm{dom}(F_\gamma)$, writing $\delta = f(\gamma)$,

1. (*Writes.*) $\mathrm{dom}(F_\delta) = \mathrm{dom}(F_\gamma)$, $\delta(m)$ and $\gamma(m)$ differ at most in $\sigma_m|_{d_n}$ for every $m \in \mathrm{dom}(F_\gamma)$ with $m \neq n$, and $\delta(n)$ and $\gamma(n)$ differ in $\sigma$ alone;
2. (*Reads.*) two states agreeing on $\sigma_n$ and on the restrictions $\sigma_m|_{d_n}$ for every $m \in \mathrm{dom}(F_\gamma)$ are carried by $f$ to states agreeing on the same two.

An effect function $e$ is *confined to $n$* when every iterator $i \in \mathrm{reach}(e)$ either instantiates a component (Definition 52) or has both its state map $\mathrm{pr}_1 \circ i$ and every inverse it yields confined to $n$.

An instantiation writes the entry O-Insert writes, at the one name it draws, and nothing else; the O-Retire it yields as its inverse writes the $\tau$ of that name and nothing else. An application of either kind therefore writes no control field of a fiber already present, save that one $\tau$, and reads none at all.

Clause (1) permits a write outside the fiber's own table, and there is exactly one kind: the value at a declared key lives in the provider's table, so a component operating on a coeffect it declared moves $\sigma_m|_{d_n}$ for the $m$ providing it. Clause (2) is why a component may read the values it declared as well: an effect function that reads no table but $\sigma_n$ would be unable to use its own coeffects. What it may neither read nor write is a table outside the two declarations, any control field, or anything no table holds, which is what keeps a component from branching on the lifecycle state of a fiber it did not declare.

The context paradigm fixes the form of an effect function — a sequence of stages, each a coeffect operation, a provision, or an instantiation — and confinement is a consequence of that form.

**Definition 56.** A stage of Definition 30 *lifts along the coeffect projection*: it acts on the one table that holds the binding at its key, over every fiber and not the Active ones alone as Definition 51 reads the tables, an extension landing in the table of the fiber acting, and it leaves the rest of the state as it stands. The *context-mediated* iterators for $n$ form the least set $\mathfrak{I}^{\mathcal{A}}_{\Gamma}(n)$ of iterators on $\Gamma$ that contains the unit and, each continuation drawn from Nothing and the members, contains three iteration forms: the lift of an operation stage at a key of $d_n \cup p_n$, the lift of a provision stage at a key of $p_n$, and an instantiation (Definition 52). Every fiber's effect function is required to lie in $\mathfrak{I}^{\mathcal{A}}_{\Gamma}(n)$ at that fiber.

**Lemma 57.** A member of $\mathfrak{I}^{\mathcal{A}}_{\Gamma}(n)$ is confined to $n$, and it lies in $\mathfrak{I}^{d_n \cup p_n}_{\Gamma}$ (Definition 37) at the projection Definition 51 fixes, which is the witness Definition 48 requires of it.

*Proof.* For confinement, by induction on the construction. An instantiation is the exception Definition 55 carves out. The lift of a stage at $k$ writes the binding at $k$ and nothing else: at $k \in p_n$ that binding lies in $\sigma_n$, by disjointness of provisions, and at $k \in d_n$ it lies in some $\sigma_m|_{d_n}$, which is clause (1); the inverse it yields, the lift of the operation's inverse or the restriction at $k \in p_n$, writes the same binding or removes it from $\sigma_n$. For clause (2), a stage reads the binding at its key and its presence, both determined by $\sigma_n$ together with the $\sigma_m|_{d_n}$, and writes what it read into the same two parts.

For membership, the induction of Lemma 39 carries over stage by stage: at an operation or provision stage the argument there applies as it stands, the lift moving the tables that jointly carry $\sigma_\gamma^{d_n \cup p_n}$ as the stage moves the projection and moving nothing else, so respect and witness at $\simeq_{d_n \cup p_n}$ on $\Sigma$ read as the same conditions on $\Gamma$ at the projection of Definition 51; an instantiating iteration adds an entry holding an empty table and yields the O-Retire writing one $\tau$, both invisible to $\simeq_{d_n \cup p_n}$, so its clauses hold outright. □

### 4.3. Metatheory

This section establishes the metatheory of the calculus: that every rule preserves the well-formedness of the registry (Section 4.3.1); that temporal and spatial composability hold in their global form, one fiber's guarantee surviving whatever the other fibers do in between (Section 4.3.2, Section 4.3.3); that the system quiesces (Section 4.3.4); and that it quiesces where a load of the same configuration from scratch would have left it (Section 4.3.5).

Every property below is a property of a sequence of steps, so we index the steps and read the fields of a state off that index.

**Definition 58.** Index the steps by $t$, so that $\gamma^t$ is the state the first $t$ of them reach, and write

$$\mathrm{step}^t := r(n) \tag{51}$$

for the step taken at $\gamma^t$: the rule $r$ it applies, one of the nine, and the name $n \in \mathfrak{N}$ it applies that rule at. The sequence starts at a $\gamma^0$ with $\mathrm{dom}(F^0) = \varnothing$, so every fiber comes into existence by an O-Insert, whether the orchestrator's or one an iteration takes (Definition 52). A field of $\gamma^t$ carries the index as a superscript, so that $\theta_n^t$, $\omega_n^t$, $\sigma_n^t$, $g_n^t$, and $i_n^t$ are the lifecycle state, committed view, table, accumulator, and remaining iterator of $n$ at $\gamma^t$, and $F^t$ and $\sigma^t$ the registry and coeffect context of $\gamma^t$ itself, the $F_\gamma$ and $\sigma_\gamma$ of Definition 50 read there. Predicates take the state as their argument and everything else as a subscript, so $\mathrm{installed}_n^t$, $\mathrm{target}_n^t$, $\mathrm{relied}_n^t$, and $\mathrm{quiet}^t$ are the predicates of Definition 49, Definition 53, and Definition 54 at $\gamma^t$. An *episode* of $n$ is a maximal interval $[b, u]$ of indices throughout which $\mathrm{installed}_n^t$ holds. It *opens* at $b$, where $b > 0$ and $\neg\,\mathrm{installed}_n^{b-1}$, the empty $F^0$ leaving no fiber installed at the outset; it *closes* at $u$ when $\mathrm{installed}_n^u$ and not $\mathrm{installed}_n^{u+1}$, which a final episode need not do.

Every rule of Section 4.2 concludes in the shape $\gamma \longrightarrow \delta[\cdots]$, where the premises compute $\delta$ from $\gamma$ and leave it as $\gamma$ where they compute nothing, and the bracket edits named fields of the registry. The two halves are named separately, and both are maps on all of $\Gamma$. The *state map* of a step taken at $\gamma^t$ by a rule acting on $n$ is

$$\Psi^t := \begin{cases} \mathrm{pr}_1 \circ i & \text{at L-Iter, L-Finish, and a landing L-Divert} \\ g & \text{at L-Unload} \\ \mathrm{id}_\Gamma & \text{at every other rule} \end{cases} \tag{52}$$

where $i$ and $g$ are the iterator and the accumulator that $\theta_n^t$ carries, and the *edit* $\mathrm{edit}^t : \Gamma \to \Gamma$ is the bracket read as a function, assigning to the fields it names the values the premises computed at $\gamma^t$. Both are therefore fixed by $\mathrm{step}^t$ together with $\gamma^t$ and defined at every state, which is what lets Theorem 68 and Lemma 78 evaluate them away from $\gamma^t$. Each step factors as

$$\gamma^{t+1} = \mathrm{edit}^t(\Psi^t(\gamma^t)) \tag{53}$$

At L-Unload, for instance, $\mathrm{edit}^t$ is $[\theta_n \mapsto \mathsf{Inactive}]$, and at O-Remove it is the removal $\setminus n$, which is why the second half is an edit rather than an assignment. The fields divide along the same seam: the *tables* $\sigma_m$, which no $\mathrm{edit}^t$ writes once the O-Insert creating $m$ has set it empty, and the *control fields* $\theta_m, \tau_m, \pi_m, d_m, p_m, e_m$ together with $\mathrm{dom}(F_\gamma)$, which no $\Psi^t$ writes save through the primitive of Definition 52.

A rule reads the control fields to decide whether it applies, so the relation two whole states are compared at has to keep them. It is Definition 33 over the registry conjoined with agreement on the registry's domain and on every control field of every fiber:

$$\gamma \simeq \delta \quad := \quad \sigma^K_\gamma \simeq \sigma^K_\delta \wedge \mathrm{dom}(F_\gamma) = \mathrm{dom}(F_\delta) \wedge \forall n,\, c \in \{\theta, \tau, \pi, d, p, e\}.\ c(\gamma(n)) \simeq c(\delta(n)) \tag{54}$$

A field of function type, as $e_n$ and the $g$ inside $\theta_n$ are, is compared as Definition 34 compares maps and iterators, and a field of any other type by equality. The results below compare states at the coarser readings Definition 51 gives, $\simeq_K$ where every table is in question and $\simeq_{d_n \cup p_n}$ where one fiber's is, and the three are nested rather than crosswise, $\simeq$ implying $\simeq_K$ and $\simeq_K$ implying $\simeq_S$ at every $S$. Lemma 60 establishes the first once for all nine rules.

Table 1 is the nine rules of Section 4.2 read as such writes. The accumulator, the committed view, and the remaining iterator are constituents of $\theta_n$, so the third column records the writes to them as well, and $h$ there names the inverse the iteration of the fourth column yields, $\mathrm{id}_\Gamma$ where L-Divert aborts that iteration. Where a $\Psi^t$ built from an iterator instantiates a fiber (Definition 52), that instantiation carries the writes of the O-Insert row at the name it draws, and an L-Unload whose accumulator retires one carries those of the O-Retire row. Every case analysis below is a lookup in the table, and five lookups recur often enough to name.

| **rule** | $\boldsymbol{\theta^t_n}$ | $\boldsymbol{\theta^{t+1}_n}$ | $\boldsymbol{\Psi^t}$ | **control fields edited** |
|---|---|---|---|---|
| O-Insert | undefined | $\mathsf{Inactive}$ | $\mathrm{id}_\Gamma$ | $\mathrm{dom}(F_\gamma)$ |
| O-Retire | unconstrained | unchanged | $\mathrm{id}_\Gamma$ | $\tau_n$ |
| O-Remove | $\mathsf{Inactive}$ | undefined | $\mathrm{id}_\Gamma$ | $\mathrm{dom}(F_\gamma)$ |
| L-Begin | $\mathsf{Inactive}$ | $\mathsf{Reloading}(e_n, \mathrm{id}_\Gamma, \omega)$ | $\mathrm{id}_\Gamma$ | $\theta_n$ |
| L-Iter | $\mathsf{Reloading}(i, g, \omega)$ | $\mathsf{Reloading}(i', g \circ h, \omega)$ | $\mathrm{pr}_1 \circ i$ | $\theta_n$ |
| L-Finish | $\mathsf{Reloading}(i, g, \omega)$ | $\mathsf{Active}(g \circ h, \omega)$ | $\mathrm{pr}_1 \circ i$ | $\theta_n$ |
| L-Divert | $\mathsf{Reloading}(i, g, \omega)$ | $\mathsf{Unloading}(g \circ h, \omega)$ | $\mathrm{id}_\Gamma$ or $\mathrm{pr}_1 \circ i$ | $\theta_n$ |
| L-Leave | $\mathsf{Active}(g, \omega)$ | $\mathsf{Unloading}(g, \omega)$ | $\mathrm{id}_\Gamma$ | $\theta_n$ |
| L-Unload | $\mathsf{Unloading}(g, \omega)$ | $\mathsf{Inactive}$ | $g$ | $\theta_n$ |

Table 1 | The rules as writes on the fiber $n$ they act on, where $\mathrm{step}^t$ is that rule applied at $n$.

**Lemma 59.** Reading Table 1 together with Definition 55, for every step $t$ and all fibers $m, n$ present at $\gamma^t$:

1. a table moves only inside a $\Psi^t$: $\sigma^{t+1}_m \neq \sigma^t_m$ only where step $t$ acts on $m$, or acts on an $n \neq m$ with $\mathrm{dom}(\sigma^t_m) \cap d_n \neq \varnothing$, in which case the two tables differ in values at keys of $d_n$ alone and $\mathrm{dom}(\sigma_m)$ is unchanged;
2. $\omega_n$ comes into existence only where $\mathrm{step}^t = \text{L-Begin}(n)$ and ceases only where $\mathrm{step}^t = \text{L-Unload}(n)$, so $\omega^t_n$ is constant for $t$ in an episode of $n$;
3. $\Psi^t = g^t_n$ only where $\mathrm{step}^t = \text{L-Unload}(n)$, and no other step applies $g_n$ to the state;

4. $\neg\,\mathrm{installed}_n^t \wedge \mathrm{installed}_n^{t+1} \Rightarrow \mathrm{step}^t = \text{L-Begin}(n)$, and $\mathrm{installed}_n^t \wedge \neg\,\mathrm{installed}_n^{t+1} \Rightarrow \mathrm{step}^t = \text{L-Unload}(n)$;
5. $\pi_n$, $d_n$, $p_n$, and $e_n$ come into existence with the entry of $n$ and are never written again, and $\tau_n$ is monotone, written only at $\top$ and only by an O-Retire.

*Proof.* Let step $t$ apply $r$ at $n$. By Definition 58 it factors as $\mathrm{edit}^t \circ \Psi^t$, where $\mathrm{edit}^t$ writes the fields the fifth column of Table 1 names and nothing else, and $\Psi^t$ is $\mathrm{id}_\Gamma$, an application of one of $n$'s iterations, or the accumulator $g_n^t$, which is a composite of the inverses those iterations yielded. Each of the three is confined to $n$ by Lemma 57, so $\Psi^t$ writes no field of a fiber present at $\gamma^t$ but $\sigma_n$ and the values other tables hold at keys of $d_n$, their domains untouched, together with the entry an instantiation adds and the $\tau$ its inverse writes. The two halves therefore partition the writes, and each clause is that partition read at one field. One reading of the second and third columns is used twice: Inactive is the one lifecycle state carrying no committed view, L-Begin the one rule leading out of it, and L-Unload the one rule leading into it, while every other row carries the $\omega$ of its premise into its conclusion unchanged.

(1) An $\mathrm{edit}^t$ writes no table, the fifth column naming none, and what a $\Psi^t$ writes outside $\sigma_n$ is values at keys of $d_n$ in the tables holding them, the domains unchanged. So $\sigma_m$ can move only inside a $\Psi^t$, at the acting fiber or at the keys of $d_n$ its table holds.

(2) $\omega_n$ is a constituent of $\theta_n$, which only an $\mathrm{edit}^t$ writes and only at the fiber the step acts on, so by the reading above $\omega_n$ comes into existence at an L-Begin of $n$ and ceases at an L-Unload of $n$. An episode of $n$ is an interval on which $\mathrm{installed}_n$ holds, hence one throughout which $\omega_n$ is defined, so neither rule falls in its interior.

(3) The fourth column, where an accumulator appears at L-Unload alone: the other rules take a forward map $\mathrm{pr}_1 \circ i$ or $\mathrm{id}_\Gamma$, and no $\mathrm{edit}^t$ applies a map to the state at all.

(4) $\mathrm{installed}_n$ is $\theta_n \neq \mathsf{Inactive}$, and by the reading above L-Begin and L-Unload are the only rules whose premise and conclusion differ in whether $\theta_n$ is Inactive. A step acting on some $m \neq n$ writes no $\theta_n$, and the entry an instantiation adds is at a name not present at $\gamma^t$.

(5) No row of the fifth column names a $\pi$, $d$, $p$, or $e$; those come into existence with the entry O-Insert adds, which its conclusion writes, as does the O-Insert an instantiation takes. Only O-Retire writes a $\tau$, at $\top$, whether taken by the orchestrator or as the inverse of an instantiation (Definition 52); O-Insert sets $\tau = \bot$ at a name not already present, so no step returns a $\tau$ to $\bot$.□

Three further lookups say what the rules cannot see. The first is that they read the state only through the observations above, so that the whole calculus descends to $\Gamma/\simeq$.

**Lemma 60. ($\simeq$-invariance.)** Let $\gamma \simeq \gamma'$ as read above. Then a rule of Section 4.2 applies at $\gamma$ acting on $n$ if and only if it applies at $\gamma'$ acting on $n$, and the states the two applications reach are again related by $\simeq$.

*Proof.* Every premise of Section 4.2 is of one of four kinds, and each reads a constituent the relation keeps. A premise matching $\theta_n$ or $\tau_n$ against a pattern, and the premise $\forall m.\ \pi_m \neq n$ of O-Remove, read control fields. The premises $(d, p, e) \in \mathfrak{C}_\Gamma$ and $\forall m.\ p \cap p_m = \varnothing$ of O-Insert read $d$, $p$, and $e$. A premise mentioning $\mathrm{target}_n$ or $\mathrm{relied}_n$ reads $\tau_n$, the committed views inside the $\theta_m$, and $\mathrm{dom}(\sigma_\gamma)$, which Definition 50 computes from the $\theta_m$ and the $\mathrm{dom}(\sigma_m)$, and Definition 33 relates two coeffect contexts only where their domains agree. The remaining premises read $\mathrm{dom}(F_\gamma)$. Two $\simeq$-related states have $\simeq$-related $\sigma_\gamma$, the relation comparing every table and the control fields deciding which of them $\sigma_\gamma$ unions, and no premise reads a value $\sigma_\gamma(k)$ otherwise than up to $\simeq_k$, so no premise separates two $\simeq$-related states.

For the conclusion, $\gamma^{t+1} = \text{edit}^t(\Psi^t(\gamma^t))$ by Definition 58. The values an $\text{edit}^t$ assigns are the constituents of the premises it matched, related at the two states by the paragraph above and by the clause $e \simeq e$ of Definition 37, which relates the triples an iterator yields at related states. And $\Psi^t$ carries $\simeq$-related states to $\simeq$-related states: it is $\text{id}_\Gamma$, an iteration of $e_n$, or the accumulator inside $\theta_n$, and the latter two respect $\simeq_{d_n \cup p_n}$ by Lemma 57, which $\simeq$ implies, while confinement leaves every binding outside the two declarations and every control field as they stand. □

The names a state carries are read by two of those observations, $\text{dom}(F_\gamma)$ and the indexing of the control fields, and the rule that draws a name draws any name not already in use (Definition 52). Reading the results below up to $\simeq$ therefore also calls for reading them up to a renaming, which is the discipline of Section 4.1 cashed out.

**Lemma 61. (Equivariance.)** Let $\chi : \mathfrak{N} \to \mathfrak{N}$ be a bijection and let $\chi \cdot \gamma$ be the state carrying the registry $F_\gamma \circ \chi^{-1}$, with every name occurring in a $\pi_m$ or an $\omega_m$ replaced by its image. Then $\chi \cdot \gamma$ is a state, well formed where $\gamma$ is, and $\text{step}^t = r(n)$ carries $\gamma^t$ to $\gamma^{t+1}$ if and only if $r(\chi(n))$ carries $\chi \cdot \gamma^t$ to $\chi \cdot \gamma^{t+1}$.

*Proof.* A premise reads a name only by comparing it with another, whether directly, as in the freshness $n \notin \text{dom}(F_\gamma)$ of O-Insert and the $\forall m.\ \pi_m \neq n$ of O-Remove, or through a table of names, as $\text{target}_n$ and $\text{relied}_n$ read the $\pi_m$ and the $\omega_m$. A bijection preserves each such comparison. The only names a rule writes are the $\pi$ that O-Insert sets and the $\omega$ that L-Begin sets, both taken from what its premises read, so the writes commute with $\chi$; an effect function writes no name at all, drawing one only through the primitive of Definition 52, which Definition 55 confines to the entry that primitive adds. Well-formedness (Definition 63) is four conditions comparing names with names. □

A sequence and its renaming therefore take the same rules in the same order and reach states differing by $\chi$ alone. Two sequences agreeing save in the names their instantiations draw are accordingly identified, and the results below are read up to the renaming that identifies them.

The second lookup is that an entry stripped of everything but its name is invisible to the rules, which is what lets Definition 52 retire a fiber where the state it recovers has none, and Lemma 79 remove the fibers a deleted episode instantiated.

**Lemma 62. (Vestigial entries.)** Call $n$ *vestigial* at $\gamma$ when $\tau_n = \top$, $\theta_n = \mathsf{Inactive}$, $\sigma_n = \varnothing$, and no $m$ has $\pi_m = n$; a vestigial entry satisfies $\gamma \simeq_K \gamma \setminus n$. If $n$ is vestigial at $\gamma$ then for every rule and every $m \neq n$:

1. a rule applying at $\gamma$ acting on $m$ applies at $\gamma \setminus n$ acting on $m$, and the states the two reach differ in the entry at $n$ alone, which stays vestigial;
2. conversely a rule applying at $\gamma \setminus n$ acting on $m$ applies at $\gamma$, unless it is an O-Insert drawing the name $n$ or claiming a key of $p_n$.

*Proof.* A vestigial $n$ contributes to no observation a premise of a rule acting on $m \neq n$ reads. It is not Active, so $\sigma_n$ enters no $\sigma_\gamma$ and $n$ is the provider of no key, leaving $\gamma \vDash d_m$ and $\text{target}_m$ unmoved; $\text{installed}_n$ fails, so $n$ contributes no disjunct to $\text{relied}_m$; no $\pi_{m'}$ names $n$, so the premise $\forall m'.\ \pi_{m'} \neq m$ of an O-Remove of $m$ is unmoved; and $\theta_n$, $\tau_n$, and $\pi_n$ are read by rules acting on $n$ alone. The two premises clause (2) excepts are the two the removal relaxes, an absent name being fresh and an absent provision meeting every other. By Lemma 59 no rule acting on $m \neq n$ writes a field of $n$ save values at keys of $d_m$ that $\sigma_n$ holds, of which the empty $\sigma_n$ holds none, so the entry survives vestigial. □

#### 4.3.1. Preservation

Definition 50 fixes the shape of a registry, and the rules have to be checked against it before the results below can add to it. This subsection identifies the invariant the rules preserve, of which the first clause is that shape and the rest what those results assume.

**Definition 63.** A registry $F_\gamma$ is *well formed* when, for all $m, n \in \mathrm{dom}(F_\gamma)$ and all $k \in K$,

1. $\pi_n \in \mathrm{dom}(F_\gamma) \cup \{\mathsf{root}\}$;
2. $m \neq n \Rightarrow p_m \cap p_n = \varnothing$;
3. $\mathrm{installed}_n(\gamma) \Rightarrow \omega_n$ is total on $d_n$ and valued in $\mathrm{dom}(F_\gamma)$;
4. $\mathrm{installed}_n(\gamma) \wedge k \in d_n \wedge \omega_n(k) = m \Rightarrow \mathrm{installed}_m(\gamma)$.

Clause (1) is the tree of Definition 50 read one edge at a time, keeping a parent pointer landing in the registry. The acyclicity that definition also requires needs no clause, since the fiber a pointer names is introduced before the fiber naming it.

**Theorem 64. (Preservation.)** If $F^t$ is well formed then so is $F^{t+1}$, whichever rule step $t$ applies. Each clause is established at $\gamma^{t+1}$ from all four at $\gamma^t$.

*Proof.* Let step $t$ act on $n$.

(1) By Table 1 only O-Insert and O-Remove write a $\pi$ or $\mathrm{dom}(F_\gamma)$. O-Insert has $\pi_n \in \mathrm{dom}(F^t) \cup \{\mathsf{root}\}$ as a premise, which is the clause for the fiber it adds, and it leaves every other $\pi$ alone while enlarging $\mathrm{dom}(F_\gamma)$. O-Remove has $\forall m.\ \pi_m \neq n$, so no surviving $\pi_m$ names the fiber it takes away.

(2) The last premise of O-Insert is $\forall m.\ p_n \cap p_m = \varnothing$, which is the clause for the fiber it adds, and by Table 1 no other rule writes a $p$ or enlarges $\mathrm{dom}(F_\gamma)$. Two consequences are used below: $\mathrm{dom}(\sigma_m) \subseteq p_m$ by Definition 48, so distinct tables are disjoint and $\sigma_\gamma$ is a function; and $k \in p_m \cap p_{m'}$ forces $m = m'$, so $k$ has at most one possible provider.

(3) By Lemma 59(2) the only rule that writes an $\omega_n$ is L-Begin, whose premise $\omega = \mathrm{target}_n^t \neq \bot$ makes it total on $d_n$ and valued in $\mathrm{dom}(F^t)$, target naming providers. By Table 1 the only rule that shrinks $\mathrm{dom}(F_\gamma)$ is O-Remove, whose premise $\theta_n^t = \mathsf{Inactive}$ gives $\neg\,\mathrm{installed}_n^t$, whence by clause (4) at $\gamma^t$ no $m$ has $\omega_m^t(k) = n$ for a $k \in d_m$ while $\mathrm{installed}_m^t$; and $n$ itself carries no $\omega$.

(4) By Lemma 59(2) and (4) the clause can fail at $\gamma^{t+1}$ only where some installed has fallen, some $\omega$ has been written, or a fiber some $\omega$ names has left $\mathrm{dom}(F_\gamma)$. The last is an O-Remove, whose removed fiber is not installed and hence, by clause (4) at $\gamma^t$, is named by no $\omega_m^t$ of an installed $m$. The first is an L-Unload of $n$, whose premise $\neg\,\mathrm{relied}_n^t$ reads

$$\forall m \neq n,\, k \in d_m.\ \mathrm{installed}_m^t \Rightarrow \omega_m^t(k) \neq n$$

and which writes no $\omega_m$ for $m \neq n$ and leaves $\neg\,\mathrm{installed}_n^{t+1}$, so the clause holds of $n$ as well. The second is an L-Begin of $n$, writing $\mathrm{target}_n^t$, whose values are the providers of the keys of $d_n$ and hence $\mathsf{Active}$ at $\gamma^t$; the step alters no other fiber's $\theta$, so they are installed at $\gamma^{t+1}$ too. $\square$

The guard on L-Unload is what carries clauses (3) and (4). The premise $\forall m.\ \pi_m \neq n$ of O-Remove speaks only of parent pointers; what keeps a committed view from naming a removed fiber is the guard, imposed several steps earlier and for a different reason. Two things follow. A name freed by O-Remove may be reissued by O-Insert, since no stale committed view can name it; and a fiber may be removed as soon as it is Inactive, without a separate check that nobody depends on it.

#### 4.3.2. *Temporal Composability*

Local temporal composability reverts one sequence of effects with one accumulator (Section 3.1.2). The registry holds one accumulator per fiber and the fibers interleave: between the moment $n$ composes an inverse onto $g_n$ and the moment $g_n$ runs, other fibers have moved the state. Whether $g_n$ still undoes what it was built to undo there is what the global form of the guarantee asserts, and the condition it turns on is that the intervening steps commute with $g_n$.

**Definition 65.** Two iterators $i, j$ over $\Gamma$ are *independent* when they are so in the sense of Definition 42, reading $\simeq$ on maps, triples, and continuations as Definition 34 does, and an instantiating iteration (Definition 52) as agreement of the component it names. Fibers $m$ and $n$ are *entangled* when one's provision meets the other's declarations, $p_m \cap (d_n \cup p_n) \neq \varnothing$ or $p_n \cap (d_m \cup p_m) \neq \varnothing$. A sequence of steps is *pairwise independent* when for every two names $m \neq n$ it ever holds — one for each fiber the orchestrator inserts and each fiber an iteration instantiates — either $e_m$ and $e_n$ are independent, or $m$ and $n$ are entangled and every key at which operations of both occur is commutative (Definition 44).

Independence in this sense is what trace theory takes as primitive: commuting actions generate an equivalence on sequences under which reordering two adjacent independent actions preserves the endpoint [46], and Lemma 78 is that reordering for these rules. Quantifying over names rather than iterators is what keeps two fibers of one component in scope: such a pair requires that component's effect function to be independent of itself, which is to require that $\mathfrak{M}(i)$ be commutative. Clause (1) of Definition 42 is what Theorem 68 uses and clause (2) what Theorem 80 needs in addition: reordering the steps of two fibers evaluates an iterator at a state the other fiber moved, and commuting the maps does not by itself say that the iterator yields the same inverse and the same continuation there. Checking clause (1) calls for no more than the iterations themselves, since Lemma 41(1) carries commutation from the generators to the monoids they generate.

The paradigm supplies both disjuncts:

**Lemma 66. (Pairwise independence.)** Every sequence of steps is pairwise independent.

*Proof.* Every key is commutative, its coeffect carrying the proof as its witness (Definition 46), which settles the second disjunct at an entangled pair. A pair that is not entangled has each member's provision outside the other's every key, which is the hypothesis $P_1 \cap S_2 = P_2 \cap S_1 = \varnothing$ of Theorem 47 read at the underlying members of Definition 30, so those members are independent, the commutativity of every shared operation key again supplied by the witness; independence transfers along the lift of Definition 56, whose transformations move the tables as the stages move the projection, an instantiating iteration adding an entry no table map reads. Two fibers of one component fall under the same two cases, Theorem 47 holding at $i_1 = i_2$ as well. $\square$

Entangled fibers are the pairs independence cannot cover, and could not be expected to: a consumer's operation acts on the very value the provider's extension installs, so the two orders differ at every state the binding is absent from, whichever equivalence the difference is read up to. What stands in for independence there is the rules themselves, which never interleave the two maps in the order that separates them. Throughout the argument, a lift applied where its precondition fails produces no transition, per the convention of Section 3.2.1, so a map meeting a state its key has left is read as the identity.

**Lemma 67. (Entangled steps.)** Let an episode of $n$ open at $b$, and let step $t \geq b$ in the episode act on an $m \neq n$ entangled with $n$. Then

1. where $p_m \cap (d_n \cup p_n) \neq \varnothing$, $\Psi^t = \mathrm{id}_\Gamma$;
2. otherwise $g_n^t(\Psi^t(\gamma^t)) \simeq_K \Psi^t(g_n^t(\gamma^t))$;
3. where moreover $\theta_n^t = \mathsf{Reloading}(-,-,-)$, $\Psi^t = \mathrm{id}_\Gamma$ in either case.

*Proof.* (1) A key of $p_m \cap p_n$ would put two registered provisions in conflict with the premise of O-Insert, so some $k \in p_m \cap d_n$, and $m$ is the one registered fiber whose provision carries $k$. The premise of $n$'s L-Begin at $b-1$ resolves $k$ to an Active provider, so $\omega_n^b(k) = m$ and $\theta_m^{b-1} = \mathsf{Active}(-,-)$; $\omega_n$ holds $m$ for as long as the episode is open (Lemma 59(2)) while $\mathrm{installed}_n$ holds, which is $\mathrm{relied}_m(\gamma^t)$ at every such $t$. The guard therefore blocks every L-Unload of $m$ there, so $m$ never reaches Inactive, and a fiber standing at Active or Unloading is acted on only by L-Leave, O-Retire, and the blocked L-Unload, of which the first two have $\Psi^t = \mathrm{id}_\Gamma$ (Table 1).

(2) Here $p_m \cap (d_n \cup p_n) = \varnothing$ and some $k \in p_n \cap d_m$. $\Psi^t$ is one of $m$'s iterations or the accumulator $g_m^t$, in either case a composite of the maps Definition 56 admits for $m$: lifts of operation maps, forward or inverse, at keys of $d_m \cup p_m$, extensions and restrictions at keys of $p_m$, instantiations, and the O-Retires those yield. The constituents of $g_n^t$ are the inverses Definition 56 admits for $n$: lifts of operation inverses at keys of $d_n \cup p_n$, restrictions at keys of $p_n$, and O-Retires. Commute the constituents of $\Psi^t$ past those of $g_n^t$ one pair at a time. A pair at distinct keys is a pair of key-local maps and commutes; an instantiation or an O-Retire writes a fresh entry or a control field, which no table map reads, and commutes with every constituent in sight; a pair of operation maps at one shared key is covered by that key's commutativity, which the witness of its coeffect supplies (Definition 46), read at the tables through the lift of Definition 56. What remains is an operation map of $m$ at a key $k \in p_n \cap d_m$ against the extension or restriction of $n$ at $k$. Such a map exists in $\Psi^t$ only under a committed view of $m$ resolving $k$ (Lemma 59(2)), whose L-Begin required an Active provider of $k$; the commitment pins that provider for as long as $m$ is installed, by the argument of (1) read at $m$, and two registered provisions cannot share $k$, so the provider is $n$ and $n$ was Active within the open episode. Its transition had therefore finished by $t$, and $g_n^t$ carries the restriction at $k$ the extension $m$ resolved yielded, composed to the left of $n$'s operation inverses at $k$ by the LIFO order of Definition 18. On one side the operation map of $m$ commutes leftward, past constituents at other keys and, by commutativity of $k$, past $n$'s operation inverses at $k$, until the restriction absorbs it, a write to the value at $k$ followed by the removal of $k$ being the removal alone; on the other side it meets a state $g_n^t$ has removed $k$ from and produces no transition. Both composites therefore agree at $\gamma^t$.

(3) The provider case is (1). In the consumer case, a consumer of $n$ is not installed while $n$ is Reloading: its commitment resolving a key to $n$ would have held $\mathrm{relied}_n$ and blocked the L-Unload closing $n$'s previous episode, and a new commitment requires an Active provider. An uninstalled fiber is acted on only by L-Begin and the orchestration rules; the orchestration rules have $\Psi^t = \mathrm{id}_\Gamma$, and L-Begin is inapplicable, the key $n$ provides lying outside $\mathrm{dom}(\sigma_{\gamma^t})$ and the target of a fiber declaring it therefore $\perp$. □

With every pair so covered, the single-accumulator invariant of Theorem 7 survives the interleaving, in the form that gives temporal composability its content: running an inverse withdraws the fiber's contribution and nothing else.

**Theorem 68.** (**Recovery exactness.**) Let an episode of $n$ open at $b$, let $u \geq b$ lie in it, and let $t_1 < \cdots < t_l$ be the indices in $[b, u)$ at which the acting fiber is not $n$. Then

$$g_n^u(\gamma^u) \simeq_K \left(\Psi^{t_l} \circ \cdots \circ \Psi^{t_1}\right)\left(\gamma^b\right) \tag{55}$$

That is, applying $n$'s accumulator at $\gamma^u$ leaves every fiber's table where those same steps would have left it from $\gamma^b$, the control fields lying outside the comparison. Reading the right side as the state reached had $n$ never begun assumes that no fiber $n$ instantiates take a step in $[b, u)$, since a fiber $n$ instantiates is one that would not be there to take it.

*Proof.* By induction on $u$, over the indices $u$ with $u + 1$ in the episode. At $u = b$ the step at $b - 1$ is an L-Begin, the episode opening by Definition 58, so $g_n^b = \mathrm{id}_\Gamma$ by Table 1, the index set is empty, and the claim is $\gamma^b \simeq_K \gamma^b$. Two facts are used at each step. An $\mathrm{edit}^t$ writes control fields alone, and the two rules that write $\mathrm{dom}(F_\gamma)$ leave the tables as they stand, an O-Insert adding an entry with an empty table and an O-Remove taking one away by its premise, so

$$\gamma^{t+1} \simeq_K \Psi^t(\gamma^t)$$

and, for every fiber $m$, every map in $\mathfrak{M}(e_m)$ carries $\simeq_K$-related states to $\simeq_K$-related states, since $\simeq_K$ implies $\simeq_{d_m \cup p_m}$, at which Lemma 57 makes such a map respect the relation, and since confinement leaves it moving no binding outside the two declarations, so that the keys outside the interface stay as related as it found them; an instantiation adds an empty entry by Definition 52.

Let step $u$ act on $n$. Since the episode is open at $u$ and $u + 1$, Lemma 59(4) excludes an L-Begin and an L-Unload of $n$, and O-Insert and O-Remove read a $\theta_n$ that $\mathrm{installed}_n^u$ denies, leaving two cases. Where the rule is L-Iter, L-Finish, or a landing L-Divert, Table 1 gives $\Psi^u = \mathrm{pr}_1 \circ i_n^u$ and $g_n^{u+1} = g_n^u \circ h$ for the inverse $h$ that iteration yields. The witness condition of Definition 37 reads $h(\Psi^u(\gamma^u)) \simeq_{d_n \cup p_n} \gamma^u$, which is $\simeq_K$ once Lemma 57 adds that neither map moves a binding outside the two declarations, and the instantiating iteration is the case where the two states differ in an entry with an empty table that $\simeq_K$ does not compare. Since $g_n^u$ carries $\simeq_K$ by the paragraph above,

$$g_n^{u+1}(\gamma^{u+1}) \simeq_K (g_n^u \circ h)(\Psi^u(\gamma^u)) \simeq_K g_n^u(\gamma^u)$$

Where the rule is L-Leave, an aborting L-Divert, or an O-Retire of $n$, Table 1 gives $\Psi^u = \mathrm{id}_\Gamma$ and $g_n^{u+1} = g_n^u$, so the same equation holds with $h = \mathrm{id}_\Gamma$. Either way the induction hypothesis carries over with the index set unchanged, which is the computation of Theorem 7 one step at a time.

Let step $u$ act on $m \neq n$. Then $g_n^{u+1} = g_n^u$ by Table 1, and $\Psi^u \in \mathfrak{M}(e_m)$, or $\Psi^u = \mathrm{id}_\Gamma$ where the rule is an orchestration rule. Where $m$ and $n$ are not entangled, Lemma 66 makes $e_m$ and $e_n$ independent, and clause (1) of Definition 42, read at the finer $\simeq$ of Definition 58 and hence at $\simeq_K$, commutes $g_n^u$ with $\Psi^u$; where they are entangled, Lemma 67 commutes the two at $\gamma^u$, its first case outright. Either way

$$g_n^u(\gamma^{u+1}) \simeq_K g_n^u(\Psi^u(\gamma^u)) \simeq_K \Psi^u(g_n^u(\gamma^u))$$

which is the induction hypothesis with $\Psi^u$ appended, $\Psi^u$ carrying $\simeq_K$-related states to $\simeq_K$-related states by the paragraph above. □

**Corollary 69.** (**Terminal recovery.**) Let an episode of $n$ open at $b$ and close at $u$. Then, with $t_1 < \cdots < t_l$ as in Theorem 68,

$$\gamma^{u+1} \simeq_K \left(\Psi^{t_l} \circ \cdots \circ \Psi^{t_1}\right)\left(\gamma^b\right) \tag{56}$$

In particular $\sigma_n^{u+1} = \varnothing$, which is the premise an O-Remove of $n$ carries.

*Proof.* By Lemma 59(4) step $u$ is an L-Unload of $n$, whose $\Psi^u$ is $g_n^u$ by Lemma 59(3), so $\gamma^{u+1} \simeq_K g_n^u(\gamma^u)$ and Theorem 68 applies. For the table, the fiber enters the episode with $\sigma_n$ empty, and the keys of $p_n$ enter $\mathrm{dom}(\sigma_n)$ only by $n$'s own extensions (Definition 56), of which the right side applies none, so the right side leaves $\sigma_n$ empty; Definition 33 relates two coeffect contexts only where their domains agree, so a table it relates to the empty one is empty. □

What the two results compare is the tables, so what they assert is bounded by what the keys of a state bind, and inside each binding by the $\simeq_k$ the key's operations induce: each binding is restored only up to what its key's equivalence forgets, so a monotone allocator is not rewound, a heap's layout `free` does not restore, and a message already sent stays sent. This is the bound Section 3.3.2 takes on Theorem 7 and for the same reason, the physical state not being recoverable as it stood; a location the system reifies at no key lies outside the calculus altogether (Definition 56), and Section 6.1 is where a system decides what to reify.

The results above therefore assume nothing of the sequence. The witness and respect conditions of $\mathfrak{I}_\Gamma^{d \cup p}$ are Lemma 57, pairwise independence is Lemma 66, and both rest on the components alone: Definition 56 fixes the form of every effect function, the effect function's witness holds each returned inverse to reverting, and the coeffect's witness holds each key to commutativity (Definition 46), an interface property Definition 31 turns into a design procedure.

#### *4.3.3. Spatial Composability*

Local spatial composability holds a component to its own specification, activating it only where its dependencies are provided and classifying every context change against them (Section 3.2.2). The global form adds what quantifies over other fibers: a provider withdraws a binding only after every dependent that resolved it has deactivated, and the resolution a transition installs its effects against does not shift under it. Two properties of the coeffect side deliver the two, and they are proved together, being two halves of one invariant, namely the fixity of $\omega_n$ over an episode that Lemma 59(2) establishes. The ordering theorem is what that fixity delivers over the part of the episode in which $n$ is Active and then Unloading, and the coherence theorem what it delivers over the part in which $n$ is installing its effects.

**Theorem 70. (Ordering.)** A fiber begins a transition only where its dependencies are provided:

$$\mathrm{step}^t = \text{L-Begin}(m) \Rightarrow \gamma^t \vDash d_m \tag{57}$$

Let further $[b', u']$ be an episode of $m$ with $\omega_m^{b'}(k) = n$ for some $m \neq n$ and $k \in d_m$, let $[b, u]$ be the episode of $n$ containing $b'$, and let $t$ range over $[b', u']$. Then

1. $\omega_m^t(k) = n$;
2. $b < b'$, and $u' < u$ if $[b, u]$ closes;
3. $k \in \mathrm{dom}(\sigma_n^t)$, and $\sigma_n^t(k)$ moves only by operations at $k$ of fibers declaring $k$.

*Proof.* The first claim is the premise $\mathrm{target}_m^t \neq \bot$ of L-Begin, which by Definition 53 gives $\gamma^t \vDash d_m$.

(1) is Lemma 59(2).

For (2), the L-Begin at $b'-1$ writes $\omega_m^{b'} = \mathrm{target}_m^{b'-1}$, whose values are providers, so $\theta_n^{b'} = \mathsf{Active}(-,-)$; the L-Begin at $b-1$ leaves $\theta_n^b = \mathsf{Reloading}(-,-,-)$, so $b \neq b'$ and hence $b < b'$, both episodes opening by Definition 58. Let $[b,u]$ close and suppose $u \leq u'$. Then $u \in [b', u']$, so $\mathrm{installed}_m^u$ and, by (1), $\omega_m^u(k) = n$; that is $\mathrm{relied}_n^u$, which the L-Unload at $u$ denies. Hence $u' < u$.

For (3), $n$ is the provider of $k$ at $\gamma^{b'}$, so $k \in \mathrm{dom}\big(\sigma_n^{b'}\big)$. No L-Unload of $n$ falls in $[b', u']$: where $[b,u]$ closes it falls at $u > u'$ by (2), and where it does not, Lemma 59(4) leaves $n$ with no L-Unload at all. Since $\theta_n^{b'} = \mathsf{Active}(-,-)$, Table 1 therefore leaves L-Leave as the only rule $n$ can be acted on by within $[b', u']$, and its $\Psi^t$ is $\mathrm{id}_\Gamma$, so $n$ withdraws nothing and $\mathrm{dom}(\sigma_n)$ is constant there by Lemma 59(1). What a step of another fiber may move is values at keys its own declarations name (Definition 56), so a write to $\sigma_n(k)$ is an operation at $k$ of a fiber with $k$ in its specification. ☐

A transition spread over steps could otherwise install effects computed against a resolution that has changed under it, and two premises prevent that. L-Iter and L-Finish carry $\mathrm{target}_n(\gamma) = \omega$, so a transition proceeds only while its committed view is still its target view, and L-Divert carries the negation, so any change to the target view takes the fiber out of the transition. The two directions of change are not distinguished: a component whose dependency has gone and one whose dependency has been replaced leave by the same route, because a target view that has become $\perp$ and one that has become some other fiber are equally unequal to $\omega$.

The landing alternative of L-Divert is what stops this from being a guarantee about every step: the iteration it lands installs an effect computed against a resolution that no longer holds. What the rules deliver is therefore a disjunction, and the second branch is what makes the first safe.

**Theorem 71.** **(Resolution coherence.)** Let an episode $[b,u]$ of $n$ open at $b$ with $\omega_n^b = \omega$. Then $\theta_n$ is $\mathsf{Reloading}(-,-,-)$ on an initial interval $[b,r]$ of the episode, and every iteration of the transition runs against the one resolution $\omega$:

$$\forall t \in [b,r].\ \mathrm{step}^t \in \{\text{L-Iter}(n), \text{L-Finish}(n)\} \Rightarrow \mathrm{target}_n^t = \omega \tag{58}$$

Where the fiber leaves that interval, so that $r < u$, exactly one of the following holds:

1. $\mathrm{step}^r = \text{L-Finish}(n)$ and $\theta_n^{r+1} = \mathsf{Active}(-,\omega)$;
2. $\mathrm{step}^r = \text{L-Divert}(n)$, and the episode closes at some $u > r$ with $\gamma^{u+1} \simeq_K \big(\Psi^{t_l} \circ \cdots \circ \Psi^{t_1}\big)\big(\gamma^b\big)$ as in Corollary 69.

*Proof.* The L-Begin at $b-1$ writes $\mathsf{Reloading}$, and by Table 1 it is the one rule leading into that lifecycle state; its premise $\theta_n = \mathsf{Inactive}$ and Lemma 59(4) put any second application of it outside the episode. So $\mathsf{Reloading}$ occupies an initial interval $[b,r]$ of $[b,u]$ and is not re-entered. The first claim is then the premise $\mathrm{target}_n(\gamma) = \omega'$ that Table 1 gives L-Iter and L-Finish, together with $\omega' = \omega$ by Lemma 59(2).

For the dichotomy, $\mathrm{step}^r$ is a rule whose premise has $\theta_n = \mathsf{Reloading}(-,-,-)$ and whose conclusion does not, of which Table 1 offers L-Finish and L-Divert; the first lands in $\mathsf{Active}(-,\omega)$ and the second in $\mathsf{Unloading}(-,\omega)$, from which Lemma 59(4) makes an L-Unload the only exit and Corollary 69 supplies the equation. The iteration a landing L-Divert contributes is one of $n$'s own, hence among the maps that accumulator withdraws. Where instead $r = u$, the sequence ends with the transition still in flight and the first claim is all that is asserted. ☐

#### 4.3.4. *Progress*

A guard that defers a provider's withdrawal until its dependents are gone delivers Theorem 70 only if it eventually releases. One relation on the fibers of a registry carries the argument.

**Definition 72.** The *precedence relation* on the names of a registry is

$$n \prec m := p_n \cap d_m \neq \varnothing \tag{59}$$

so that $n$ may provide a key $m$ declares. It reads $d$ and $p$ alone, which by Lemma 59(5) come into existence with a fiber's entry and are never written again.

Theorem 73 and Theorem 80 are established on the hypothesis that $\prec$ is acyclic, which is an assumption and not something the definition delivers, $n \prec n$ holding of a component that declares a key it provides itself. What $\prec$ orders is the two fibers' activations and not their lifetimes: $n \prec m$ says that $n$ has to become $\mathsf{Active}$ before $m$ can, whereas that a provider outlives its consumer is Theorem 70(2), a theorem about the guarded calculus.

A fiber's target view answers to the fiber that created it as well as to its providers. What a creator writes is $\tau_n$, through the primitive of Definition 52, and $\tau$ is monotone by Lemma 59(5). A creator can therefore turn its child's target view at most once over that child's whole existence.

Progress is a claim that some rule applies, so it is formulated over the rules a host must offer: L-Begin, L-Leave, L-Unload, the landing rules L-Iter and L-Finish, and L-Divert. It appeals to the aborting alternative of L-Divert nowhere, which Section 4.4 puts to use.

**Theorem 73. (Progress.)** Assume $\prec$ acyclic, $\operatorname{len}(e_n) \leq K$ for every $n$, and the set $N$ of names the sequence ever holds (Definition 65) finite; and let every step apply a lifecycle rule. Write $S(n)$ for the number of steps acting on $n$ and

$$V(n) := \left|\{t : \mathrm{target}_n^t \neq \mathrm{target}_n^{t+1}\}\right| \tag{60}$$

for the number of times its target view turns. Then

1. (*No deadlock.*) $\neg\,\mathrm{quiet}^t$ implies that some lifecycle rule applies at $\gamma^t$;
2. (*Termination.*) $S(n) \leq (K+3)(V(n)+1)$, and both $V(n)$ and $\sum_n S(n)$ are finite.

Consequently every maximal sequence of lifecycle steps ends in a quiescent state.

*Proof. No deadlock.* Let $\neg\,\mathrm{quiet}^t$, so some fiber $n$ satisfies neither clause of the $\mathrm{quiet}$ of Definition 53. Reading Table 1 against the four kinds it can then be:

- $\theta_n^t = \mathsf{Inactive}$ with $\mathrm{target}_n^t \neq \bot$: L-Begin applies;
- $\theta_n^t = \mathsf{Reloading}(-,-,\omega_n)$ with $\mathrm{target}_n^t = \omega_n$: whichever of L-Iter and L-Finish the value of $i_n^t(\gamma^t)$ selects applies;
- $\theta_n^t = \mathsf{Reloading}(-,-,\omega_n)$ with $\mathrm{target}_n^t \neq \omega_n$: L-Divert applies, landing that iteration rather than aborting it;
- $\theta_n^t = \mathsf{Active}(-,\omega_n)$ with $\mathrm{target}_n^t \neq \omega_n$: L-Leave applies.

Let no fiber be of any of these kinds, leaving some $m_0$ with $\theta_{m_0}^t = \mathsf{Unloading}(-,-)$. Construct $m_0, m_1, \ldots$ as follows: given $m_j$ in $\mathsf{Unloading}$, either $\neg\,\mathrm{relied}_{m_j}^t$, in which case L-Unload applies to $m_j$ and the construction stops, or there are $m_{j+1} \neq m_j$ and $k_j$ with $\mathrm{installed}_{m_{j+1}}^t$ and $\omega_{m_{j+1}}^t(k_j) = m_j$. In the latter case

$$k_j \in d_{m_{j+1}} \cap \mathrm{dom}\left(\sigma_{m_j}^t\right) \subseteq d_{m_{j+1}} \cap p_{m_j}$$

the second membership being Theorem 70(3) at the episode of $m_{j+1}$ that $t$ lies in, so that $m_j \prec m_{j+1}$. Moreover $\mathrm{target}^t_{m_{j+1}} \neq \omega^t_{m_{j+1}}$: an Unloading fiber is outside the union defining $\sigma_\gamma$, so $k_j$ at $\gamma^t$ is unprovided or provided by a fiber other than $m_j$. Were $m_{j+1}$ in Active or Reloading it would then be of one of the four kinds excluded, so it is in Unloading and the construction continues. The $m_j$ are $\prec$-increasing, hence distinct by acyclicity, and $\mathrm{dom}(F^t)$ is finite, so the construction stops.

*Termination.* Two claims bound $S(n)$.

(A) Over a maximal interval on which $\mathrm{target}^t_n$ is constant at $\omega^*$, at most $K+3$ steps act on $n$. Reading the $\theta_n$ columns of Table 1, from Active$(-,\omega)$ with $\omega \neq \omega^*$ the fiber takes an L-Leave and an L-Unload and then, if $\omega^* \neq \bot$, an L-Begin and at most $\mathrm{len}(e_n) \leq K$ landings; from Reloading against an $\omega \neq \omega^*$ it takes an L-Divert in place of the L-Leave, and from any other state a suffix of that sequence. No further L-Divert or L-Leave falls in the interval, the $\omega$ that the L-Begin writes being $\mathrm{target}^t_n = \omega^*$ itself, and at Active$(-,\omega^*)$ and at Inactive with $\omega^* = \bot$ no rule applies at all.

(B) If $\mathrm{target}^t_n \neq \mathrm{target}^{t+1}_n$ and step $t$ acts on $m$, then either $m \prec n$ or step $t$ writes $\tau_n$. By Definition 53 the value of $\mathrm{target}_n$ is a function of $\tau_n$, of the lifecycle states, and of the domains of the providers' tables, never of a bound value; a provider satisfies $k \in \mathrm{dom}(\sigma_m) \cap d_n$ and hence $m \prec n$, and a table's domain changes only at a step acting on its own fiber by Lemma 59(1). Acyclicity gives $m \neq n$ in the first case, and the monotonicity of Lemma 59(5) admits the second at one $t$ per fiber.

By (A) the interval count bounds $S(n)$ as $S(n) \leq (K+3)(V(n)+1)$, and by (B) each turn of $\mathrm{target}_n$ either consumes a step of a fiber strictly $\prec$-below $n$ or is the one turn $\tau_n$ affords, so $V(n) \leq 1 + \sum_{m \prec n} S(m)$. Since $\prec$ is acyclic and $N$ is finite, the recursion

$$B(n) := (K+3)\left(2 + \sum_{m \prec n} B(m)\right)$$

is well founded and defines $B$ with $S(n) \leq B(n)$; hence $V(n)$ is finite and $\sum_n S(n) \leq \sum_n B(n)$. By (1) a sequence that cannot be extended is quiescent. □

Finiteness of $N$ is assumed rather than derived, and one condition on the components delivers it. The components a host holds are finitely many programs given before anything runs, so if no component can instantiate, however indirectly, a fiber of a component that instantiates one of its own, the instantiations form a tree of bounded depth, and $\mathrm{len}(e_n) \leq K$ bounds its branching. What the assumption rules out is a component that instantiates itself without bound.

The target records the providing fiber rather than a boolean, and under the single-source discipline of O-Insert the two drive the same transitions, a key having one possible provider there. The view supplies the vocabulary of the results above, Theorem 70 and Theorem 71 both speaking of the resolution a fiber activated against, and it is what makes those results survive the scoped resolution of Section 3.2.3, under which one key resolves to different providers in different realms and the provisions no longer force the view. The implementation carries that scoping and holds the view in `fiber.committed` (Section 5.1.3).

#### 4.3.5. *Confluence*

The results so far are about individual fibers. The property that characterizes the system as a whole is that its dynamic history leaves no trace: whatever sequence of activations and deactivations a running system has been through, the state it quiesces at is the one the same insertions and retirements would have produced had each component that ends up active been loaded once, in dependency order, and none ever unloaded. The lifecycle relation is confluent, and the normal form it converges on is the statically assembled one. This is the analogue, for dynamic composition, of the consistency with a from-scratch evaluation that change propagation establishes for incremental computation [47].

The claim is about $\longrightarrow$ alone. Orchestration steps are inputs, and two sequences given different inputs land in different places for no interesting reason; what is at issue is whether the lifecycle rules, which are nondeterministic in which fiber steps next and in which exit a $\mathsf{Reloading}$ fiber takes, can be made to disagree. Which fiber provides a key is not among the choices, a key having one possible provider (Definition 50), so the schedule picks orders and exits and nothing else.

Three lemmas are needed first. The first fixes the set of fibers that end up $\mathsf{Active}$ without reference to any sequence of steps, which is what makes it a function of the input rather than of the schedule.

**Definition 74.** A fiber is *supported* at $\gamma$ when it is not retired, the fiber instantiating it is supported, and every key it declares is provided by a supported fiber. The *support relation* on $\mathrm{dom}(F_\gamma)$ is the union of the two relations those clauses read,

$$m \lhd n := m \prec n \lor \pi_n = m \tag{61}$$

and where it is well founded (Lemma 75) we write $A$ for the *support set*, the fibers supported at $\gamma$:

$$n \in A := \neg\tau_n \land (\pi_n = \mathsf{root} \lor \pi_n \in A) \land \forall k \in d_n.\ \exists m \in A.\ k \in p_m \tag{62}$$

where $\pi_n = \mathsf{root}$ marks a fiber the orchestrator inserted and $\pi_n$ otherwise the fiber whose activation instantiates $n$. The clauses read no field but $\tau, \pi, d, p$. Both halves relate a fiber to one immediately below it, a parent rather than an ancestor and a direct provider rather than a transitive one, since that is what the clauses read; where the results below want an order they take the transitive closure, whose minimal elements, maximal elements, and linearizations are those of $\lhd$.

The clauses refer to $A$ itself, so the definition is a recursion along $\lhd$, and it is the following that makes it one with a solution.

**Lemma 75.** (**Support is well founded.**) Let $\prec$ be acyclic and let $\gamma$ be reached by a sequence of steps. Then $\lhd$ is well founded, and $A$ is the one solution of Definition 74, a function of $\tau$, $\pi$, $d$, and $p$ alone.

*Proof.* Order the names of $\mathrm{dom}(F_\gamma)$ by the index of the O-Insert that introduced each, which Definition 58 supplies by starting the sequence at an empty registry. The parent half of $\lhd$ descends in that index: an O-Insert has $\pi \in \mathrm{dom}(F_\gamma)$ as a premise, so a parent pointer names a fiber introduced earlier, and iterating it reaches the whole ancestry of a name in finitely many steps. A cycle therefore has to use $\prec$, and since $\prec$ is acyclic it has to mix the two, which needs some $m$ to declare a key that a fiber of $m$'s own subtree may provide. Such a fiber is instantiated by an activation of $m$ or of one of $m$'s descendants, hence at a step after the L-Begin of $m$; that L-Begin has $\gamma \vDash d_m$ as a premise, so a fiber providing the key is Active already before it, and clause (2) of Definition 63 leaves the key no second possible provider. The fiber that would close the cycle is therefore never introduced, and the edge is absent from $\mathrm{dom}(F_\gamma)$. A well-founded recursion has one solution, and the clauses read the four fields alone. $\square$

The last clause reads $p$, the keys a component may provide, whereas the target reads $\mathrm{dom}(\sigma_\gamma)$, the keys its fibers have installed, and Definition 48 relates the two by $\mathrm{dom}(\sigma_n) \subseteq p_n$ alone. The support set therefore over-approximates the Active fibers in general, and the condition that closes the gap is the following.

**Definition 76.** A component $(d, p, e)$ is *total on its provision* when an activation of it that finishes has installed every key of $p$, so that $\mathrm{dom}(\sigma_n) = p_n$ at every Active fiber instantiating it.

This is a condition on the components alone, mentioning no lifecycle state and no step, and independence (Lemma 66) already bounds how far it can fail: were a component to install a key only at context states another component's effects reach, its forward map would not commute with that component's, so the keys a fiber installs are fixed by its component rather than by the schedule. What totality adds is that the fixed set is all of $p$ rather than a proper subset of it.

**Lemma 77. (Support at quiescence.)** Let $\prec$ be acyclic, let $\mathrm{quiet}(\gamma)$, and let every component of $\gamma$ be total on its provision (Definition 76). Then the support set is the set of Active fibers:

$$A = \{n : \theta_n = \mathsf{Active}(-,-)\} \tag{63}$$

*Proof.* Write $A'$ for the right-hand side. The quiet of Definition 53 leaves Inactive and Active as the only states and reads

$$n \in A' \iff \mathrm{target}_n(\gamma) \neq \bot$$

By Definition 53 the right side holds exactly when $\neg\tau_n$ and every $k \in d_n$ lies in $\mathrm{dom}(\sigma_\gamma)$, and $\mathrm{dom}(\sigma_\gamma) = \bigcup_{m \in A'} p_m$ by Definition 76. The middle clause is the one the target no longer carries, and instantiation supplies it: a fiber with $\pi_n \neq \mathsf{root}$ is instantiated only by an activation of $\pi_n$, and if $\pi_n \notin A'$ then $\pi_n$ is not Active, so its accumulator has run and retired $n$ by Definition 52, giving $\tau_n$. Hence $A'$ satisfies the clauses of Definition 74, and Lemma 75 gives them one solution, so $A = A'$. $\square$

**Lemma 78. (Transposition.)** Let $F^t$ be well formed and let steps $t$ and $t+1$ act on distinct fibers $m$ and $n$.

1. If both apply an activation rule, namely L-Begin, L-Iter, or L-Finish, $e_m$ and $e_n$ are independent (Definition 65), and step $t+1$ is applicable at $\gamma^t$, then step $t$ is applicable at the state step $t+1$ produces from $\gamma^t$, and the two orders reach the same $\gamma^{t+2}$.
2. If step $t$ applies an activation rule at $m$, step $t+1$ an orchestration rule at $n$, and step $t$ does not instantiate $n$, then the same holds of the two.

*Proof.* For (1), by Table 1 the step of $m$ writes $\theta_m$ and, within $\Psi^t \in \mathfrak{M}(e_m)$, the tables at keys of $d_m \cup p_m$. It therefore leaves $\theta_n$ and $i_n$ alone, and by clause (2) of Definition 42 leaves the inverse and the continuation that $i_n$ yields alone as well, so only the premises of step $t+1$ that mention $\mathrm{target}_n$ remain to be checked. Its retirement half cannot fall, no activation rule writing a $\tau$. Its resolution half cannot move either: $\mathrm{target}_n$ reads lifecycle states and table domains, of which $\Psi^t$ moves $\mathrm{dom}(\sigma_m)$ alone (Lemma 59(1)); step $t+1$ being applicable at $\gamma^t$ puts every $k \in d_n$ in $\mathrm{dom}(\sigma^t)$, and clause (2) of Definition 63 makes the fiber providing such a $k$ the only one that can, so $k \notin p_m$ and no domain at a key of $d_n$ moves. The same argument in the other direction leaves step $t$ applicable. Finally $\Psi^t \in \mathfrak{M}(e_m)$ and $\Psi^{t+1} \in \mathfrak{M}(e_n)$ commute by clause (1) of Definition 42, and the two edits write control fields of distinct fibers, so the composite is the same in either order.

For (2), the orchestration step has $\Psi^{t+1} = \mathrm{id}_\Gamma$ by Table 1, so the two state maps commute outright, and its $\mathrm{edit}^{t+1}$ writes $\tau_n$ or $\mathrm{dom}(F_\gamma)$ at $n$ alone, which the activation step neither reads nor writes: the premises of the latter read $\theta_m$, $i_m$, $\tau_m$, and $\mathrm{target}_m$, and an O-Insert of a fresh $n$ moves no target, a fresh fiber providing nothing, whereas an O-Retire or O-Remove of $n$ leaves $\sigma_\gamma$ where it was, $n$ being Inactive in the one case and unaffected in its table in the other. So step $t$ remains applicable. Conversely each premise of the orchestration step is either read at $n$, which step $t$ does not write, or is one of the two premises of O-Insert that a smaller registry only relaxes, whence its applicability at $\gamma^{t+1}$ gives its applicability at $\gamma^t$; here step $t$ not instantiating $n$ is what keeps $n$ present at $\gamma^t$ where O-Retire and O-Remove require it. □

**Lemma 79. (Deletion.)** Let every component be total on its provision (Definition 76), let the sequence of steps reach a quiescent $\gamma^T$, let $[b, u]$ be an episode of $n$ that closes, let no episode of any $m$ with $n \prec m$ close in the sequence, and let no fiber $n$ instantiates during $[b, u]$ have an episode. Write $R$ for the names those instantiations draw. Then deleting the steps that act on $n$ in $[b, u]$, together with every step acting on a name of $R$, leaves a sequence of steps reaching a state $\simeq_K$-equal to $\gamma^T$ and $\simeq$-equal to it outside $R$.

*Proof. The deleted steps leave the state where they found it.* Let $t_1 < \cdots < t_l$ be the steps of $[b, u]$ that act on fibers other than $n$. Corollary 69 reads

$$\gamma^{u+1} \simeq_K \left(\Psi^{t_l} \circ \cdots \circ \Psi^{t_1}\right)\left(\gamma^b\right)$$

whose right side is what the surviving steps of $[b, u]$ produce on their own, $\gamma^{b-1} \simeq_K \gamma^b$ and their edits writing control fields of fibers other than $n$ that the deletion does not touch. By Table 1 the deleted steps of $n$ edit no field but $\theta_n$, which Lemma 59(4) restores to Inactive at $u$ and which it held at $\gamma^{b-1}$.

*An invariant carries the suffix.* Write $\gamma'^t$ for the state the surviving steps reach at the point corresponding to $t$. We claim, for every $t > u$, that $\gamma^t \simeq_K \gamma'^t$, that every name of $R$ is vestigial at $\gamma^t$ and absent from $\gamma'^t$, and that the two states agree on every field of every name outside $R$. At $t = u + 1$ this is the paragraph above together with Definition 52, which leaves each name of $R$ retired by the accumulator that ran at $u$, Inactive and holding an empty table, the fibers of $R$ having no episode by hypothesis. The induction step is Lemma 62(1) applied at each name of $R$ in turn: a step acting outside $R$ has the same premises at the two states, reaches states again $\simeq_K$-equal, and leaves the entries of $R$ vestigial. A step acting on a name of $R$ is one of the deleted ones, and Lemma 62(2) is why it has to be deleted rather than kept, an O-Retire or O-Remove of an absent name having no fiber to act on; by (1) again such a step moves no field outside $R$, so dropping it preserves the invariant. Hence the final states are $\simeq_K$-equal, and equal outside $R$.

*No surviving step loses a premise.* A step acting on $m \notin R \cup \{n\}$ reads $n$ only through $\mathrm{target}_m(\gamma)$ or $\mathrm{relied}_m(\gamma)$. The first depends on $n$ when $m$ declares a key $n$ provides, hence $n \prec m$, and when $n$ instantiated $m$, which puts $m \in R$. In the first case $m$'s episode does not close, by hypothesis, so it is open at $\gamma^T$, where quiet gives $\omega_m = \mathrm{target}_m^T$ and Lemma 77 puts its values among the Active fibers, which $n$ is not; since a key has at most one possible provider, $n$ provided no key of $d_m$ at $m$'s L-Begin either. The second reads $n$ only through the values of $\omega_n$, and deleting the episode can only make relied false, which relaxes the guard on L-Unload rather than blocking it. What such a step reads of a name of $R$ is covered by the invariant. □

**Theorem 80. (Confluence.)** Let a sequence of steps reach a quiescent $\gamma^T$, let every component be total on its provision (Definition 76), and let $A$ be as in Definition 74. Then

1. (*Canonical form.*) $\gamma^T$ is reached, up to the names whose entries the reduction withdraws, from $\gamma^0$ by a sequence that takes the same orchestration steps in their original order, those at a fiber the orchestrator inserted preceding every lifecycle step and each of the rest following the step that instantiated the fiber it acts on, and that takes, for an enumeration $n_1, ..., n_k$ of $A$ linearizing $\lhd$, one episode of each $n_i$ in that order.
2. (*Confluence.*) Any two such sequences from $\gamma^0$ taking the same orchestration steps reach states related, after a renaming as in Lemma 61, by the $\simeq$ of Definition 58 and hence by $\simeq_K$.

*Proof.* For (1), the episodes of the sequence are of two kinds: those that close and those still open at $\gamma^T$, which by $\text{quiet}^T$ and Lemma 77 are one episode of each fiber of $A$.

*Closing episodes go first, by induction on their number.* At each stage pick a closing episode of a fiber $n$ that is $\lhd$-maximal among the fibers whose episodes still close; one exists by Lemma 75 and the finiteness of $N$. The three hypotheses of Lemma 79 are then met. No $m$ with $n \prec m$ has a closing episode, by maximality. And no fiber $n$ instantiates during $[b, u]$ has an episode: such a fiber is retired by the accumulator that ran at $u$ (Definition 52) and by Lemma 59(5) stays retired, so its target view is $\bot$ and Lemma 77 puts it outside $A$, whence it has no episode open at $\gamma^T$; and $\lhd$ relates it to $n$ through its parent pointer, so by maximality it has no closing one either. The lemma removes the episode, together with the steps of the names it instantiated, leaving $\gamma^T$ where it was up to those names. The measure drops by one, so no closing episode remains.

*A fiber outside A takes no lifecycle step.* It has no open episode at $\gamma^T$, by Lemma 77 and $\text{quiet}^T$, and no closing one now remains, so it has no episode at all and is Inactive throughout; L-Begin is the only rule that applies there, and applying it would open an episode.

*Orchestration steps go next.* An orchestration step at a fiber the orchestrator inserted moves one place earlier past a lifecycle step of a different fiber by Lemma 78(2), which applies because a step of a fiber of $A$ instantiates no such name: instantiations draw fresh names, whereas the name here is one an O-Insert of the original sequence introduced. With a lifecycle step of the same fiber there is nothing to exchange, an O-Insert of $n$ already preceding every step of $n$ and an O-Retire or O-Remove of $n$ applying only outside $A$, which takes no lifecycle step. Moving each to the front in turn preserves their relative order. An orchestration step at a fiber some activation instantiated cannot go to the front, its premises requiring that fiber to be present, so it stays where the instantiation put it; it acts outside $A$ by the paragraph above and therefore commutes with everything between it and the instantiation by the same clause of Lemma 78.

*Episodes are sorted and made contiguous, by induction on $|A|$.* Let $n_1$ be $\lhd$-minimal in $A$. Then $d_{n_1} = \varnothing$ and $\pi_{n_1} = \text{root}$, since Definition 74 puts a provider of a key of $d_{n_1}$ and the fiber instantiating $n_1$ in $A$ while $\lhd$ puts both below $n_1$. So $\text{target}_{n_1}$ reads no field of another fiber and, no orchestration step remaining to write $\tau_{n_1}$ and no fiber below $n_1$ remaining to retire it, is constant. Every step acting on $n_1$ is an activation step, no episode closing, and its remaining premises read $\theta_{n_1}$ and $i_{n_1}$, which by Table 1 only $n_1$ writes; each is therefore applicable at every earlier state, and Lemma 78 moves it one place earlier without moving the endpoint. Its independence hypothesis is met because no fiber entangled with $n_1$ takes a step in the region crossed — $d_{n_1} = \varnothing$ leaves $n_1$ no provider, and a fiber declaring a key of $p_{n_1}$ has target $\perp$ until $n_1$ is Active, so its steps all follow the last step of $n_1$ — and Lemma 66 makes every other pair independent. The number of steps of other fibers preceding a step of $n_1$ drops by one at each application, so the episode of $n_1$ becomes an initial contiguous block. The argument repeats on $A \setminus \{n_1\}$ over the suffix that follows the block, where $n_1$ is Active throughout and takes no further step, so it too contributes a constant target; the providers a later $n_j$ declares lie in earlier blocks and take no step in the suffix, so the entanglement argument above holds at every stage. The enumeration this produces linearizes $\lhd$ by construction.

For (2), both sequences reduce by (1) to a canonical one, and the two reductions run over the same $A$ up to a renaming. Definition 74 reads $\tau$, $\pi$, $d$, and $p$, of which the last three are written once with a fiber's entry (Lemma 59(5)), so what has to be seen is that the same names come into existence carrying the same $d$, $p$, and $\pi$, and that the same names are retired. Insertions the two sequences share by hypothesis. Instantiations they share as well: an activation of a fiber of $A$ instantiates, at each of its iterations, the component the iterator names there, which the interleaved steps hold fixed — clause (2) of Definition 42 for a fiber not entangled with the activating one (Lemma 66), and Lemma 67(3) leaving an entangled fiber no $\Psi \neq \mathrm{id}_\Gamma$ while the activation runs — so the tree of instantiations below an $A$-fiber is a function of that fiber's component; the names those instantiations draw are not shared, and it is here that Lemma 61 is applied, matching the two trees by a bijection. And a retirement is either an orchestration step, shared, or the O-Retire an accumulator takes, which retires exactly the names the same activation instantiated. Two enumerations linearizing $\lhd$ differ by transpositions of incomparable episodes; Lemma 78 leaves each endpoint unchanged up to the $\simeq$ of Definition 58, and Lemma 60 carries the steps that follow across that relation, so the two canonical sequences agree. With the termination of Theorem 73, the lifecycle relation therefore has unique normal forms. □

The theorem is what licenses reasoning about a Cordis application as though it were statically assembled. An orchestrator that adds a component, removes it, replaces a provider, and undoes the replacement is guaranteed to arrive at the state it would have obtained by writing the final composition down at the outset, and a component author reasoning about which coeffects are in scope may reason about the quiescent state alone. It also delimits the guarantee: it speaks of the state, not of the emissions the system produced along the way, which is the distinction Section 6.1 draws between an acquisition, tracked inside the boundary, and an emission, which crosses it.

### 4.4. Extensions

We give four extensions of the calculus, each realized by the implementation of Section 5 and each leaving the results of Section 4.3 intact.

**Asynchrony.** The rules are synchronous: the state map $\Psi^t$ of a step (Definition 58) is applied whole at that step, and the environment moves only between one map and the next. In an asynchronous host the iterations and the inverses yield futures, so a map in flight runs to completion whether or not it is still wanted, and the aborting alternative of L-Divert is not one such a host can offer. Such a host is *inertial*: of L-Divert it takes the landing alternative alone, and a fiber whose target view turns during an iteration deactivates after that iteration lands, from Unloading, holding the inverse it produced. Inertia is therefore a restriction on which alternative of L-Divert a host may take; every result of Section 4.3 quantifies over all sequences of steps and so covers the inertial ones, and Theorem 73 appeals to the aborting alternative nowhere, so a host bound by inertia still quiesces. An inverse in flight calls for no counterpart of inertia: the rules never decline a deactivation, Unloading being left by L-Unload alone with no premise on the target view, and a step of another fiber falling within the accumulator's application commutes with the inverses not yet applied, by the arguments Theorem 68 rests on (Definition 65, Lemma 67, Lemma 66), so an application spread over an interval reaches the state the one-step application reaches, up to $\simeq_K$. A deactivation may also chain straight back into an activation: the accumulator runs whatever the target view has become, and from Inactive an L-Begin may immediately follow, which is the mutual chaining of `reload` and `unload` in the implementation (Section 5.1.3).

**Failure.** The effects a component installs reach outside the context that tracks them, and a location they reach may refuse: a port already bound, a file that is not there, a peer that does not answer. Refine the iterator so that an iteration may *raise* an error in place of yielding a triple, $\Gamma \to \mathsf{Either}(\Xi, \Gamma \times (\Gamma \to \Gamma) \times \mathsf{Maybe}(\mathfrak{I}))$ for a set $\Xi$ of errors, the witness constraining the Right case alone, a raise having nothing to undo. A raise exits Reloading by the route of an aborting L-Divert whose premise on the target view is dropped, the iteration rather than the environment choosing the abort: the fiber routes into Unloading with the accumulator built up to the failing iteration, arrives at Inactive having installed nothing (Corollary 69), and the exit writes the error as an *outcome* on the fiber. The outcome withholds re-entry: L-Begin is read as requiring an error-free fiber, so an effect function that raised is not retried against an unchanged environment, and quiet admits a failed fiber whatever its target view; the failure likewise stays on the fiber rather than propagating to its parent, leaving siblings running. A retry is a revision: the reinserted fiber of the Configuration paragraph starts without an outcome. Preservation and recovery hold unchanged, a raise leaving by the same Unloading route every deactivation takes. Confluence excludes failed fibers, and has to: whether an iteration raises depends on the state it meets, so one schedule may fail a fiber where another completes it, and the two quiescent states then differ in that fiber's lifecycle state and, by Corollary 69, in nothing else. The `FAILED` state of the implementation carries this outcome (Section 5.1.3).

**Isolation.** Section 4.1 reads every key at one shared realm and names the relaxation a calculus carrying realms would make: provisions disjoint within a realm rather than outright, each declared key resolved against the realm of the fiber declaring it. For realms fixed at a fiber's insertion, the relaxed calculus is the present one read at a larger key set. Take the key set to be $K \times R$, each pair $(k, r)$ carrying the value set $\mathcal{V}_k$ and the operations $\mathcal{A}_k$ of its underlying key, so that $\simeq_{(k,r)}$ is $\simeq_k$, a key commutative in the sense of Definition 44 is commutative at every realm, and the coeffect at a pair inherits the witness of its underlying key (Definition 46). A fiber inserted under a realm table $\rho$ then declares and provides pairs, a key $k$ of its interface standing for $(k, \rho(k))$; the last premise of O-Insert, read at pairs, is disjointness within a realm, two fibers providing one key in different realms provide different pairs, and the one shared realm is the diagonal $(k, k)$, a key outside $\mathrm{dom}(\rho)$ resolving to its own realm (Definition 24). Keys are as atomic to the rules as names: no rule computes one, inspects its structure, or relates two of them by anything but equality, so the rules and the results of Section 4.3 apply at $K \times R$ as they stand. The reading stops at isolate itself (Definition 25), which reassigns $\rho$ on a running context: under the pairing a reassignment moves a declaration from one pair to another, and $d_n$ and $p_n$ are written once with a fiber's entry (Lemma 59(5)), so a fiber whose realm turns at runtime is one whose interface has changed, a revision the Configuration paragraph carries. Interception (Definition 26) calls for no extension: metadata adjusts how a binding is used rather than what a key resolves to, is consulted when the binding is accessed, and is discarded with the context that carries it, so no premise reads it and no field of a fiber holds it.

**Configuration.** A component of the implementation takes a configuration, and instantiation binds the payload into the effect function the fiber runs (Section 5.1.3). The calculus therefore carries configuration inside $e$: one such component bound to two payloads is two components of Definition 48, differing in their effect functions. The declarative layer of Section 5.2.1 further lets the orchestrator revise a running fiber, replacing its configuration, reassigning its realms, or disabling and later re-enabling it. Each revision is a composite of the rules: disabling is an O-Retire, and every other revision retires the fiber, lets the lifecycle rules deactivate it, removes the entry, children before parent as O-Remove requires, and reinserts the fiber at the same name, with the new effect function, the new realm pairs, or, at a re-enablement, the component unchanged; the name may be reissued, no stale committed view naming a removed fiber (Section 4.3.1). Dependents follow unprompted: the guard on L-Unload orders the withdrawal after the deactivations it causes, and the target-view comparison reactivates each dependent once the reinserted fiber provides the keys again. Re-enablement takes the composite rather than a rule writing $\tau$ back to $\perp$, for two reasons: Theorem 80 rests on a fiber an accumulator retires staying retired (Lemma 59(5)), and Definition 53 reads no parent pointer, so a fiber un-retired after the accumulator of the fiber that instantiated it has run would activate with its creator gone. The implementation keeps the same division: re-enabling an entry instantiates a fresh fiber, the entry being the identity that survives revision and the fiber the identity of one enablement. The composite is held to its endpoint rather than to its steps: by Theorem 80 the system quiesces where a load of the revised configuration from scratch would have left it, and that endpoint is what the loader's shorter routes answer to, a new payload handed to a component that reloads only on a material change, a realm moved without reloading its provider (Section 5.2.1).

## 5. Implementation and Case Study

This section presents *Cordis*, which realizes the formal models of Section 3 as a practical programming abstraction. Cordis is a *meta-framework* of spatiotemporal composability: unlike application frameworks that target a specific domain (e.g., web routing, ORM, UI rendering), it prescribes no concrete scenario; its sole responsibility is to supply universal dynamic composition semantics. The implementation is layered into three tiers: (1) the *core library* (Section 5.1) implements the effect and coeffect systems directly; (2) the *component loader* (Section 5.2) extends the core with configuration reconciliation and hot module replacement; and (3) application frameworks such as *Koishi* (Section 5.3) build domain-specific functionality on top of the former two tiers.

### 5.1. Core Library

Table 2 summarizes the correspondence between theoretical constructs and their runtime counterparts. In particular, we use the runtime names introduced below throughout this section, reserving the theoretical symbols for the formal correspondence. We also write `@@name` for a framework-internal symbol key, so the brackets in `ctx[@@store]` denote symbol-keyed access to an opaque slot on the context, rather than indexing into a string-keyed map.

| **Theory (Section 3, Section 4)** | **Implementation** |
|---|---|
| $\Gamma_\infty$ | `ctx`, the first-class context |
| $\gamma \in \Gamma$ | the context tree together with everything the running system has touched |
| $\mathfrak{E}_\Gamma$, $\mathfrak{I}_\Gamma$ | Effect callback returning / yielding inverses |
| $\text{effect}_\Gamma(e)$ | `ctx.effect(callback)` |
| $\Sigma$, $\Sigma^{\text{iso}}$, $\Sigma^{\text{inter}}$ | `ctx[@@store]`, `ctx[@@isolate]`, `ctx[@@intercept]` |
| $\text{get}(k)$, $\text{set}(k, v)$ | `ctx.get(key)`, `ctx.set(key, value)` |
| $\text{isolate}(k, r)$ | `ctx.isolate(key, realm)` |
| $\text{intercept}(k, \nu)$ | `ctx.intercept(key, metadata)` |
| $\langle d, p, e, \pi, \sigma, \tau, \theta \rangle$ | `fiber`, the instantiation of a component in $\mathfrak{C}_\Gamma$ |
| $\text{dom}(F_\gamma)$ | enumerated through `ctx.registry` |
| $n : \mathfrak{N}$ | `fiber.uid` |
| $d : \mathfrak{D}_\Gamma$ | `fiber.inject` |
| $p : \mathfrak{P}_\Gamma$ | the component's `provide` |
| $e : \mathfrak{I}_\Gamma^{d \cup p}$ | `fiber.apply` |
| $\pi : \mathfrak{N}$ | `fiber.parent.fiber.uid`, the fiber owning the context it was instantiated on |
| derived realization (Definition 23) | `fiber.ctx`, the child context the fiber runs in |
| $\theta$ (Definition 49) | `fiber.state`, the lifecycle state, whose `LOADING` is Reloading and whose `FAILED` carries the error outcome of Section 4.4 |
| recover, accumulator $g$ | `fiber.dispose`, the accumulator |
| $\omega$ (Definition 49) | `fiber.committed`, the committed view |
| $\text{provider}_k(\gamma)$ | an `Impl` whose provider fiber is `ACTIVE` |
| $\text{target}(\gamma, n)$ | `fiber.target`, recomputed by `refresh` (Algorithm 5), where $\perp$ is `INACTIVE` |
| Future, inertia (Section 4.4) | `fiber.inertia`, the handle of the transition in flight |
| O-Insert, O-Retire (Definition 52) | `ctx.use` and the inverse of its `callback` (Algorithm 4) |
| O-Remove | the fiber dropped from its runtime, with `uid` cleared |
| L-Begin, L-Iter, L-Finish | `execute`'s iteration loop (Algorithm 1) |
| L-Divert | the `guard` failing at an iteration boundary (Algorithm 1), or `reload` chaining into `unload` |
| L-Leave | `refresh` marking the fiber `UNLOADING` (Line 10) |
| L-Unload | `unload` and its inertial chaining (Algorithm 5) |
| guard on L-Unload | `unload` awaiting the notified dependents (Line 25) |
| failure (Section 4.4) | the error recorded on the fiber, with its `target` set to $\perp$ |

Table 2 | Theory-to-implementation correspondence

The remainder of this section builds the core library from the bottom up. Section 5.1.1 realizes revertible effects, the sole primitive through which a context is mutated; Section 5.1.2 realizes reactive coeffects over it; Section 5.1.3 composes both into the component lifecycle; and Section 5.1.4 exposes the context-level operations built on them.

### *5.1.1. Effect Tracking*

This section realizes revertible effects (Section 3.1). Every context mutation in Cordis flows through a single primitive, `ctx.effect`: coeffect provision, component instantiation, and every other context-mutating operation reduces to a `ctx.effect` call, so any operation performed through the context is automatically tracked and reverted upon component unloading. Operationally, `ctx.effect` is the realization of $\text{effect}_\Gamma^{\text{iter}}$ (Definition 18): it takes a callback of type $\mathfrak{I}_\Gamma$ and lifts it to $\mathfrak{I}_{\partial\Gamma}$, yielding a dispose closure that, when invoked, reverts the effect. Cordis accepts both $\mathfrak{E}_\Gamma$ and $\mathfrak{I}_\Gamma$ through this one operation (ad-hoc polymorphism); we take the iterator form as representative, since a plain effect function is the degenerate iterator that yields a single inverse. What the operation does not check is the witness that $\mathfrak{E}_\Gamma^*$ carries: the callback supplies an inverse, and that the inverse reverts the effect it accompanies is an obligation on the component author rather than a property the runtime verifies. Theorem 68 is where the calculus appeals to it, and Section 6.1 is where the obligation is delimited. The witness of a coeffect (Definition 46) is unchecked in the same way: that the operations published at a key commute is an obligation on the component providing it, discharged by the representation choice of Section 3.4.2.

Algorithm 1 shows the construction of `ctx.effect`. We write $f \circ g$ for the disposer that runs $f$ after $g$, and `id` for the no-op; prepending each new inverse therefore yields LIFO recovery.

**Algorithm 1** Effect tracking

```
1  async function execute(callback, guard)
2      iter ← callback()
3      inverse ← id
4      while guard()
5          (value, done) ← await iter.next()
6          if value then inverse ← value ∘ inverse
7          if done then break
8      return inverse
9  function effect(ctx, callback)
10     armed ← true
11     task ← execute(callback, () ↦ armed)
12     async function dispose()
13         if not armed then return
14         armed ← false
15         recover ← await task
16         recover()
17     ctx.dispose ← dispose ∘ ctx.dispose
18     return dispose
```

The engine `execute` drives the callback as an effect iterator ($\mathfrak{I}_\Gamma$, Definition 17) and folds the inverse yielded at each step into a single composite. Before each step it consults a caller-supplied `guard`; once the guard trips, iteration stops and only the inverses accumulated so far remain. This is the step-boundary interruption of Section 4.2.2: the $\mathsf{Maybe}(\mathfrak{I})$ continuation is realized by the iterator's `done` flag together with `guard`.

`ctx.effect` is a thin wrapper over `execute` that adds two things. First, self-disposal: the guard reports the `armed` flag, and the returned `dispose` flips `armed` to false, which simultaneously halts any in-flight iteration and makes recovery fire at most once. Firing twice would apply an inverse at a state no application of the effect produced, where nothing holds it to reverting anything. Second, parent composition: `dispose` is prepended to the enclosing context's accumulated inverse `ctx.dispose`, so a child effect's inverse is itself an effect on the parent, which is the recursive structure of $\partial^2\Gamma$. The component level (Section 5.1.3) reuses the same `execute` with a guard that tests the stability of `fiber.target` instead of `armed`.

### *5.1.2. Coeffect Operations*

This section realizes reactive coeffects (Section 3.2). All coeffect operations act on three symbol-keyed slots that each context carries:

- `@@store`: the value store $\sigma : (r : R) \rightharpoonup \mathcal{V}_r$ from realm symbols to typed values;
- `@@isolate`: the realm table $\rho : \mathrm{Map}(K, R)$ from coeffect keys to realm symbols;
- `@@intercept`: the interception table $\iota : (k : K) \to \mathcal{M}_k$ assigning each key its metadata.

The first two compose into the two-layer resolution $k \to \rho(k) \to \sigma(\rho(k))$: `ctx.get(key)` (Algorithm 2) reads the realm symbol $\rho(k)$ from `@@isolate`, then the bound value $\sigma(\rho(k))$ from `@@store`. The $\rho$ indirection lets isolation redirect a key to an independent binding, whereas `@@intercept` is consulted only when a binding is accessed, adjusting how it is used rather than what it resolves to. We realize these operations in two parts: (1) *provision and notification*, which install or withdraw bindings and propagate the change to dependents; and (2) *isolation and interception*, which reshape how a key resolves.

**Provision and notification.** Since $\mathrm{set}(k, v)$ has type $\mathfrak{E}_\Sigma$ (Section 3.1), coeffect provision is a `ctx.effect` call and inherits its automatic tracking and recovery. Algorithm 2 implements `ctx.set(key, value)`, the concrete $\mathrm{set}(k, v)$: the callback binds a value into the store under the realm symbol $\rho(k)$, and the returned dispose function removes it. Both installation and removal invoke `notify` to propagate the change to dependent components.

**Algorithm 2** Coeffect operations

```
1  function get(ctx, key)
2  |   realm ← ctx[@@isolate][key] ▷ ρ(k)
3  |   return ctx[@@store][realm] ▷ σ(ρ(k))
4  function set(ctx, key, value)
5  |   function callback()
6  |   |   realm ← ctx[@@isolate][key] ▷ ρ(k)
7  |   |   ctx[@@store][realm] ← value ▷ σ[ρ(k) ↦ v]
8  |   |   notify(ctx, [key])
9  |   |   return function()
10 |   |   |   delete ctx[@@store][realm] ▷ σ \ ρ(k)
11 |   |   |   notify(ctx, [key])
12 |   return ctx.effect(callback)
```

Algorithm 3 propagates each binding change to dependents by testing, for each live fiber, whether a changed key appears in its `fiber.inject` and resolves to the same realm; if so, it calls `refresh` (Section 5.1.3) to re-evaluate that fiber against the new state, and it returns the fibers it re-evaluated so that a caller can wait for them. This is the reactive classification of Definition 22: a change that flips satisfaction activates or deactivates the fiber, and `refresh`'s idempotence renders a neutral change harmless. The interaction of this re-evaluation with diverse control flows is developed in Section 5.1.3.

**Algorithm 3** Reactive notification

```
1  function notify(ctx, keys)
2  |  affected ← ∅
3  |  for fiber in all_fibers do
4  |  |  for key in keys do
5  |  |  |  if key ∈ fiber.inject and fiber.ctx[@@isolate][key] = ctx[@@isolate][key] then
6  |  |  |  |  refresh(fiber)
7  |  |  |  |  affected ← affected ∪ {fiber}
8  |  |  |  |  break
9  |  return affected
```

A binding counts as available to a dependent only while the fiber that installed it is `ACTIVE`, so `refresh` resolves each declared key against an active provider rather than against the store alone. This is the *provided by* relation of Definition 53, and it is what makes a withdrawal visible to dependents one step before it happens: a provider that has entered `UNLOADING` has stopped providing, so its dependents recompute an unsatisfied target view and begin their own teardown while its bindings are all still in place.

**Isolation and interception.** The two operations do structurally the same thing: each derives a child context that adjusts one inherited table for `key`, leaving the parent untouched, so recovery is implicit: discarding the child context suffices, with no explicit inverse to run. `ctx.isolate(key, realm)` overrides the realm mapping $\rho$ with `realm`, or a freshly generated symbol by default (realizing $\mathrm{isolate}$, Definition 25), so two contexts that assign different symbols to the same key resolve to independent bindings. `ctx.intercept(key, metadata)` merges `metadata` into the interception table $\iota$ (realizing $\mathrm{intercept}$, Definition 27): following that definition, the new metadata is combined with whatever the context already carries for `key` and takes priority over it.

### *5.1.3. Component Lifecycle*

A component is instantiated as a *fiber* by `ctx.use`. This section gives the fiber (introduced in Section 5.1) operational meaning as the inertial state machine of Section 4.4. Two fields drive the algorithm below: `fiber.parent`, the parent context of `fiber.ctx` that forms the component hierarchy (the recursive structure of $\Gamma_\infty$, Section 3.3.1), and `fiber.inertia`, a handle to the in-flight asynchronous transition (or null if idle).

Algorithm 4 shows component instantiation. A *component* pairs a coeffect specification `component.inject` ($d$) with an effect function `component.apply`; instantiation binds the component's `config` into `fiber.apply` (Line 9), the config-applied effect function ($e$) that the lifecycle then runs. The `callback` function (Line 2) is the effect tracked in the *parent* fiber: when executed, it initiates the child's lifecycle by calling `refresh` (Algorithm 5); when reverted, it forces the child's target to $\bot$ and triggers `unload`. This is the instantiation primitive of Definition 52, with `callback` as its O-Insert and the closure `callback` returns as its O-Retire: an instantiation is an ordinary tracked effect of the parent, so unloading a parent cascades to its children.

**Algorithm 4** Component instantiation

```
1  function use(ctx, component, config)
2      function callback()
3          refresh(fiber)
4          return function()
5              fiber.target ← ⊥
6              unload(fiber)
7      fiber ← Fiber(parent: ctx, inject: component.inject)
8      fiber.ctx ← ctx[fiber ↦ fiber]
9      fiber.apply ← () ↦ component.apply(fiber.ctx, config)
10     ctx.effect(callback)
11     return fiber
```

Algorithm 5 realizes the inertial state machine of Section 4.4, in which `reload` and `unload` are inertial: once entered, a transition runs to completion before the system responds to a target-state change. It uses two auxiliary lookups over the coeffect store: `resolve(inject)` returns the bindings the declared keys currently resolve to, and `provided(fiber)` returns the keys whose binding this fiber installed. The `refresh` function recomputes `fiber.target` from the coeffect store and, if the fiber is not already in a transition, initiates either a `reload` or `unload` task[2]. The `reload` function records the current target and executes the component's effect function `apply`. Upon completion, it checks whether the target still matches: if so, the fiber enters `ACTIVE`; if not (regardless of whether the new target is $\perp$ or a different set of providers), it chains into `unload`. Symmetrically, `unload` reverts all tracked effects in LIFO order and then either enters `INACTIVE` or chains into `reload`. This mutual recursion implements the inertial property: once a transition begins, it completes before any new transition can start.

**Algorithm 5** Component lifecycle

```
1  function refresh(fiber)
2      target ← target(γ, n)
3      if target = fiber.target then return
4      fiber.target ← target
5      if fiber.inertia then return
6      if target ≠ ⊥ then
7          fiber.state ← LOADING
8          fiber.inertia ← create_task(reload(fiber))
9      else
10         fiber.state ← UNLOADING ▷ out of service before any inverse is scheduled
11         fiber.inertia ← create_task(unload(fiber))
12 async function reload(fiber)
13     target_0 ← fiber.target
14     fiber.committed ← resolve(fiber.inject) ▷ commit the view
15     recover ← await execute(fiber.apply, () ↦ fiber.target = target_0)
16     fiber.dispose ← recover ∘ fiber.dispose
```

[2] `create_task` schedules an async function to run concurrently and returns a handle to it (stored in `fiber.inertia`). We write it explicitly for language independence: with eager scheduling (e.g., TypeScript promises), the call is implicit and the returned promise is the handle, whereas with lazy scheduling (e.g., Python coroutines, Rust futures) the host must spawn the task for it to progress.

```
17 │ if fiber.target = target₀ then
18 │ │ fiber.state ← ACTIVE
19 │ │ notify(fiber.ctx, provided(fiber))
20 │ │ fiber.inertia ← null
21 │ else
22 │ │ fiber.state ← UNLOADING
23 │ │ fiber.inertia ← create_task(unload(fiber))
24 async function unload(fiber)
25 │ await all(notify(fiber.ctx, provided(fiber)).map(f ↦ f.await())) ▷ drain dependents
26 │ await fiber.dispose()
27 │ fiber.dispose ← id
28 │ fiber.committed ← ⊥
29 │ if fiber.target = ⊥ then
30 │ │ fiber.state ← INACTIVE
31 │ │ fiber.inertia ← null
32 │ else
33 │ │ fiber.state ← LOADING
34 │ │ fiber.inertia ← create_task(reload(fiber))
```

`fiber.target` is computed by resolving each declared key against the current coeffect store and tupling the `uid` of the fiber that provides it, so it is a digest of $\mathrm{target}(\gamma, n)$ (Definition 53). Identifying a binding by its provider rather than by its value is what makes a single comparison against the recorded target sufficient: a `uid` is drawn fresh and never reused, so a provider that is replaced cannot be mistaken for the one it replaced, even when the two provide equal values. Since `notify` (Section 5.1.2) recomputes the target on every coeffect change, a fiber reloads precisely when one of its declared keys comes to be provided by a different fiber. A provider that overwrites its own binding in place is therefore not observed; a component that wants its replacement to propagate withdraws the binding and installs it afresh.

The algorithm operates at two complementary levels. At the *transition* level, `reload` and `unload` check the target at completion, enabling inertial chaining across transitions. At the *iteration* level within each transition, the effect execution (Algorithm 1) checks the target at each iteration boundary, enabling partial rollback within a single transition. These two mechanisms correspond to the inter-transition chaining of Section 4.4 and the intra-transition staleness check that Theorem 71 rests on.

Three lines carry the coeffect ordering of Theorem 70, and where each of them sits is what makes the ordering hold. `reload` commits the resolved view at Line 14 and `unload` discards it only after every inverse has run, so a fiber reads the same bindings for as long as it is loaded, its own teardown included. `refresh` marks the fiber `UNLOADING` at Line 10 before the transition task is created, which is the L-Leave step: the fiber stops providing, and the dependents recompute against that before any of its inverses is scheduled. `unload` then waits at Line 25 for each notified dependent to reach `INACTIVE`, which is the guard on L-Unload; `notify` admits a dependent only when its declared key resolves to the same realm symbol as the provider's, which is the runtime form of the guard's demand that the dependent see the key *from this fiber* rather than merely declare it. The wait sits ahead of the whole recovery rather than inside one of the inverses being waited on, since `fiber.dispose` initiates a fiber's effects concurrently and a wait placed within one of them would leave the rest unordered. Termination follows Theorem 73: a fiber only ever waits on dependents that have already stopped being satisfiable, and a dependent that is itself a provider waits the same way for its own, so the provider graph is traversed on demand rather than analyzed in advance.

#### *5.1.4. Context Access*

The coeffect operations of Section 5.1.2 form a *reflective* API: a coeffect is written with `ctx.set(key, value)` and read with `ctx.get(key)`, both keyed by name. Cordis layers a second, more native way to extend and consume the context on top of this reflective API: *property access*. A component can access a coeffect as the property `ctx[key]`, as if it were native structure of the context, rather than through a method call. In TypeScript, Cordis realizes this with a `Proxy` whose `get` trap mediates every property access. Algorithm 6 shows how a context resolves such an access to a coeffect, atop the primitive `get` of Section 5.1.2.

**Algorithm 6** Proxy-mediated context access

```
1  function resolve(ctx, key)
2      fiber ← ctx.fiber
3      repeat
4          if key ∈ fiber.committed then return fiber.committed[key]
5          if key ∈ fiber.inject then throw INACTIVE_ACCESS
6          if fiber = root then throw UNDECLARED_ACCESS
7          fiber ← fiber.parent.fiber
```

Algorithm 6 walks the fiber chain upward from the accessing context: at the first fiber whose committed view binds `key`, the access is authorized and that binding is returned; if the walk reaches a fiber that declares `key` without having committed it, the fiber is not loaded and the access fails; and if it reaches the root without any declaration, the access is rejected as undeclared. This is where the proxy differs from the bare `ctx.get`: `ctx.get(key)` is a lookup against the store that returns the bound value or nothing and never fails, whereas the proxy resolves against the accessing fiber's own view and enforces the coeffect specification $d$ at the point of use. Reading the view rather than the store is also what Theorem 70 rests on, since it is what keeps a dependency readable to a component whose teardown was triggered by that dependency going away.

This rejection is a *runtime* check performed at the point of access. Because a component's coeffect specification $d$ is declared statically, the same violation is in principle detectable at compile time, by resolving each `ctx[key]` against the declared $d$ before execution; Section 6.4 discusses how a host language's type-level dependency declarations and compile-time metaprogramming can carry out exactly this mediation.

### 5.2. Component Loader

The core library equips *component developers* with imperative primitives for dynamic composition, such as `ctx.effect`, `ctx.use`, and `ctx.set`. A separate concern arises for *application orchestrators*, who assemble pre-existing components into a running system and adjust the composition over its lifetime. The *component loader* addresses this concern by introducing a declarative configuration layer: the orchestrator specifies the desired composition as a persistent data structure, and the loader translates changes to this specification into the corresponding imperative fiber operations.

### *5.2.1. Declarative Configuration*

Section 4 decomposes a running system into fibers, each an instantiation of one component. Everything an instantiation needs can be declared, so an orchestrator can describe a whole system as a *declarative configuration*: a persistent record that the loader realizes as fibers and keeps in step with them.

**Entries.** A configuration consists of *entries*. Each entry specifies a fiber and manages it, and the binding runs in both directions: the loader responds to a change in an entry's fields by adjusting the fiber, and a component that revises its own configuration or disables itself has the change written back to its entry.

**Definition 81.** An *entry* declares a single fiber, recording:
- `id` — a stable identifier, used as the reconciliation key when its group's child list changes;
- `url` — the URL of the component module to instantiate;
- `isolate` — an isolation annotation applied to the entry's context;
- `intercept` — an interception annotation applied to the entry's context;
- `config` — the configuration bound into the component to form its effect function `apply`;
- `disabled` — whether the entry is administratively turned off.

An entry can serve as a faithful specification because what supports a fiber is exactly what an entry records. The support set of Definition 74 reads $\tau$, $\pi$, $d$, and $p$ and nothing else, and an entry gives all four: `disabled` gives $\tau$, the entry's parent in the tree gives $\pi$, and `url` selects the component which declares $d$ and $p$. The fields the support set leaves unread are the fiber's runtime state, which an instantiation does not need either, and Lemma 77 identifies the support set with the Active fibers of a quiescent state (Definition 53) as far as each component installs every key it declares (Definition 76).

These entries form a *configuration tree* that is the authoritative record of what the system loads. An entry may be a leaf mapping to a single fiber, or its component may in turn load further components, making the entry a branch node. Cordis provides components for such grouped and nested loading: `@cordisjs/group` takes a list of child entries as its configuration and loads them as a subgroup, and `@cordisjs/include` loads an external configuration file (YAML or JSON) and grafts its entries in as a nested subtree. Both are ordinary components resting on the instantiation primitive of Definition 52 (Algorithm 4), so a nested tree stays within the calculus and the results below hold of it.

**Reconciliation.** When an entry's record changes, the loader reconciles incrementally rather than tearing the fiber down and rebuilding it wholesale. Reconciling this way is sound for reasons the metatheory supplies.

- Theorem 80 makes the quiescent state a function of the final configuration alone: whatever instantiations and retirements the loader performs on the way, and in whatever order, the system quiesces where a load of the final configuration from scratch would have left it. Which components end up loaded is read off the declarations only as far as each of them installs every key it declares (Definition 76); a component that declares a key and installs it under some configurations alone is one the loader can still reconcile, but the set of loaded components then answers to those configurations as well.
- Theorem 73 proves that the system does quiesce, so a reconciliation is complete once its instantiations and retirements have been issued.
- Corollary 69 puts a departing fiber's contribution to the state at nothing, so rebuilding one entry withdraws what its fiber installed and leaves the fibers around it as they were.

- Theorem 70 lets the entries be instantiated together, with no load order for the orchestrator to arrange: a fiber whose declared keys are not yet provided waits at its L-Begin, and one whose provider leaves is deactivated ahead of it. A dependency therefore constrains when a fiber activates rather than when its module is fetched and evaluated, so the loader loads modules concurrently, where bringing up a large configuration spends its time.

On top of the fiber that an entry declares, the loader dispatches on which of the entry's fields changed and applies the least disruptive operation for each.

- `id`, `url` — rebuilds the entry, since its identity or its component has changed;
- `isolate` — reassigns the entry's realms (Algorithm 7);
- `intercept` — updated in place, as interception metadata is consulted at read time and needs no reload;
- `config` — handed to the component, which decides how to apply the new payload, typically by diffing it against the previous one and reloading only on a material change. In particular, an `@cordisjs/group` entry's `config` is its list of child entries, so it applies the update as a keyed diff over child `id`s, creating, removing, or updating each child; since updating a surviving child re-enters this same per-field dispatch, group reconciliation and entry update recurse together down the tree;
- `disabled` — unloads the fiber when set and reloads it when cleared.

**Managed realms.** Isolation in the core derives a child context overriding the realm table $\rho$ at one key (Section 5.1.2), which suffices while the context tree stands still. An entry may be moved between groups at runtime, so the loader manages realms of its own, and the `isolate` field selects between two scoping rules per key. A value of `true` selects a *local* realm, private to the entry and tagged by its `id`, which the entry carries with it wherever it moves; a string selects a *global* realm shared by every entry naming that string, so moving such an entry changes which entries it shares a binding with rather than which realm it belongs to. A realm is discarded once no entry names it.

Reassigning an entry's realms turns on which keys changed realm, whether the entry is itself the provider at a changed key, and which dependents to notify. The middle question is the hard one, since a realm symbol may be shared by several fibers of which only one is the provider. The loader answers it with *delimiters*: one symbol $\delta_k$ per key, under which each context stores a tag of its own. A delimiter is written on a context and inherited by its descendants, so the entry's tag and the provider's agree exactly when the two were derived within one isolate scope for $k$, which is the case in which the binding at $k$ is the entry's own and has to move with it.

**Algorithm 7** Isolation realm reassignment

1. **function** patch_isolation(entry, $\rho'$)
2. $\quad \rho \leftarrow$ entry.ctx[`@@isolate`]
3. $\quad$ store $\leftarrow$ entry.ctx[`@@store`]
4. $\quad \Delta \leftarrow \{k \mid \rho(k) \neq \rho'(k)\}$ ▷ *keys whose realm changes*
5. $\quad$ **for** $k$ **in** $\Delta$ **do**
6. $\quad\quad$ entry.ctx[$\delta_k$] $\leftarrow$ fresh tag
7. $\quad\quad$ diff[$k$] $\leftarrow$ ($\rho(k)$, $\rho'(k)$, entry.ctx[$\delta_k$], store[$\rho(k)$].fiber.ctx[$\delta_k$])
8. $\quad$ entry.ctx[`@@isolate`] $\leftarrow \rho'$
9. $\quad$ reload(entry.fiber)
10. $\quad$ **for** $k$ **in** $\Delta$ **do**

```
11  |  | (s_1, s_2, d_1, d_2) ← diff[k]
12  |  | if d_1 = d_2 and store[s_1] and not store[s_2] then ▷ the binding is the entry's own
13  |  |  | store[s_2] ← store[s_1]
14  |  |  | delete store[s_1]
15  | function affected(fiber, k)
16  |  | (s_1, s_2, d_1, d_2) ← diff[k]
17  |  | return fiber.ctx[@@isolate][k] ∈ {s_1, s_2} and (fiber.ctx[δ_k] = d_1) ≠ (d_2 = d_1)
18  | notify(entry.ctx, Δ, affected) ▷ in place of the realm test of Algorithm 3
```

The test turns on one property of delimiters. The tag under $\delta_k$ is written on the entry's context and inherited by every context derived from it, and it is drawn afresh at each reassignment, so for a context $\gamma'$

$$\gamma'[\delta_k] = d_1 \quad \Longleftrightarrow \quad \gamma' \text{ is derived from the entry's context} \tag{64}$$

Write $\text{own}(\gamma')$ for that condition, of which $d_2 = d_1$ is the instance at the provider. The reassignment moves the contexts satisfying own from $s_1$ to $s_2$ and leaves the others where they are, and by the loop above it moves the binding to $s_2$ exactly when the provider satisfies own. A dependent sees the binding while its own realm at *k* is the realm the binding sits in. Where own agrees on the dependent and the provider, both move or neither does, so the dependent sees the binding afterwards exactly when it saw it before. Where own separates them, one side moves and the other stays, so the dependent gains or loses the binding. The inequality is that separation, and the membership test drops the dependents resolving *k* in neither realm, which no part of the move reaches.

#### 5.2.2. *Hot Module Replacement*

Hot module replacement (HMR) applies the revertible-effect pattern at the *module* level: when source files change, typically during development, the system replaces the affected modules in-place without restarting the process. Because a fiber already bounds all of its component's effects and coeffects, a module that is itself a component can be replaced through fiber operations alone: disposing the old fiber recovers everything the component installed, and a new fiber instantiated from the reloaded module reinstalls it. HMR therefore needs no developer-annotated acceptance boundaries, as opposed to Webpack [48] or Vite [49] HMR.

The `@cordisjs/hmr` component provides the HMR *engine*, which operates in three phases.

**Phase 1: Module classification.** The engine takes two inputs: the *stashed* set (file URLs whose contents have changed since the last reload) and the *externals* set (modules that cannot be hot-replaced and instead trigger a full restart). Writing `get_imports(url)` for the modules that `url` directly imports, it classifies the changes' dependency subgraph, marking each module `accepted` or `declined`:

**Algorithm 8** Module classification

```
1 function classify(stashed, externals)
2  | accepted ← stashed
3  | declined ← externals
4  | pending ← ∅
5  | for url in stashed do
```

```
6        pending ← pending ∪ (get_imports(url) \ (accepted ∪ declined))
7    repeat
8        progress ← false
9        for url in pending do
10           if get_imports(url) ∩ accepted ≠ ∅ then
11               accepted ← accepted ∪ {url}
12               pending ← pending \ {url}
13               progress ← true
14           else if get_imports(url) ⊆ declined then
15               declined ← declined ∪ {url}
16               pending ← pending \ {url}
17               progress ← true
18           else
19               pending ← pending ∪ (get_imports(url) \ (accepted ∪ declined))
20   until not progress
21   declined ← declined ∪ pending
22   return (accepted, declined)
```

Seeded with the imports of the stashed files, the fixed point accepts a module once one of its imports is accepted and declines one once all of its imports are declined; any module left undecided, caught in an import cycle, defaults to declined.

**Phase 2: Stale-entry detection.** Using `accepted` and `declined`, the engine then filters the component entries down to the *stale* ones, whose dependency tree reaches a changed module. It walks each entry's tree with `get_dependencies`, which collects the transitive imports of a module while respecting `declined` as a boundary:

**Algorithm 9** Stale-entry detection

```
1  function get_dependencies(root, declined)
2      deps ← ∅
3      function traverse(url)
4          if url ∈ deps or url ∈ declined then return
5          deps ← deps ∪ {url}
6          for child in get_imports(url) do traverse(child)
7      traverse(root)
8      return deps
9  function detect(entries, accepted, declined)
10     stale_entries ← ∅
11     for entry in entries do
12         tree ← get_dependencies(entry.url, declined)
13         if tree ∩ accepted ≠ ∅ then
14             accepted ← accepted ∪ tree
15             stale_entries ← stale_entries ∪ {entry}
16     return stale_entries
```

An entry is stale exactly when its tree intersects `accepted`; that tree is then folded into `accepted`, so every stale module along it is invalidated in the next phase.

**Phase 3: Transactional reload.** Finally, the engine reloads the stale entries. It invalidates the accepted modules' caches[3], backing up each removed module to enable rollback, then re-imports each stale entry's component module by its `url` and swaps in a fresh fiber:

**Algorithm 10** Transactional module reload

```
1   function reload(ctx, accepted, stale_entries)
2       backup ← invalidate_caches(accepted)
3       try
4           for entry in stale_entries do
5               entry.fiber.dispose()
6               entry.fiber ← ctx.use(import(entry.url), entry.config)
7       catch error
8           restore_caches(backup)
9           for entry in stale_entries do
10              entry.fiber.dispose()
11              entry.fiber ← ctx.use(backup[entry.url], entry.config)
12          throw error
```

The transactional guarantee ensures that the system never enters a half-reloaded state: if any module fails to import (e.g., due to a syntax error), the caches are restored and every stale entry is rebuilt from `backup[entry.url]`, the previous component whose cache was just restored, undoing the swaps already made.

### 5.3. Case Study: Koishi

Koishi is an open-source chatbot application framework built on Cordis[4]. Over four years of development, it has accumulated over 4000 community-contributed plugins[5], ranging from instant-messaging (IM) adapters and database drivers to administrative consoles and end-user features. Its scale and diversity make it a representative validation of Cordis's dynamic composability in a production setting.

**Expressiveness and generality of the meta-framework.** Koishi runs as a server-side bot whose every feature is realized as a plugin over the context primitives of Section 5.1; Koishi itself contributes only the chatbot-domain vocabulary. The same model reappears in a wholly different runtime: Koishi's web console is a second, independent Cordis application whose plugins compose the primitives of the browser and its user interface rather than those of the server. The disparate settings above establish two properties of the model of Section 3. (1) It is expressive: its primitives suffice to carry a complete production system, the host framework supplying only domain vocabulary. (2) It is general: it fixes how effects and coeffects compose while leaving their meaning to each application, and so presupposes neither a particular domain nor a particular runtime.

**Temporal composability without cognitive overhead.** The plugin systems surveyed in Section 1.2.1 cannot unload an individual extension's effects without restarting the extension host. Koishi routinely performs this operation: an orchestrator disables a plugin from the console and its effects are reverted in place; during development, the HMR engine re-applies edited plugins on save while preserving cache state and live connections elsewhere in the system. Cordis makes such removal not merely possible but effortless for the plugin author. Because effects performed through the context are tracked and their inverses composed automatically (Section 3.1), even an inexperienced author obtains ordered cleanup for a plugin's context-mediated effects without writing an uninstall path. This achieves the locality of concern whose absence Section 1.2.1 identifies: correctness that would otherwise rest on each author's diligence is instead discharged once, by the abstraction.

[3]On Node.js, this means clearing the caches of both the ES module and CommonJS module systems, since a module imported through the ES loader can appear in both.

[4]Koishi currently uses Cordis v3. This paper presents Cordis v4, which refines the effect and coeffect semantics and redesigns the loader; the core compositional model is shared across both versions.

[5]Koishi uses the term *plugin* for the concept this paper formalizes as *component*.

**Spatial composability across an open ecosystem.** In contrast to the plugin systems of Section 1.2.1, where inter-plugin dependencies are largely absent, Koishi's ecosystem exhibits a genuine dependency topology: IM adapters provide access to each messaging platform, database drivers provide persistent storage, and functional plugins declare these as coeffects and access them. Reconfiguring a provider at runtime, such as switching the storage backend or reconnecting an adapter, reactivates only the dependents whose resolved dependency changed (Section 3.2); a plugin whose dependency is unavailable stays inactive until it appears, without erroring. What the case study substantiates is that this composition holds across independently authored code: a plugin and its dependencies are typically written by different authors who coordinate on nothing beyond the coeffect that connects them, so reactive coeffects keep the assembly consistent across an open ecosystem of independent contributors.

**Threats to validity.** The evidence here is drawn from a single ecosystem in a single host language, so it cannot separate the merits of the paradigm from those of its TypeScript realization or of Koishi's particular domain, and it is observational rather than a controlled comparison against an alternative architecture. What the case study establishes is thus an existence-and-adoption result rather than a quantitative one; measuring the abstraction's overhead and its effect on developer productivity against a baseline remains future work.

## 6. Discussion

The formal model and implementation presented in the preceding sections introduce a programming paradigm for dynamic composability. This section examines how the paradigm extends to broader engineering concerns, and discusses the design tensions and open problems.

### 6.1. System Boundary

Every effect in Section 3.1 carries an inverse, and what that inverse amounts to is settled by the *system boundary*. The boundary divides the environment a system runs against into two parts. (1) A location lies *inside* when the system is able to modify it exclusively and to restore the state before that modification, so an operation on it is tracked in $\Gamma$ and can be reverted later. (2) A location lies *outside* when either ability fails, so an operation on it acts as $\mathrm{id}_\Gamma$ and is therefore neither tracked nor reverted. This section develops the properties of this boundary and their consequences for recovery.

**Boundaries from coeffects.** A coeffect moves the boundary by reifying an external location: it confines every access to that location to a set of operations it provides, each of which it can supply an inverse for, so operations that acted as $\mathrm{id}_\Gamma$ come to be tracked in $\Gamma$ and reverted. The boundary is therefore drawn per location rather than per medium, since both aforementioned abilities are properties of a location, and reification changes how a location is accessed while leaving its medium as it was. For example, a memory region lies inside when the system alone writes it, and outside when other processes write it too; a file lies inside when only the system can reach it, as with a scratch file under a private path, and outside when it is a path other programs read or write. Moving the boundary is itself a trade-off, between whether the environment provides revertible semantics for a location and what supplying those semantics costs on every access. We take up the co-design this suggests in Section 6.7.

**Acquisition and emission.** An operation that reaches outside the boundary generally proceeds in two stages. (1) In the *acquisition* stage, the operation obtains access and installs a record inside the boundary: `open` installs a descriptor that `close` removes, `malloc` reserves a block that `free` releases, `fork` starts a child process that `kill` terminates. The record itself is part of the coeffect that reifies the location, e.g. an entry in a map it keeps, and installing that entry is a revertible effect. That record is at the same time the *channel* along which data can leave. (2) In the *emission* stage, the operation pushes data through that channel, as with the bytes a `write` hands to the file or the datagram a `send` puts on the wire, and the push acts as $\mathrm{id}_\Gamma$, leaving the data where other parties may read and write it. The two stages therefore fall on opposite sides of the boundary: the acquisition stays inside it, whereas the emission crosses to the outside.

**Withholding and compensation.** A system that must nonetheless recover from an emission has two approaches available. One is to withhold an emission until the state that produced it is certain to persist, which is the *output commit problem* of rollback-recovery [50]. The other is *compensation* [51]: an action that restores the state up to an equivalence the application supplies, coarser than the $\simeq$ of Definition 33, as in deleting a file that was created or refunding a charge that was made. Such actions compose in the same LIFO order as inverses do, so the composition of Section 3.1 transfers to them. The metatheory does not: the commutation of Definition 65 is proved against $\simeq$ and has to be re-established against the coarser one.

### 6.2. Service Multiplexing

Dynamic component platforms such as OSGi [52] organize composition around *services*: units of functionality that a provider publishes under an interface and a consumer binds to. The Cordis coeffect model echoes this notion, with a service corresponding to the interface behind a key. Components that `provide` a service are its *providers*, and components that `inject` a service are its *consumers*. A single service may be implemented by multiple providers, and this multiplicity can be realized in two forms. (1) *Exclusive binding*: several implementations share one interface but at most one is bound at a time; the orchestrator selects which implementation is bound, and switching between them requires unloading one provider and loading another, momentarily perturbing every consumer's dependency. (2) *Service broker*: a central service that acts as the entrypoint for the interface is injected by both the backing providers and the consumers, so that multiple providers coexist and the broker dispatches each request among them. Compared to exclusive binding, the broker absorbs this perturbation: updating a backing provider leaves the broker in place, so consumers see no change to their dependency and no reload is triggered.

The service broker underlies three capabilities: load balancing, rolling updates, and cross-process invocation.

**Load balancing.** When several providers coexist, the broker distributes requests among them according to a configurable policy (e.g., round-robin, least-loaded, latency-weighted) or an explicit target named by the consumer. Because providers are ordinary components, they can be added or removed to scale capacity up or down; each provider registers with the broker through a revertible effect, so unloading it reverts the registration and drops it from the broker's routing set automatically.

**Rolling updates.** Upgrading a service implementation at runtime reduces to a controlled provider transition [53, 54]. To carry out the transition, the new provider is loaded as an additional fiber and registers with the broker; once it becomes `ACTIVE`, traffic is gradually shifted from the old providers to the new one (e.g., by adjusting selection weights), and the old providers are unloaded once they no longer carry in-flight requests. This provider transition turns what is traditionally an infrastructure-level operation (e.g., container orchestration, blue-green deployment) into an application-level composition pattern.

**Cross-process invocation.** The service broker can also be applied across process boundaries [55]. Each process hosts its own Cordis context with local providers; a coordinating component links them, treating each as a remote provider. Cross-process service access is mediated by an RPC mechanism that preserves the interface, making the distribution transparent to consumers. One caveat is that a cross-process call incurs latency and may fail mid-flight, so exposing it synchronously would block the caller. An interface intended to be exposed across processes must therefore be designed against an asynchronous contract.

### 6.3. Access Control and Sandboxing

Given an application assembled from independent components, securing the application calls for two complementary mechanisms: (1) constraining what dependencies a component may access, and (2) sandboxing untrusted code from the host environment. Cordis supports the first through dependency declarations and interception; the second requires an external sandbox.

**Capability-based access control.** The dependency access mechanism (Section 5.1.4) already constitutes a form of access control over proxy-mediated properties: a component can only access dependencies it has declared; an undeclared access raises an error. This is structurally similar to capability-based security [56–58], where authority is conferred by possession of a reference rather than by ambient authority. The `inject` declaration acts as a capability request, and the context proxy acts as a capability mediator. Since these requests are declared statically, the complete set of proxy-mediated capabilities a component requires is known before it runs, letting the orchestrator review and approve them at load time rather than discovering accesses as they happen.

This mediation generalizes to fine-grained policy through the interception mechanism. Access-control metadata can be carried by contexts or declared by components (Definition 26), and the provider consults it when the dependency is invoked to decide whether a request is permitted. For example, a filesystem dependency may carry metadata declaring which paths a component may read or write, and the provider checks each call against the metadata. Because this interception lives on the context rather than in either party's code, an orchestrator can adjust it to constrain any component's access to a dependency without modifying the provider, e.g., granting read-only database access to a community component whereas a core component retains full access. Moreover, since interception affects only how a dependency is invoked, not whether it is satisfied, it can be installed, reconfigured, or removed at runtime without triggering any reload or perturbing the dependency graph.

**Sandboxing untrusted components.** When a component's code cannot be trusted, language-level access control is insufficient, since a malicious component with access to the host runtime can reach the underlying objects directly, rendering such checks moot. Sandboxing requires an execution boundary beyond the reach of language-level means, such as software fault isolation [59], a separate language runtime, a sandboxed process, or a virtualized container [60]. Whatever the mechanism, the untrusted component runs in its own sandboxed context and reaches host-provided dependencies through a bridge, generalizing the cross-process invocation of Section 6.2: the same transparency argument renders this bridged access indistinguishable from local injection. On the host side, the bridge is an ordinary fiber whose capabilities can be attenuated by the access control described above.

### 6.4. Language Independence and Selection

Although Cordis is implemented in TypeScript, the context paradigm is language-agnostic: spatiotemporal composability is defined only by its two composability dimensions, and thus can be realized in any language that meets certain requirements along both. We analyze these requirements along each dimension in turn.

**Temporal composability.** At its most basic, temporal composability requires *closures*: a revertible effect pairs an action with an inverse, and that inverse must be captured as a value, along with the state it restores, so it can be replayed on teardown. Beyond this, a component's code and the side effects of loading it must be introducible and retractable at runtime.

How a language meets this second requirement depends on its execution model. In managed runtimes, this takes the form of a programmatic module registry, where a loaded module can be evicted from the registry and garbage-collected once unreferenced; Node.js, for instance, exposes such a registry.[6] Native code exposes no module registry, so introduction and retraction take the form of explicit dynamic linking and unlinking (e.g., `dlopen`/`dlclose` on Unix, `LoadLibrary`/`FreeLibrary` on Windows) [61], i.e., loading object code into a running process and later detaching it. WebAssembly takes one path or the other depending on its embedder: a module instance is reclaimed by the host's collector under a managed embedder (e.g., a JavaScript host), or released when a native embedder drops it (e.g., Wasmtime). Across these mechanisms, the revertible effects model treats loading as an effect on the context, with inverses that undo the registration of symbols, types, or handlers the module introduced.

**Spatial composability.** Spatial composability requires a mechanism for components to declare their dependencies and for the runtime to provide and inject these dependencies. This reduces to a dependency injection (DI) problem [39], which manifests at two levels that differ across languages: how dependencies are *typed* and how their access is *mediated*.

At the type level, the language should provide a way for developers to express well-typed dependency access. A consumer obtains a coeffect by reading its key from the context, so the context type (Section 3.2.1) must record each key's coeffect. Typeclasses (Haskell) [62] and traits (Rust) [63] achieve this by letting a provider extend the context type from its own module through an `instance` or `impl` [64]. TypeScript's module augmentation [65] likewise lets a provider module merge declarations into the context type.

At the runtime level, dependency access must be dynamically mediated: the coeffect behind a key may change as providers are loaded and unloaded, and may be resolved differently across contexts. The language therefore needs a way to interpose on access transparently, leaving the consumer's code unchanged, e.g., via JavaScript's `Proxy` object [66] or Python's descriptor protocol (`__get__`) [67]. Absent such a primitive, runtime reflection [68, 69] can mediate access dynamically, at the cost of type safety and developer experience.

[6] CommonJS exposes the module cache via `require.cache`; ES modules provide no public eviction API, though modules can still be managed through engine-internal interfaces.

Across both levels, metaprogramming facilities supply the typing and the mediation together. Annotations [70] and decorators attach metadata to a declaration, which a processor expands into the accessor that mediates access; compile-time metaprogramming (e.g., Rust procedural macros, Scala macros [71], Zig comptime) emits, for each dependency, a typed declaration together with such an accessor, dispensing with a general-purpose interception primitive.

### 6.5. Mutual Dependencies and Component Granularity

In the reactive coeffect model, a dependency cycle simply leaves the involved components permanently inactive: given two components $A$ and $B$, if $A$ requires a key provided by $B$ and $B$ a key provided by $A$, neither's satisfaction predicate can ever become true. Unlike deadlock in concurrent systems, which depends on the schedule and must be detected as it happens, this condition is predictable from the dependency declarations alone, so a runtime can report it when components are loaded.

In practice, most apparently mutual dependencies can be decomposed into finer-grained components that eliminate the cycle. Consider two components: a server (providing a network interface) and an access controller (enforcing authorization policies). The two components interact bidirectionally: the access controller mediates requests arriving at the server, and the server exposes an endpoint for modifying access-control policies. A monolithic design would make each component depend on the other. However, the two interaction directions are logically independent concerns. Decomposing them yields four components: server-core, access-control-core, request-mediation (depending on both cores to apply access control to incoming requests), and policy-management (depending on both cores to expose policy modification via the server). Through this approach, the cycle is eliminated because neither core depends on the other; only the integration components depend on both.

This decomposition is always possible in principle, since every bidirectional interaction can be factored into independent unidirectional bindings, but it increases the number of components: in the general case, given $n$ mutually interacting components, the number of integration components can grow quadratically with $n$, since each pair of interacting components may require a distinct component for each direction of interaction. This does not affect correctness or runtime performance (components are lightweight), and finer granularity can be beneficial: users gain the ability to load only the specific integration bindings they need, effectively increasing the system's composability. However, it may affect developer experience: more components require more configuration, more naming, and more cognitive overhead in understanding the dependency graph.

Mitigating this granularity cost is an engineering concern rather than a theoretical one. Practical strategies include package bundling (i.e., grouping related fine-grained components into a single installable unit), convention-based wiring (i.e., automatically connecting components whose names or types match a pattern), and scaffold tooling (i.e., generating boilerplate integration components from declarative specifications). These strategies preserve the formal guarantees of the acyclic model while reducing the authoring burden to something closer to the monolithic case.

### 6.6. Dependency Typing and Versioning

In the formal model, a dependency link is established purely by key identity: a component providing key $k$ satisfies any component declaring $k$ in its dependency set. The type family $\mathcal{V}_k$ ensures type-level agreement within a single compilation unit, but this guarantee breaks down when components are developed and built independently, which is a common scenario in component ecosystems. This breakage leads to two distinct problems.

**Interface drift.** A provider may modify the interface associated with $k$ (adding fields, changing method signatures, altering behavioral contracts) between versions, while a consumer compiled against an earlier interface continues to declare the same key $k$. The dependency is satisfied at the coeffect level ($k \in \mathrm{dom}(\sigma)$), yet the runtime value no longer conforms to the consumer's expectations, leading to type errors, method-not-found failures, or silent behavioral divergence [72].

**Key collision.** Two independently developed providers may use the same key name $k$ to denote entirely unrelated interfaces. Since key identity alone establishes the link, a consumer expecting one provider's interface will accept the other's value without any compatibility check. Unlike interface drift, where the provider and consumer at least share a common lineage, key collision involves no relationship whatsoever between the expected and actual types, making the resulting failures unpredictable and difficult to diagnose.

Both problems point to the same gap: the coeffect model provides only *nominal* linking (by key name) but no *versioned* or *structural* linking (by interface compatibility) [73]. We discuss three approaches to the gap, from most infrastructure-coupled to most language-agnostic.

**Key namespacing.** Extending the key space from $K$ to $K \times P$, where $P$ identifies the interface-defining package, eliminates key collision by construction: independently developed interfaces with the same local name occupy distinct keys. This is the most direct solution but also the most coupled: it embeds the package namespace into the formal model itself, making the system dependent on an external package registry for key identity.

**Peer dependencies.** A lighter coupling is to declare version constraints through the host-language package manager [74]. This is the approach Cordis currently adopts. Component dependencies are semantically *peer dependencies*: a component does not bundle its dependencies internally but expects the runtime context to supply them. Package managers with peer dependency support (e.g., npm) can enforce version compatibility: if the version of the package providing a key falls outside a consumer's declared peer range, the incompatibility is caught at install time rather than surfacing as a runtime failure. However, this approach has two limitations: (1) it depends on providers faithfully adhering to semantic versioning, which is an unenforceable convention; (2) package managers typically resolve each dependency to a single version, which prevents loading components from multiple versions of the same package within one application.

**Structural compatibility.** A fully language-agnostic approach would replace the membership check $k \in \mathrm{dom}(\sigma)$ with a compatibility predicate that verifies the provider's actual interface structurally subsumes the consumer's expectation. This is analogous to structural subtyping [75]: a provider satisfies a consumer if the provided interface is a subtype of the required interface. The challenge lies in defining this predicate language-agnostically: structural compatibility is straightforward for record types (width subtyping) but becomes complex for behavioral contracts (e.g., pre/postconditions [76], effect specifications [22]), and undecidable once parametric polymorphism introduces bounded quantification [77].

These three approaches address different aspects of the problem. Designing a unified dependency model that combines these approaches while preserving the dynamic composition guarantees of the coeffect model remains an open problem.

### 6.7. Co-Design with Languages and Operating Systems

Section 6.4 identifies the minimum a host language must supply for the context paradigm. This section takes up the converse question, what a language or operating system co-designed with the paradigm can offer beyond that minimum.

**Co-design with languages.** A language designed around the context paradigm can improve on a library in two respects: the semantics it gives to contexts, and the primitives it gives to effects and coeffects.

Such a language can make the context implicit again while preserving the context semantics of Section 3.3. An imperative language already runs every statement against an implicit context, and that single context neither tracks effects nor resolves coeffects. The context paradigm instead distinguishes multiple contexts, where an operation either modifies the context it runs against or derives another from it (Definition 23). An in-place realization modifies the ambient context, just as an imperative language does. A derived realization instead introduces a separate context, for which the language must provide a construct. Making the context implicit brings both an ergonomic and a safety benefit. (1) In a library realization, every function involving effects or coeffects takes the context as an ordinary argument or a receiver, as in Section 5.1. Where the language supplies the context implicitly, functions no longer need to take it. (2) Every context carries its own lifecycle state and committed view (Section 4.1). A library realization passes a context as an ordinary variable, so a component may reach another component's context by mistake, through a closure or a global variable. An effect it installs there then leaks out of its own lifecycle, and a coeffect it reads there escapes its dependency specification. Making the context implicit closes both.

Such a language can also make effects and coeffects known to its compiler. (1) For effects, an effect iterator (Definition 17) allocates a closure at every step to hold the inverse together with the state it restores. With syntax for performing an effect, a compiler can emit a single state machine for the whole iteration and hold those inverses in its frame. (2) For coeffects, the coeffect specification can be admitted into the type system, with two benefits. First, a dependency cycle is reported at compile time instead of being left to the runtime (Section 6.5). Second, a dependency can be compared by the structure of its type rather than by key identity alone, as row types do [28], which is type-level support for the structural compatibility of Section 6.6.

**Co-design with operating systems.** Section 1.2.3 observes a coarse-grained substitute for dynamic composability, where the operating system supplies temporal composability at the granularity of a process, and the container orchestrator above it supplies spatial composability at the granularity of a service. An operating system co-designed with the paradigm would support fine-grained composition, by making the coeffect specification a component declares the whole of what it can reach, and by providing its own resources as coeffects.

Such an operating system can supply the sandbox that Section 6.3 defers to a mechanism outside the language. It does so by bounding a component to the dependencies it declares, supplying them when the component is loaded and leaving nothing else reachable from within it, as a WebAssembly module receives its imports from its embedder at instantiation [78]. It can also provide the coeffect isolation and interception of Section 3.2.3 as abilities of its own, binding a key differently for each component and mediating the accesses it supplies.

Such an operating system can also provide its own resources as coeffects. A resource lying outside the boundary is made revertible where the runtime records each acquisition against the component that made it (Section 6.1), and every runtime keeps a record of its own. An operating system that provides the resource as a coeffect keeps that record once, since it is the party that hands the resource out and can attribute it to the component that asked. Memory and file descriptors are the immediate candidates, and tracking them for the sake of recovery has been done at the kernel interface [79, 80]. Furthermore, an operating system can make revertible some of the operations Section 6.1 can only withhold or compensate for. A system that performs a write to persistent storage transactionally can roll it back [81], and one built on copy-on-write or immutable storage reaches an earlier state by moving a pointer [82, 83].

# 7. Related Work

Dynamic composability intersects several established research areas. We survey the most relevant lines of work and distinguish our contribution from each of them.

## 7.1. Effect and Coeffect Systems

Section 2 reviewed effects and coeffects as the theoretical pillars underlying our work. We first situate the monadic effect systems now common in industrial practice, then survey three research lines that extend effects and coeffects in directions relevant to Cordis: recasting algebraic effects as capabilities, giving effects a reversible semantics, and unifying effects and coeffects under a single graded discipline.

**Monadic effect systems.** One family of libraries encodes effects in the type systems of existing general-purpose languages, representing them as monadic values that a runtime executes. ZIO in Scala [84] models a computation as `ZIO[R,E,A]` and Effect-TS in TypeScript [85] as `Effect<A,E,R>`, a generic type whose parameters describe its result, its typed errors, and the services its context must supply; the fp-ts library [86] encodes the same error and requirement channels through Reader-based monad transformers. Two traits separate these systems from Cordis. First, the tracking costs a monadic embedding: a program obtains it only by being written inside the effect type, whereas Cordis tracks effects as an overlay over ordinary host code. Second, a requirement is discharged by interpretation, an installed service that supplies its operations, and when that service is withdrawn what its operations performed remains in place; Cordis instead pairs each effect with an inverse and re-resolves requirements as providers come and go (Section 3.1, Section 3.2).

**Algebraic effects as capabilities.** Algebraic effects (Section 2.1) make effect operations visible to the type system. The extension closest to our work is Brachthäuser et al.'s Effekt language, which reinterprets effect types as *capabilities* [87, 88]: an effect type expresses what a computation requires from its context rather than what side effects it may produce. This perspective, like ours, treats the context as a mediator of capabilities. Cordis and Effekt differ in two respects. (1) In purpose, algebraic effects make effects visible to enable *modular interpretation*, giving one operation many handler semantics, whereas Cordis makes them visible to enable *tracking and reversion*, pairing every context transformation with an inverse. (2) In setting, Effekt disciplines effects statically at the type level, defaulting to scope-based reasoning in which capabilities are second-class and confined to their lexical scope, and recovering first-class use through boxing, which lifts that restriction by tracking captured capabilities in types; Cordis instead disciplines effects at runtime, aiming at complete resource recovery on component removal; Section 6.7 takes up what a language that made the context second class in this sense would offer.

**Reversible effect semantics.** A parallel line gives effects a reversible semantics rather than an interpretive one. Heunen et al. [89] model side effects in a reversible setting by adapting Hughes' arrows to *dagger arrows* and *inverse arrows*, capturing effects such as serialization and mutable store whose operations admit inverses. This is the formal account closest to our revertible effects: both pair each effect with the means to undo it rather than discharging it through a handler. The two differ in where reversibility resides, and in how much of it they demand. Heunen et al. work in a denotational, categorical setting where reversibility is a global property, guaranteed by construction since every computation is invertible, and the inverse is two-sided and recovered from the categorical structure. Cordis tracks inverses at runtime and requires less of them: not that the whole computation be reversible, but that each atomic effect admit a one-sided inverse, supplied by the caller at the point of application rather than derived, from which the inverse of any composite follows by composition (Section 3.1).

**Graded types as unified effects and coeffects.** Orchard et al. [90] proposed *graded modal types* as an umbrella notion encompassing both effect reasoning (via graded monads) and coeffect reasoning (via graded comonads), realized in the Granule language, demonstrating that a single type system can track both what a computation does and what it needs; more recent work extends coeffects to imperative Java-like languages [91, 92] and to call-by-push-value [93]. All of these operate at the type level: effects and coeffects are static annotations checked at compile time over lexically fixed scopes. Our contribution is orthogonal to this analysis: we lift the same two notions to runtime mechanisms, which lets Cordis handle dynamic composition. Temporal retraction and spatial dependency are re-resolved as the set of loaded components evolves, instead of being settled once over a fixed program text.

### 7.2. Programming Paradigms

The context paradigm (Section 3.3) mediates every effect and coeffect through an explicit context. This section first compares it with the functional and imperative treatments of side effects, and then with two established paradigms, one sharing our terminology and the other our treatment of crosscutting concerns.

**Explicit threading and implicit mutation.** Purely functional languages make effects explicit in types: the `State` monad $S \to (A, S)$ [23] threads the environment through every computation, securing equational reasoning at the cost of the threading itself, every function on the call path accepting and returning the state whether or not it touches it; the monadic effect systems of Section 7.1 are this pole in industrial form. Imperative languages leave effects and dependencies implicit at the call site, so reading what a call does to the system means reading its implementation transitively, and moving or removing a call may silently break distant invariants. The context paradigm takes the traceability of the first treatment and the ergonomics of the second: effects and coeffects pass through a context the component holds, so each operation is attributable to the context it was invoked on and hence to the component, and everything else stays ordinary host code. The paradigm is in this sense an overlay, realizable atop a language of either style: it fixes each operation's denotation and leaves its realization to the host language, in place where the host mutates and derived where it stays pure (Definition 23).

**Context-oriented programming.** COP [94, 95] equips a language with *layers*—partial method and class definitions that are activated and deactivated at runtime according to the execution context, so that behavior adapts without the base code naming its context dependencies [96]. COP and Cordis coincide in treating context as a first-class, runtime-mutable entity and in activating and deactivating behavior dynamically, but the resemblance is nominal. In COP, "context" denotes the ambient execution situation (e.g., location, user, mode), and activation changes method dispatch within a dynamically scoped extent; a layer neither tracks the side effects it induces nor reverts them, and activation is not governed by dependency satisfaction. In Cordis, the context is the $\Gamma_\infty$ entity mediating effects and coeffects: activation runs a component's revertible effects and is driven by reactive coeffect satisfaction (Section 3.2), and deactivation reverts them in full. COP varies what behavior runs; Cordis composes and reverts what effects and dependencies a component installs. Their difference is one of trade-off. COP folds activation into the host language's method dispatch, gaining dynamically-scoped layer extents at the cost of language specificity, whereas Cordis, as a language-agnostic overlay, resolves activation reactively over a shared context. Cordis can thus express as a coeffect only COP's global, value-driven fragment: context-dependent selection among implementations, but not dynamically-scoped activation.

**Aspect-oriented programming.** AOP [97, 98] modularizes a crosscutting concern into an *aspect*: a *pointcut* that quantifies over *join points* selected in the base program, and *advice* woven in at each. Cordis addresses the same problem of contextual behavior that would otherwise scatter across components, but its analogue of an aspect is a coeffect: a shared point of mediation many components declare a dependence on, so that crosscutting behavior can be reshaped there without editing any of them. The two paradigms then differ on two axes. (1) *Declaration versus obliviousness*: an AOP pointcut is oblivious and quantified, matching arbitrary join points whose code is unaware it is advised, whereas Cordis confines crosscutting to the coeffects each component declares, so its reach is exactly that declared surface. This yields determinacy and traceability: an application orchestrator can inspect and govern what cross-cuts a component at the configuration layer, without reading or analyzing its source, whereas an AOP concern is legible only through the aspects that quantify over it. (2) *Lifecycle integration*: a crosscutting change in Cordis is carried by a component's effects, reverted when the component unloads and propagated reactively to its dependents, so it is one move within the dynamic composition model; dynamic-AOP systems [99, 100] can also weave and unweave at runtime, but as a standalone operation, neither bound to a component's lifecycle nor triggering re-resolution among the advised code.

### 7.3. Temporal Composability

Temporal composability concerns replacing or removing a component in a running program while recovering the effects it installed. Prior approaches divide by how they treat a departing component's state and effects: carrying state forward to a successor version, recovering effects through developer-authored cleanup, reversing effects automatically within a scope fixed in advance, or reclaiming resources from a record the runtime accumulates by interposing on an interface.

**Stateful forward migration.** A broad family of systems replaces components in a running program without downtime by carrying their state forward across versions. All observe the same timing discipline: a component may be swapped only once it reaches a safe, interaction-free point. Kramer and Magee established this criterion as *quiescence* [53], which Vandewoude et al. later relaxed to the less disruptive *tranquility* [54]; our rolling-update pattern (Section 6.2) enforces it by draining in-flight requests before unloading a provider. Dynamic software updating (DSU) then migrates state forward through hand-written transformation functions: Hicks et al.'s general-purpose DSU for C [101], Stoyle et al.'s type-safe update points via con-freeness analysis [102], and Hayden et al.'s Kitsune [103] all map old-version data to new-version representations, inheriting heap objects, open files, and connections in place while re-initializing whatever is left unmigrated. The same discipline extends to persistent state: Overeem et al. [104] convert a running event store's data between schema versions through hand-written upgrade operations while keeping the system available. Erlang/OTP [15] takes the same stance at the process level, migrating state through `code_change/3` and recovering from faults by restarting supervised processes rather than reverting their effects; JavaScript's Hot Module Replacement (e.g., webpack [48], Vite [49]) does the same at the module level, handing state forward through the `module.hot` or `import.meta.hot` API across a reload. Compared with Cordis's module replacement (Section 5.2), these approaches migrate in-memory state more gracefully: Cordis reverts the old component's tracked effects and reapplies the new component's from a clean slate, so a component's own in-memory state does not survive a reload unless placed in a longer-lived dependency, and layering DSU-style forward migration atop revertible effects is future work. Cordis's approach is nonetheless more general in two respects: it needs no hand-written migration functions of the kind DSU and HMR require, and it supports unloading a component entirely and recovering its resources, not merely updating one in place.

**Developer-authored recovery.** A second family recovers a component's effects through cleanup or compensation logic that the developer writes by hand. Plugin lifecycle conventions (e.g., OSGi [52], Eclipse's extension points, IntelliJ and VSCode) delegate cleanup to developer-written unload callbacks; the Command pattern [105] encapsulates an operation together with an `undo` method for undo/redo stacks; the saga model [51] structures a long-lived transaction as steps each paired with a compensating action; algebraic effect handlers can attach finalizers that run on teardown [106]; and event sourcing [107] retracts state by appending compensating events rather than executing an inverse at all. In all of them the inverse is an unenforced duty, decoupled from the operation, so that a forgotten one leaks resources silently (as documented empirically in Section 1.2.1). React's `useEffect` hook [108] comes closest to pairing an effect with its inverse structurally, returning a cleanup the runtime invokes before each re-execution and on unmount. Its shortfall is composability: a hook may be called only at the top level of a component or another hook, never inside a conditional, loop, or nested function, and its effect body accepts neither an async function nor an iterator. Effects thus cannot be assembled from other effects or interleaved with control flow, leaving nothing from which a composite inverse could be derived. Cordis effects carry no such restriction: they are ordinary operations that compose freely and may run asynchronously, and require a hand-written inverse only for each atomic effect, from which the inverse of any composite is derived by composition, so that assembling existing effects requires writing no inverses at all. This structural pairing of every effect with its inverse makes complete recovery an invariant of the system rather than a matter of developer discipline.

**Statically scoped reversal.** A third family reverses effects automatically, by construction, but confines reversal to a scope fixed in advance. Software transactional memory [109, 110], descended from hardware transactional memory [111], records a read/write log so that a group of memory operations either commits or aborts, rolling memory back to its pre-transaction state. Reversible computing, from Landauer and Bennett's thermodynamic analyses [112, 113] to reversible languages such as Janus [114], goes further and makes every step of a whole computation globally invertible. Reversible process calculi build backtracking into the semantics itself: RCCS [115] carries a memory alongside each process and admits a step to be taken back when the past it leads to is causally equivalent, and Phillips and Ulidowski [116] derive reversible operators for CCS, ACP, and CSP uniformly while preserving their forward operational semantics. Their causal-consistency criterion is the concurrent counterpart of the order Cordis's recovery follows, an accumulator applying a component's own inverses in last-in-first-out order and the guard of Section 4.2.2 deferring a provider's withdrawal until its consumers have deactivated (Theorem 70). The reach, however, is fixed by the semantics, every action performed remaining undoable, whereas a Cordis component supplies an inverse for each atomic effect and its accumulator brings the context back to where its composition began. Linear types [117], RAII [4], and Rust's ownership system [63] tie a resource's release to a lexical region. Each fixes the scope and reach of reversal statically; Cordis, by contrast, fixes no such scope in advance: it reverts arbitrary context operations over a component's lifecycle, and treats lexical resource management as complementary, appropriate for local resources within a single component. Verification supplies the same pairing at the granularity of a data structure. Kim and Rinard specify and verify an inverse for every state-changing operation of a collection of set and map implementations, together with the conditions under which two operations commute, reasoning on abstract state so that orders leaving equivalent rather than identical structures count as commuting [118]. Both of the ingredients Section 3.4 requires of a coeffect are therefore mechanically checkable at the interfaces most keys publish, and their reason for preferring an inverse to a saved copy of the state is the one Section 3.1 acts on. Their inverses do not compose, each being verified for one operation, whereas a Cordis accumulator carries the inverses of a whole lifecycle.

**Interposed reclamation.** A fourth family reclaims what a component acquired without the component itself supplying the inverses, by recording its acquisitions at an interface the runtime controls. Nooks [79] wraps every call crossing the boundary between the Linux kernel and its loadable extensions, so that the kernel objects an extension touches pass through an object tracker whose record tells the recovery manager what to release when the extension fails; shadow drivers [80] tap the same calls from the other side, recording the requests and configuration that determine a driver's state so that a restarted instance can be restored to it. Akeso [119] obtains the record by compiler instrumentation instead, dividing kernel execution into nestable recovery domains that log their state changes and cross-thread dependencies, and rolling a faulting request back together with every domain that depends on it. Reclamation thus follows from a record the runtime maintains rather than from cleanup the developer remembers to write, which makes this family the closest systems-level precedent for revertible effects. It differs from Cordis in vocabulary and in reach. The platform fixes what can be recorded, whether as release code per kernel object type, one shadow per driver class, or an inverse per instrumented allocator, so a component may hold only resources the platform already knows how to release; a Cordis component instead introduces effects of its own and supplies an inverse for each atomic one (Section 3.1). Reclamation is likewise bounded by a request that commits or a restart of the same extension, whereas Cordis reverts over a component's whole lifetime and propagates removal to its dependents, which release their own effects in turn (Section 3.2).

### 7.4. Spatial Composability

Spatial composability concerns how a component's dependencies on others are declared and bound. Prior mechanisms divide by how binding responds to change: wiring dependencies once at initialization, reacting to the availability of whole components, or propagating change at the granularity of individual values.

**Initialization-time dependency wiring.** Two established mechanisms wire components together at initialization time. Dependency injection frameworks [39] (e.g., Spring [120], Guice, Angular, Inversify) inject dependencies into components at initialization, and UI framework context (e.g., Vue.js's `provide/inject` and React's Context API) passes them along a component tree. Some support dynamic scoping (e.g., Spring's prototype/request scopes, Angular's hierarchical injectors), but neither re-resolves reactively: when a provider is replaced or removed at runtime, existing dependents are neither deactivated nor re-initialized, and none offers lifecycle management of the kind our component state machine provides. Cordis's reactive coeffects (Section 3.2) supply this: the notification mechanism triggers lifecycle transitions whenever the satisfaction predicate changes.

**Availability-reactive component models.** The closest precedent to our reactive coeffects reacts to service availability. OSGi's Declarative Services and iPOJO [121, 122] let components declare provided and required services, with the runtime automatically activating and deactivating them as services appear and disappear; iPOJO's Gravity project [122] explicitly targets autonomous runtime adaptation to changing service availability, and its provide/require model directly prefigures Cordis's `ctx.provide/ctx.get` pattern. R-OSGi [55] extends the same abstraction transparently to distributed settings via RPC, mapping network failures to service-withdrawal events, a pattern Section 6.2 discusses as an extension of the Cordis model. All these systems recover through a deactivation callback, which is limited in two ways. First, the callback is hand-written, so resource safety rests on developer discipline and a forgotten one leaks silently. Second, the callback is synchronous: should teardown require an asynchronous exchange with the departing dependency, the frameworks offer no protocol to await it, forcing a blocking wait against a reference that may already be stale. Cordis's reactive coeffects close both gaps: deactivation reverts the dependents' accumulated effects, and its inertial `Unloading` state (Section 4.4) runs asynchronous teardown to completion before acting on further change.

**Value-level reactivity.** Functional reactive programming (FRP) [123] and its modern incarnations (e.g., signals [124, 125] in SolidJS, Vue's reactivity system, Angular Signals) propagate change at a *value-level* granularity: when a signal changes, derived computations are re-evaluated synchronously or under a scheduler [126]. Cordis's reactive coeffects act at a *component-level* granularity, adding asynchronous lifecycle semantics that value-level propagation does not model. The same granularity difference runs the other way for consistency: propagating in a turn, in an order the dependency graph fixes, lets FRP require that no derived computation read a mixture of updated and stale inputs, which is *glitch freedom* [127], whereas Cordis has no counterpart of a turn, orchestration actions arriving one at a time, and guarantees only that no single transition straddles two resolutions of its coeffects (Theorem 71). The two are complementary rather than competing: a Cordis coeffect can itself carry reactive values, and a component updates on only the parts it actually consumes, refining component-level reactivity into finer-grained reactive coeffects that span both levels.

## 8. Conclusion

We have presented a formal foundation for dynamic composability by lifting the classical concepts of effects and coeffects to runtime mechanisms. *Revertible effects* address local temporal composability: every context transformation carries an inverse that the runtime holds, and both tracking and recovery preserve composition, so the context is recovered upon component removal. *Reactive coeffects* address local spatial composability: every context change is classified against a component's coeffect specification as activating, deactivating, or neutral, and the classification drives its activation and deactivation. We then unify the effect context and the coeffect context into a single context type and mediate every effect and coeffect through it, yielding a discipline we call the *context paradigm*; the mediation induces an observational equivalence up to which the effects of distinct components attain independence. Combining these mechanisms into the notion of a *component*, we give a calculus of dynamic composition whose metatheory carries spatiotemporal composability from a single component to a whole system of interleaved components. We realize this paradigm as the *Cordis* meta-framework, with a core library providing effect tracking and coeffect resolution, as well as a declarative component loader with configuration reconciliation and hot module replacement. The Koishi case study validates the design of Cordis in a production system with over 4000 community plugins.

Beyond human-curated plugin ecosystems, a compelling direction for future validation is self-evolving agent harnesses (Section 1.2.2), where an AI agent generates and replaces its own harness components continuously and with little human oversight. Applying Cordis in such a setting would validate the temporal guarantees of complete recovery under rapid component replacement, as well as the spatial guarantees of dependency coordination under frequent topological change. Such validation would demonstrate the paradigm's applicability as a foundation for recoverable, coordinated, and continuous self-evolution in agent harnesses and other autonomous systems.